\documentclass[aps,prc,showpacs,showkeys,superscriptaddress,nofootinbib,twocolumn,floatfix]{revtex4-1}
\usepackage{bm}
\usepackage{placeins}
\usepackage[]{graphicx}
\usepackage{amsfonts}
\usepackage{bbm}
\usepackage{graphicx,color,amsmath,amssymb}

\def\blfootnote{\xdef\@thefnmark{}\@footnotetext}
\usepackage{amsmath, amssymb, graphics, setspace}

\newcommand{\mathsym}[1]{{}}
\newcommand{\unicode}[1]{{}}
\def\eg{{\em e.g.}}
\def\ie{{\em i.e.}}
\def\wrt{{\em wrt}}

\newcommand{\beq}{\begin{equation}}
\newcommand{\eeq}{\end{equation}}
\newcommand{\bea}{\begin{eqnarray}}
\newcommand{\eea}{\end{eqnarray}}
\newcommand{\lsim}{\raisebox{-4pt}{$\,\stackrel{\textstyle <}{\sim}\,$}}

\newcommand{\mlog}{\mathbb{L}\text{og}}

\begin{document}

\title{$T$-matrix Approach to Quark-Gluon Plasma}
\author{Shuai Y.F. Liu and Ralf Rapp}
\affiliation{Cyclotron Institute and Department of Physics and Astronomy, Texas A{\&}M University, 
College Station, TX 77843-3366, USA}

\date{\today}

\begin{abstract}
A selfconsistent thermodynamic $T$-matrix approach is deployed to study the microscopic properties of 
the quark-gluon plasma (QGP), encompassing both light- and heavy-parton degrees of freedom in a unified 
framework. The starting point is a relativistic effective Hamiltonian with a universal color force. 
The input in-medium potential is quantitatively constrained by 
computing the heavy-quark (HQ) free energy from the static $T$-matrix and fitting it to pertinent 
lattice-QCD (lQCD) data. The corresponding $T$-matrix is then applied to compute the equation of state 
(EoS) of the QGP in a two-particle irreducible formalism including the full off-shell properties of the 
selfconsistent single-parton spectral functions and their two-body interaction. In particular, the 
skeleton diagram functional
is fully resummed to account for emerging bound and scattering states as the critical temperature 
is approached from above. We find that the solution satisfying three sets of lQCD data (EoS, HQ free 
energy and quarkonium correlator ratios) is not unique. As limiting cases we discuss a weakly-coupled 
solution (WCS) which features color-potentials close to the free energy, relatively sharp quasiparticle 
spectral functions and weak hadronic resonances near $T_{\rm c}$, and a strongly-coupled solution (SCS) 
with a strong color potential (much larger than the free energy) resulting in broad non-quasiparticle 
parton spectral functions and strong hadronic resonance states which dominate the EoS when approaching 
$T_{\rm c}$.
\end{abstract}

\pacs{25.75.Dw, 12.38.Mh, 25.75.Nq}
\keywords{Luttinger-Ward Functional, $T$-matrix, Grand Potential, Equation of State, Quark-Gluon Plasma}
\maketitle

\section{Introduction}
\label{sec_intro}
Heavy-ion collision experiments at RHIC and the LHC create the hottest matter ever made by mankind, with 
temperatures more than 8 orders of magnitudes larger than the surface temperature of the 
sun~\cite{Braun-Munzinger:2015hba}. It is widely accepted that this matter evolves through a quark-gluon 
plasma (QGP), a de-confined phase of nuclear matter where the spontaneously broken chiral symmetry is 
restored.  The success of relativistic hydrodynamics in describing light-hadron 
spectra~\cite{Teaney:2000cw,Heinz:2009xj,Gale:2013da}, 
and the surprisingly large modification of heavy-flavor (HF) 
meson spectra~\cite{Prino:2016cni} have revealed the hot QCD medium to be a strongly coupled 
system~\cite{Shuryak:2014zxa}. 
However, it currently remains an open issue what the  microscopic mechanisms underlying the small 
viscosity-to-entropy density ratio and HF diffusion coefficient are, and what relevant degrees of 
freedom of the medium go along with it. It is quite conceivable that the nearby transition from 
quark-gluon to hadronic matter plays an essential role, and that large collision rates lead to 
nontrivial spectral functions of the matter constituents. These features are not readily captured 
by perturbative or quasiparticle approaches, see, \eg, Refs.~\cite{Blaizot:2003tw,Rischke:2003mt}
for reviews. 
On the other hand, the use of lattice-QCD (lQCD) motivated potentials, specifically the 
heavy-quark (HQ) internal energy, has led to the idea of a bound-state 
QGP~\cite{Shuryak:2004tx,Mannarelli:2005pz} as a ``transition" medium, with essential contributions 
from nonperturbative interactions, \ie,  remnants of the confining force. For heavy quarks these 
ideas have been implemented within a thermodynamic $T$-matrix 
approach~\cite{Cabrera:2006wh,vanHees:2007me,Riek:2010fk,Riek:2010py,Huggins:2012dj}, thereby connecting 
the open and hidden HF sectors. This framework has met fair success in understanding pertinent 
low-momentum HF observables in ultra-relativistic heavy-ion collisions (URHICs), and has reinforced the 
need for a more rigorous determination of the underlying 2-body interaction, rather than bracketing it 
by the free and internal energies which roughly correspond to a weakly and strongly coupled scenario, 
respectively. In a lQCD-based extraction~\cite{Burnier:2014ssa}, it was found that the static potential 
is close to the free energy while the associated imaginary part is near expectations from hard-thermal-loop 
perturbation theory.  In Ref.~\cite{Liu:2015ypa} the HQ free energy was calculated within the $T$-matrix 
formalism where the underlying potential was defined as the driving kernel in the corresponding integral 
equation. It was found that, in the presence of large imaginary parts of the static quarks, the 
lQCD data support a solution where the potential rises well above the free energy. Furthermore, 
implementing this potential in a selfconsistent quantum many-body framework (the Luttinger-Ward-Baym 
(LWB) formalism)~\cite{PhysRev.118.1417,Baym:1961zz,Baym:1962sx}, a description of the equation of state 
(EoS) of the QGP was achieved where parton spectral functions become very broad, losing their 
quasiparticle nature at low momenta, and the degrees of freedom change to broad hadronic states as 
the transition temperature is approached from above~\cite{Liu:2016nwj}.    

In the present paper, we expand on our previous studies by setting up a unified LWF formalism to 
investigate the microscopic properties of light, heavy and static degrees of freedom of the QGP, 
and firmly root it in information available from thermal lQCD.
Our starting point is an effective Hamiltonian in quark and gluon degrees of freedom with a color 
interaction of Cornell potential-type including relativistic corrections. While this approach reduces to 
potential non-relativistic QCD in the HQ limit we here pursue the question in how far the interactions 
encoded in the potential approximation (including remnants of the confining force) are relevant for 
understanding bulk and spectral properties of the QGP. We also note that the vacuum potential model using 
the Cornell interaction has met with fair success in light-hadron spectroscopy (with caveats for spontaneous 
chiral symmetry breaking and its Goldstone bosons)~\cite{Godfrey:1985xj,Capstick:1986bm,Lucha:1991vn}.   
We determine the input to our Hamiltonian by systematically constraining the interaction through the 
static HQ free energies, Euclidean correlators for charmonia and bottomonia, and the EoS in the light 
sector with 2 additional effective-mass parameters for light quarks and gluons. As mentioned above, a 
key feature of this approach is to retain the full off-shell properties of one- and two-body spectral 
functions (and scattering amplitudes), which renders the emerging micro-structure of the QGP a prediction 
of the formalism. Since the latter is directly formulated in real time, transport coefficients ($\eta/s$ 
or the HF diffusion coefficients, ${\cal D}_s$)~\cite{Liu:2016ysz} and other quantities of experimental 
interest (\eg, photon and dilepton production rates) can be readily computed. 
As it will turn out, the selfconsistent solution to the 3 sets of lQCD data is not unique. We will 
therefore discuss limiting 
cases of the underlying force strength, elaborate on the pertinent consequences for QGP structure and 
indicate ways to further constrain  the ``correct" scenario. 

This paper is organized as follows. In Sec.~\ref{sec_Tmatrix} we introduce the effective Hamiltonian 
and the 3-dimensional (3D) relativistic $T$-matrix approach used in this work. In Sec.~\ref{sec_lQCD} 
we lay out how the latter can be systematically constrained via various quantities computed in lQCD, 
namely: the EoS of the QGP using the LWB formalism (Sec.~\ref{ssec_EoS}) including a matrix-log 
technique to resum the skeleton diagram (2-body interaction) contribution, the static HQ free energies 
(Sec.~\ref{ssec_freeE}), and quarkonium correlators (\ref{ssec_correlator}) including interference 
effects in the imaginary part of the potential; 
in Sec.~\ref{ssec_ansatz} we introduce our ansatz for the in-medium potential 
(Sec.~\ref{sssec_pot}) and describe the concrete procedure for carrying out the overall 
selfconsistent fit (Sec.~\ref{sssec_proc}).  
In Sec.~\ref{sec_results} we show and discuss the main numerical results 
in comparison to lQCD data, specifically for what we will denote as a ``weakly coupled 
solution" (WCS, Sec.~\ref{ssec_wcs}) and a ``strongly coupled solution" (SCS, 
Sec.~\ref{ssec_scs}); each of these solutions is elaborated in four parts, pertaining to the 
potential extraction via fits to the HQ free energy (Secs.~\ref{sssec_wcs-pot} and 
\ref{sssec_scs-pot}), Euclidean quarkonium correlator ratios and associated quarkonium 
spectral functions (Secs.~\ref{sssec_wcs-corr} and \ref{sssec_scs-corr}),
the fits to the EoS and its (change in) underlying degrees of freedom 
(Secs.~\ref{sssec_wcs-eos} and \ref{sssec_scs-eos}), and the resulting parton spectral 
functions in heavy and light sectors with corresponding 2-body $T$-matrices 
(Secs.~\ref{sssec_wcs-spec} and \ref{sssec_scs-spec}). 
In Sec.~\ref{sec_sum} we summarize our findings and outline future directions and opportunities within
our approach. In the Appendix we collect further information on more general aspects of the relativistic 
potential approach (App.~\ref{app_rel-pot}), generalized thermodynamic relations within the LWB formalism for an effective in-medium Hamiltonian (App.~\ref{app_lwb-muq}), additional relations involving the static-potential limit (App.~\ref{app_static}) and a discussion of interference effects in its imaginary part (App.~\ref{sec_imv}).

\section{Thermodynamic $T$-Matrix}
\label{sec_Tmatrix} 
Bound states are key entities of the nonperturbative physics of a quantum system, especially 
in QCD where the hadrons encode the phenomena of confinement and mass generation.
In diagram language, bound states require an infinite resummation of (ladder) diagrams, represented 
by an integral equation 
such as the 4D Bethe-Salpeter (BS) equation~\cite{Salpeter:1951sz} or a 3D reduced $T$-matrix 
equation~\cite{Blankenbecler:1965gx,Thompson:1970wt,Woloshyn:1974wm}. 
Both equations allow for a simultaneous and straightforward treatment of scattering states. As a 
resummed series, the solution of the integral equation analytically 
continues to the strongly coupled region.\footnote{\label{foot1} The series \(1+\alpha+\alpha^2\cdots=1/(1-\alpha)\) 
is convergent for strong coupling. Divergence at strong coupling is different from the \(N!\) divergence 
of a perturbative series at small coupling~\cite{Dyson:1952tj,Weinberg:1996kr}.} This equation is 
therefore well suited to study the strongly coupled QGP (sQGP) near \(T_c\) where both bound and 
scattering states are expected to be important and entangled with each other 
in the presence of strong quantum effects, \ie, large scattering rates. 
Applications of the $T$-matrix approach in media has been carried out in various contexts, mostly 
in non-relativistic many-body systems~\cite{kadanoff1962quantum,kraeft1986quantum,Pantel:2014mao} 
but also in systems 
where relativistic effects are relevant~\cite{kapusta2006finite}, \eg, the nuclear many-body 
problem~\cite{Brockmann:1990cn,SCHMIDT199057}, hot hadronic matter~\cite{Rapp:1995py} or the 
QGP~\cite{Mannarelli:2005pz,Lacroix:2012pt,Lacroix:2015uta,Liu:2016nwj,Liu:2016ysz}.

In the present work our starting point is a Hamiltonian with relativistic dispersion relations and potential,
which maps onto the Thompson scheme~\cite{Thompson:1970wt} for the 3D reduction from the BS to the $T$-matrix 
equation (as employed earlier in the HQ sector~\cite{Riek:2010fk}).
It can be written in the form
\begin{align}
&H=\sum\varepsilon_{i}(\textbf{p})\psi_{i}^\dagger(\textbf{p})\psi_i (\textbf{p})+\nonumber\\&
\frac{1}{2}\psi_{i}^\dagger(\frac{\textbf{P}}{2}-\textbf{p})\psi_{j}^\dagger(\frac{\textbf{P}}{2}+\textbf{p})V_{ij}^{a}\psi_{j}(\frac{\textbf{P}}{2}+\textbf{p}')\psi_{i}(\frac{\textbf{P}}{2}-\textbf{p}')
\label{Hqgp}               
\end{align}
where $\varepsilon_{i}(\textbf{p})=\sqrt{M_{i}^2+\textbf{p}^{2} }$ and $\textbf{P}$ is the total momentum of the 2-particle state.  
The summations over $i,j$ include  momentum, spin, color, and particle species 
(3 light-quark flavors and gluons for the bulk matter description, 
or charm and bottom flavors for pertinent correlation functions). The index ``$a$" specifies the two-body 
color channels.  In this paper, we do not account for spin-dependent interaction, which are expected to 
be subleading but can be included in the future.
For the potential, $V$, we include both color-Coulomb ($V_\mathcal{C}$) and (remnants of the) confining (``string") 
interaction ($V_\mathcal{S}$), 
\begin{equation}
V_{ij}^{a}(\textbf{p},\textbf{p}')=\mathcal{R}^\mathcal{C}_{ij}\mathcal{F}^\mathcal{C}_{a}V_\mathcal{C}(\textbf{p}-\textbf{p}')+\mathcal{R}^\mathcal{S}_{ij}\mathcal{F}
^\mathcal{S}_{a}V_\mathcal{S}(\textbf{p}-\textbf{p}')
\label{eq_potential}            
\end{equation}
Relativistic effects in the vertices of the 4D theory are included by introducing 
relativistic factors \(\mathcal{R}\)~\cite{Brown:2003km,Riek:2010fk}    
\begin{align}
&\mathcal{R}^\mathcal{C}_{ij}=\sqrt{1+\frac{p^{2}}{\varepsilon_{i}(p)\varepsilon_{j}(p)}}\sqrt{1+\frac{p'^{2}}{\varepsilon_{i}(p')\varepsilon_{j}(p')}}\label{eq_Bfator}\\
&\mathcal{R}^\mathcal{S}_{ij}=\sqrt{\frac{M_{i}M_{j}}{\varepsilon_{i}(p)\varepsilon_{j}(p)}}\sqrt{\frac{M_{i}M_{j}}{\varepsilon_{i}(p')\varepsilon_{j}(p')}}
 \ , 
\label{eq_Rfator}
\end{align}
and $\mathcal{F}^\mathcal{C,S}$ are color factors in diagonal representation; specifically, the Coulomb  
factors, $\mathcal{F}^\mathcal{C}$, 
are the standard Casimir coefficients~\cite{Shuryak:2004tx,Riek:2010fk} collected in Table~\ref{table_casimir},
while for the string factors, $ \mathcal{F}^\mathcal{S}$, we take the absolute values of the Casimir 
coefficients, to ensure a positive
definite string tension, which appears to be weaker in colored channels~\cite{Petreczky:2004pz}. 
The precise form of $V_\mathcal{C}$, $ V_\mathcal{S }$ and the parton mass values, \(M_i\), are inputs to 
the Hamiltonian that need to be constrained by the lQCD data to be discussed in the following sections.
\begin{table}[!htb]
	\centering
	\resizebox{0.3\textwidth}{!}{  
		\begin{tabular}{|c|c|c|c|}\hline
			\large\(qq\) &\large\(q\bar q\) &\large $(q/\bar q) g$ &\large $gg$  \\\hline		
			( 1/2, 3) & ( \space\space\space1, 1) &( 9/8, 3 ) &  ( 9/4, 1 )\\\hline
			(-1/4, 6)& (-1/8, 8) & ( 3/8, 6 )&( 9/8, 16)   \\\hline
			&  	& (-3/8, 15)&(-3/4, 27)  \\\hline
	\end{tabular}}
	\caption{Casimir and degeneracy factors for different color channels quoted as (Casimir factor,degeneracy).}
	\label{table_casimir}
\end{table}
\begin{figure}[!htb]
	\begin{center}
		\includegraphics[width=0.99\columnwidth]{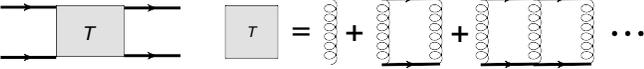}
		\caption{\(T\)-matrix resummation for ladder diagrams}
		\label{fig_Tmatrix}
	\end{center}
\end{figure}
 
The finite-temperature calculations are carried out in the Matsubara formalism where the ``bare" 
propagators for both quarks and gluons are taken as
\begin{align}
G^0_{i}(i\omega_n,\textbf{p})=\frac{1}{i\omega_n-\varepsilon_{i}(\textbf{p})} \ . 
\label{eq_G1bare}
\end{align}
We resum the ladder diagrams of the Hamiltonian by the \(T\)-matrix equation, pictorially
displayed in Fig.~\ref{fig_Tmatrix}. In the center-of-mass (CM) frame it can be written as
\begin{align}
&T_{ij}^{a}(z,\textbf{p},\textbf{p}')=V_{ij}^{a}(\textbf{p},\textbf{p}')+
\nonumber\\
&\int_{-\infty}^{\infty}\frac{d^3\textbf{k}}{(2\pi)^3}V_{ij}^{a}(\textbf{p},\textbf{k})
G^{0}_{ij}(z,\textbf{k})T^{a}_{ij}(z,\textbf{k},\textbf{p}') 
\label{eq_BSE3D}
\end{align}
where $z=i E_{n}$ is the two-body Matsubara frequency (or analytical energy variable 
$E\pm i\epsilon$), and \(\textbf{p},\textbf{p}'\) are the incoming and 
outgoing 3-momenta, respectively, for each parton in the CM frame, \ie, for total momentum \( \textbf{P}=0\);  
\(T_{ij}^{a}(z,\textbf{p},\textbf{p}')\) denotes the \(T\)-matrix between particle type \(i\) and \(j\) 
in color channel \(a\).
The two-body propagator is defined in Matsubara representation as
\begin{equation}
G^{0}_{ij}(iE_n,\textbf{k})=-\beta^{-1}\sum_{\omega_n} G_i(iE_n-i\omega_n,\textbf{k})G_j(i\omega_n,\textbf{k}) \ , 
\end{equation}
and, using a spectral representation, can be written in terms of single-particle spectral functions as
\begin{align}
&G^{0}_{ij}(z,\textbf{k})=\int_{-\infty}^\infty d\omega_1 d\omega_2\nonumber\\
& \times \frac{(1\pm n_{i}(\omega_{1})\pm n_{j}(\omega_{2}))}{z-\omega_1-\omega_2 }\rho_{i}(\omega_1,\textbf{k})\rho_{j}(\omega_2,\textbf{k}) \ 
\label{eq_G2define}
\end{align}
with the single-particle propagators
\begin{align}
&G_{i}(z)=\frac{1}{[G^0_{i}(z,k)]^{-1}-\Sigma_{i}(z,k)}=\frac{1}{z-\varepsilon_{i}(p)-\Sigma_{i}(z,k)},\nonumber\\
&\rho_{i}(\omega_,\textbf{k})=-\frac{1}{\pi}\text{Im}G_i(\omega+i\epsilon) \ .
\label{eq_G1define}
\end{align}
In Eq.~(\ref{eq_G2define}) the $ \pm $ sign refers to bosons (upper) and fermions (lower),  and $ n_i $ is 
the Bose or Fermi distribution function for parton $ i $. The in-medium selfenergies, \(\Sigma_{i}(z,k)\), 
will be selfconsistently computed through the 2-body \(T\)-matrix, as detailed below.
 
In vacuum it is sufficient to solve the \(T\)-matrix in the CM frame due to Lorentz invariance. 
However, in medium, Lorentz invariance is in general broken, although usually not by much for
the scattering amplitude at total momenta comparable to the thermal scale in non-degenerate 
media.  Thus, a standard approximation is to assume the in-medium $T$-matrix to be independent 
of $\textbf{P}$~\cite{Riek:2010fk,Riek:2010py}, which leads to 
a major simplification of the calculations.
We thus write
\begin{align}
T^{a}_{ij}(\omega_1+\omega_2,\textbf{p}_1,\textbf{p}_2|\textbf{ p}_1',\textbf{p}_2')=T^{a}_{ij}(E_{\text{cm}},p_{\text{cm}},p_{\text{cm}}',
x_\text{cm}),
\label{eq_TmInGeneralFrame}
\end{align}
where $E_{\text{cm}}$, $p_{\text{cm}}$, $p_{\text{cm}}'$ and $x_\text{cm}\equiv\cos(\theta_{\text{cm}})$ are 
functions expressed via \(\omega_1+\omega_2, \textbf{p}_1, \textbf{p}_2, \textbf{ p}_1', \textbf{p}_2'\) 
using momentum conservation 
\(\textbf{p}_1+\textbf{p}_2=\textbf{ p}_1'+\textbf{p}_2'\) to define the transformation to the CM 
frame
\small
\begin{align}
&E_{\text{cm}}=\sqrt{(\omega_1+\omega_2)^{2}-(\textbf{p}_1\textbf{+p}_2)^{2}}\nonumber\\
\label{eq_CM}
&s_{\text{on}}=(\varepsilon_{1}(\textbf{p}_1)+\varepsilon_{2}(\textbf{p}_2))^{2}-(\textbf{p}_1\textbf{+p}_2)^{2}\\
&p_{\text{cm}}=\sqrt{\frac{(s_{\text{on}}-M_i^2-M_j^2)^2-4M_i^2M_j^2}{4s_{\text{on}}}}\nonumber\\&\cos(\theta_{\text{cm}})=\frac{\textbf{p}_{\text{cm}}\cdot\textbf{p}_{\text{cm}}'}{p_{\text{cm}}p_{\text{cm}}'}.\nonumber
\end{align}
\normalsize
For \(p_\text{cm}'\), we simply change \(s_{\text{on}}(\textbf{p}_1,\textbf{p}_2)\) to 
\(s_{\text{on}}(\textbf{p}_1',\textbf{p}_2')\). 
The reason for using the on-shell \(s\) for \(p_{\text{cm}}\) is to keep the analytical properties of the 
\(T\)-matrix after the transformation. 
Also, this transformation recovers Galilean invariance in the non-relativistic limit for the off-shell 
case. The relation for \(p_{\text{cm}}\) can be derived by solving the equations originating from Lorentz 
invariants $\varepsilon_{1}(\textbf{p}_1)^2-p_1^2=M_1^2$, 
$\varepsilon_{2}(\textbf{p}_2)^2-p_2^2=M_1^2$ and 
$(\varepsilon_{1}(\textbf{p}_1)+\varepsilon_{2}(\textbf{p}_2))^{2}-(\textbf{p}_1\textbf{+p}_2)^{2} 
=(\varepsilon_{1}(\textbf{p}_\text{cm})+\varepsilon_{2}(\textbf{p}_\text{cm}))^{2}$ 
in the CM and the moving frame.
We note that this procedure does not work for the CM angle in the off-shell case. However, since we 
only need forward scattering amplitudes for our present purposes, we do not discuss this issue any 
further here.

Rotational symmetry in the CM frame implies that a partial-wave expansion remains intact, given by
\begin{equation}
X(\mathbf{p},\mathbf{p}')=4\pi\sum_{l}(2l+1)X^{l}(p,p')P_{l}(\cos(\theta)),
\label{eq_partialX}
\end{equation}
where $X= V, T$. The partial-wave expanded scattering equation becomes 
\begin{align}
&T_{ij}^{l,a}(z,p,p')=V_{ij}^{l,a}(p,p')
\nonumber\\
& \quad \quad \quad + \frac{2}{\pi}\int_{-\infty}^{\infty}k^2dkV_{ij}^{l,a}(p,k)G^{0}_{ij}(z,k)T^{l,a}_{ij}(z,k,p') \ ,
\label{eq_partialT}
\end{align}
where $l$ denotes the angular-momentum quantum number. The set of now 1D integral equations 
can be solved by discretizing the 3-momenta \(p,p',k\),
\begin{equation}
\mathbb{V}_{mn}\equiv V(k_m,k_n),\hat{\mathbb{G}}_{(2)}^{0}(z)_{mn}\equiv \frac{2\Delta k}{\pi} k_m^2G_{(2)}^{0}(z,k_m)\delta_{mn} 
\label{discrete}
\end{equation}
and invert the pertinent matrix equation~\cite{Haftel:1970zz}, 
\begin{equation}
\mathbb{T}(z)_{mn}=T(z,k_m,k_n), \mathbb{T}(z)=[\mathbbm{1}- 
\mathbb{V}\hat{\mathbb{G}}_{(2)}^{0}(z)]^{-1}\mathbb{V} \ .
\label{Tmat}		
\end{equation}
The integral over \(k\) in Eq.~(\ref{eq_partialT}) is encoded in a matrix multiplication with 
measure \(dk\). Here and in the following, we (occasionally) use the subscript ``(2)" as an abbreviation 
for ``$ ij $" to denote two-body quantities.

Once the $T$-matrices have been computed, we calculate the single-particle selfenergies by summing
over all partial waves and the pertinent two-body flavor and color channels in interactions with light 
medium partons. 
Closing the $T$-matrix with an in-medium single-parton propagator ($\pm$ for boson/fermion) in the Matsubara formalism, 
\begin{equation}
\Sigma(iw_{n})=\pm\frac{-1}{\beta}\sum_{\nu_n} T(i\omega_{n}+i\nu_{n})G(i\nu_n) 
\end{equation}
one can use spectral representations to carry out the summation
over discrete frequencies to obtain
\begin{align}
\Sigma_{i}(z,\textbf{p}_1)=\int \frac{d^{3}\textbf{p}_2}{(2\pi)^{3}}
\int^{\infty}_{-\infty} d\omega_2 \frac{dE}{\pi}\frac{-1}{z+\omega_2-E}\frac{1}{ d_i}\sum_{a,j}d^{ij}_{s}d^{ij}_{a}\nonumber\\
\times\text{Im} T^a_{ij}(E,\textbf{p}_{1},\textbf{p}_{2}|\textbf{p}_1,\textbf{p}_2)
\rho_{j}(\omega_2,\textbf{p}_2)[n_{j}(\omega_2)\mp n_{ij}(E)]\nonumber\\
\label{eq_selfEbyTfull}
\end{align}
which involves the forward-scattering amplitude, \ie, $\textbf{p}_1'=\textbf{p}_1$ and 
$\textbf{p}_2'=\textbf{p}_2$ and thus $x_{\text{cm}}=x=1$; $n_{ij}$ refers to the Bose 
or Fermi distribution appropriate for the two-body state $ ij $, but the ``\(-/+\)" sign 
refers to the bosonic/fermionic single-parton state \(i\). The \(d^{ij}_{a,s}\) are color and spin degeneracy factors of the two-body system, summarized in Table~\ref{table_casimir}. Here, we enforce 
two physical polarizations for the gluons; \(d_i\) is the spin-color degeneracy of the single parton 
\(i \). The energy, $z=\omega_1 +i \varepsilon$, is taken to be retarded in this work. 
Within the CM transformation defined via Eqs.~(\ref{eq_CM}), the integrations in Eq.~(\ref{eq_selfEbyTfull}) 
are restricted to the timelike 2-body phase, \ie, real values for $E_{\rm cm}$ (we have verified that 
\(\text{Im}T^a_{ij}(\sqrt{E^2-P^2})\) is strongly suppressed when approaching the spacelike region).
The above selfenergy expression does not include the purely real thermal Fock term~\cite{fetter2003quantum} 
which we add explicitly by calculating 
\begin{align}
&\Sigma_{i}(\textbf{p}_1)=\mp\int \frac{d^{3}\textbf{p}_2}{(2\pi)^{3}}\int^{\infty}_{-\infty} d \omega_2 
V^1_{i\bar i}(\textbf{p}_1-\textbf{p}_2)\rho_{i}(\omega_2,\textbf{p}_2)n_{i}(\omega_2) \ .
\label{eq_selfEbyTfock}
\end{align}

Finally, we recall that Eq.~(\ref{eq_selfEbyTfull}) can be expressed a functional equation of \(\Sigma\), 
\begin{align}
\Sigma= T(\Sigma)G(\Sigma)=T(\Sigma)\frac{1}{(G^0)^{-1}-\Sigma} \ . 
\label{eq_selfcEq}
\end{align}
It is equivalent to an integral equation for the full Green function, \(G\), as 
\(\Sigma=(G^0)^{-1}-G^{-1}\). The $T$-matrix depends on the selfenergy, \(T(\Sigma)\), through 
the two-body propagator, see Eq.~(\ref{eq_G2define}), in which the spectral function depends on 
the single-parton selfenergy, see Eq.~(\ref{eq_G1define}). Although it is a non-linear functional 
equation, it usually can be solved selfconsistently. The selfenergy as the solution of 
Eq.~(\ref{eq_selfcEq}) satisfies conservation laws for the Green function~\cite{Baym:1961zz}.

\section{Constraints from Lattice QCD}
\label{sec_lQCD}
The Hamiltonian given in Eq.~(\ref{Hqgp}) is the input to our approach that needs to be constrained 
by independent information. To achieve this, we will make extensive use of first-principles 
lQCD computations, where we treat the pertinent data as ``observables" in imaginary time.
Specifically, we will utilize the QGP EoS~\cite{Borsanyi:2010cj,Bazavov:2014pvz}, HQ free 
energies~\cite{Petreczky:2004pz,kaczmarek2005static,Mocsy:2013syh}, and Euclidean quarkonium 
correlators~\cite{Datta:2003ww,Jakovac:2006sf,Aarts:2007pk,Aarts:2011sm}.
In this section, we elaborate on the concrete procedure to do that, which includes theoretical 
developments to best take advantage of the comparisons within the \(T\)-matrix approach. 
In Sec.~\ref{ssec_EoS} we briefly recapitulate the LWF 
formalism~\cite{PhysRev.118.1417,Baym:1961zz,Baym:1962sx} 
to compute the in-medium single- and two-body interaction contributions to the EoS for the effective 
Hamiltonian and lay out the corresponding matrix-log technique to resum the pertinent skeleton 
diagrams~\cite{Liu:2016nwj,Liu:2016ysz}. 
In Sec.~\ref{ssec_freeE} we recall the formalism to calculate the static-quark free energy from the
$T$-matrix, where large imaginary parts turn out to play a critical role~\cite{Liu:2015ypa}. 
In Sec.~\ref{ssec_correlator} we briefly review the formalism to calculate quarkonium correlator ratios 
based on Refs.~\cite{Cabrera:2006wh,Riek:2010fk,Riek:2010py}, thereby introducing an effective way to
account for interference effects in the complex potential for quarkonium spectral functions. 

\subsection{Equation of State}
\label{ssec_EoS}
The equation of state (EoS) usually refers to the pressure as a function energy density, or, alternatively, 
as a function of temperature and chemical potential of a many-body system, \(P(T,\mu)\). It characterizes 
the macroscopic dynamics of the bulk which are ultimately
driven by the relevant microscopic degrees of freedom of the medium. Although the EoS depends on the 
interactions in the system, it is usually most sensitive to the masses of the prevalent degrees of 
freedom in the medium (which, however, may be generated dynamically through the interactions, \eg, via
bound-state formation). Therefore, comparing the calculated EoS with lQCD results is expected to
primarily constrain the ``bare" parton masses in the Hamiltonian, Eq.~(\ref{Hqgp}). 

For a homogeneous grand canonical ensemble, the EoS is encoded in the grand potential (per unit volume), \(\Omega=-P\), 
which can be calculated using diagrammatic techniques within the LWB 
formalism~\cite{PhysRev.118.1417,Baym:1961zz,Baym:1962sx}(for recent application to QCD matter, see also Refs.~\cite{Blaizot:2000fc,Blaizot:2003tw,Blaschke:2016hzu}) as spelled out in Sec.~\ref{sssec_lwb}. 
Since the QGP near \(T_c\) can be expected to be a mixture of interacting partons and their bound states, 
a nonperturbative ladder resummation for the two-body amplitudes is in order. Some care needs to be exerted 
since the ladder resummation to calculate \(\Omega\) is not the same as for the \(T\)-matrix, due to a 
double-counting when closing the external legs of the latter. This will be carried out using a matrix-logarithm 
resummation technique~\cite{Liu:2016nwj,Liu:2016ysz} detailed in Sec.~\ref{sssec_impl}.

\subsubsection{Properties of the LWB Formalism}
\label{sssec_lwb}
The diagram language of the LWB formalism leads to the following expression for grand potential, 
\begin{equation}
\Omega = \mp\frac{-1}{\beta}\sum_{n}\text{Tr}\{\ln(-G^{-1})+[(G^0)^{-1}-G^{-1}] G\}\pm\Phi
\label{Omega}
\end{equation}
where we combined spin, color, flavor and momentum summations in the trace operation, ``\(\text{Tr}\)",
while explicitly writing the Matsubara frequency sum, \(\sum_{n}\). 
Here, 
\begin{equation}
\Phi=\sum_{\nu=1}^\infty \Phi_\nu
\label{phi}
\end{equation}
denotes the Luttinger-Ward functional (LWF), and  
\begin{equation}
\Phi_\nu=\frac{-1}{\beta}\sum_{n}\text{Tr}\{\frac{1}{2\nu}(\frac{-1}{\beta})^\nu[(-\beta)^\nu\Sigma_\nu(G)]G\} \   
\label{phi_nu}
\end{equation}
and \(\Sigma_\nu(G)\) are the LWF and selfenergy at \(\nu^\text{th}\) order of the potential 
in the ``skeleton" expansion~\cite{PhysRev.118.1417}. 
These three quantities should be understood as functionals of the full single-particle propagator, \(G\). 
The full selfenergy is the sum of all 
selfenergies of order \(\nu\), \(\Sigma (G)=\sum_\nu\Sigma_\nu(G)\). The extra factor \(1/\nu\) in 
Eq.~(\ref{phi_nu}) complicates the resummation of \(\Phi(\Sigma_\nu)\) 
for ladder diagrams, to be discussed in the next section.
The factor \((-1/\beta)^\nu (-\beta)^\nu\) aims to separate out the \(-1/\beta\) temperature dependence from 
loop integrals in the selfenergy, such as \(-1/\beta\sum_{n}X_1(\omega_n)X_2(z_m-\omega_n)\). 
At \(\nu^\text{th}\) order, there are \(\nu\) loops, with the pertinent factor \((-1/\beta)^\nu\). 
After this separation, \([(-\beta)^\nu\Sigma_\nu(G)]\) only has a temperature 
dependence stemming from \(G\) and the interaction kernel, \(V\). This separation procedure is convenient 
for proving thermodynamic relations involving temperature derivatives, cf.~App.~\ref{app_lwb-muq}.

The skeleton diagram expansion for the selfenergy can be obtained via a functional derivative of \(\Phi\), 
\begin{align}
\Sigma(G)=\frac{\delta\Phi}{\delta G} \ .
\label{sigphi}
\end{align} 
The functional derivative is equivalent to cut open one \(G\) line in a closed loop~\cite{PhysRev.118.1417}. 
Since there are \(\nu\) equivalent \(G\) lines at \(\nu^\text{th}\) order, this cancels the factor \(1/\nu\) 
and recovers the full selfenergy.  With the help of Eq.~(\ref{sigphi}) one finds the 
thermodynamic potential to reach an extremum, 
\begin{align}
\frac{\delta\Omega}{\delta G}=0 \ ,
\label{eq_domegadGeq0}
\end{align}
when the functional relation  
\begin{equation}
\Sigma(G) = (G^0)^{-1}-G^{-1}
\label{sigG}
\end{equation}
is satisfied. 
In this sense, \(G\) acts like a functional order parameter for the thermodynamic potential to reach 
an extremum.

In a slight variation of the standard LWB formalism, the ``bare"  masses (or dispersion relations,
$\varepsilon(\textbf{p})$) and potential of our effective Hamiltonian depend on temperature \(T\) and
chemical potential \(\mu_q\) of the medium. These dependences represent
a macroscopic average over the micro-physics that we do not treat explicitly (such as remaining gluonic
condensates in the QGP that can induce mass terms and the nonperturbative string term in the potential).
This leads to modified expressions for several thermodynamic relations, \eg, more complicated relations for 
energy and entropy to reconstruct the pressure; this is elaborated in App.~\ref{app_lwb-muq}.

\subsubsection{Matrix Logarithm Resummation of Skeleton Diagrams}
\label{sssec_impl}
The main challenge in calculating the grand potential, $\Omega$, is to evaluate the LWF, $\Phi$. 
In our derivation we limit ourselves to the case of a 3D reduced $T$-matrix, rather than the more general 
4D BS equation discussed in Ref.~\cite{Liu:2016ysz}, expanding on what we indicated earlier in Ref.~\cite{Liu:2016nwj}.

Using the notation $\int d\tilde{p}\equiv-\beta^{-1}\sum_n\int d^{3}\textbf{p}/(2\pi )^3$
with $\tilde{p}\equiv(i\omega_n,\textbf{p})$, the $\nu^\text{th}$ order of the selfenergy appearing 
in Eq.~(\ref{Omega}) in ladder approximation can be formally written as
\begin{align}
\Sigma_\nu(G)=\int d\tilde{p} \ [VG^{0}_{(2)}V G^{0}_{(2)}\cdots V]G
\end{align}
containing $\nu$ factors of $V$. Thus, the LWF functional $\Phi$ can be expressed as
\begin{align}
\Phi=\frac{1}{2}\sum\text{Tr} &\bigg\{G\bigg[V+\frac{1}{2}V G^{0}_{(2)}V+\ldots\nonumber\\
&+\frac{1}{\nu}VG^{0}_{(2)}V G^{0}_{(2)}\ldots .V+\ldots\bigg]G\bigg\}
\label{phi2}
\end{align}
where ``Tr" denotes, as before, a 3-momentum integral and the summation over discrete quantum 
numbers, while $\sum$ denotes the sum over Matsubara frequencies including $ \beta $ factors. 
The part in brackets, $[\cdots]$,  has a structure very similar to the 
$T$-matrix resummation,  
\begin{eqnarray}
T&=&V+V G^{0}_{(2)}V+\ldots+VG^{0}_{(2)}V G^{0}_{(2)}\ldots V+\ldots
\nonumber\\
&=&  \left[\sum\limits_{\nu=0}^{\infty} \left(V G^{0}_{(2)}\right)^\nu \right] V
\nonumber\\
&=& [1-VG^{0}_{(2)}]^{-1}  V\ , 
\end{eqnarray}
except for the extra coefficients $1/\nu$. However, we can write  
\begin{align}
&V+\frac{1}{2}V G^{0}_{(2)}V+\ldots+\frac{1}{\nu}VG^{0}_{(2)}V G^{0}_{(2)}\dots V + ... 
\nonumber\\
&=  \left[\sum\limits_{\nu=1}^{\infty}\frac{1}{\nu}\left(VG^{0}_{(2)} \right)^\nu\right] [G^{0}_{(2)}]^{-1}
\nonumber\\
&= -\ln[1-VG^{0}_{(2)}] [G^{0}_{(2)}]^{-1}
\nonumber \\
&\equiv \text{Log}\,T
\label{logT}
\end{align}
where the (natural-base) logarithm is to be understood as a general matrix operation (in a 
discrete space of quantum numbers, including spin, color, flavor as well as energy-momentum), 
defined through its power 
series.\footnote{A similar expression is known for the ground-state energy
at zero temperature~\cite{blaizot1986quantum} and for cold-atom systems~\cite{PhysRevA.75.023610}.} 
It can also be tested in the case of a separable potential for which the analytical result
is known~\cite{Rapp:1993ih}. At large coupling, the perturbative series does not converge
(in the present context, we have checked this, \eg, for the HQ friction coefficient discussed 
in Ref.~\cite{Liu:2016ysz}) and does not capture the formation of bound states which are expected 
to become important at low temperatures, cf.~also footnote~\ref{foot1}.

\begin{figure}[!htb]
\begin{center}
\includegraphics[width=0.9\columnwidth]{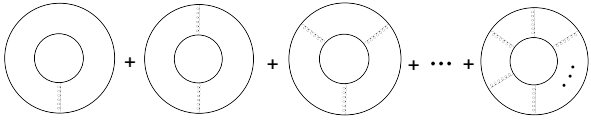}
\caption{Examples of diagrams that are resummed by the generalized \(T\)-matrix for EoS.}
\label{fig_wheel}
\end{center}
\end{figure}

The similarity between the $T$-matrix and the $\text{Log}\,T$ operation further allows to migrate
the partial-wave expansion, Eq.~(\ref{eq_partialT}), and CM approximation, Eq.~(\ref{eq_BSE3D}), 
from the $T$-matrix to the LWF. With the numerical discretization of the 3-momentum integrals as 
in Eqs.~(\ref{discrete}) and (\ref{Tmat}), we can define $\text{Log}\,T^{l,a}_{ij}$ in a given 
channel as
\begin{align}
&\mlog\mathbb{T}(z)_{mn}\equiv\text{Log}\,T(z,k_m,k_n)\nonumber\\
&\mlog\mathbb{T}(z)=-\mlog[\mathbbm{1}- 
\mathbb{V}\hat{\mathbb{G}}_{(2)}^{0}(z)][\hat{\mathbb{G}}_{(2)}^{0}(z)]^{-1}\ .
\label{eq_logTmatdiscrete}		
\end{align}
Compared to the $T$-matrix equation~(\ref{Tmat}), the only change is replacing the inverse matrix 
(with an extra factor $\mathbb{V}$) by the ``matrix-Log" operation, $\mlog\mathbb{T}$ (with an 
extra factor $[\hat{\mathbb{G}}^0_{(2)}(z)]^{-1}$). 
Standard software like Mathematica can compute this matrix function at a speed similar to a matrix 
inversion.  
With the result in a given channel, we first sum over partial waves using Eq.~(\ref{eq_partialX}) and then 
transform back from the CM to an arbitrary frame using Eq.~(\ref{eq_TmInGeneralFrame}) with 
$E_\text{cm}$, $p_\text{cm}$, $p_\text{cm}'$, and $x_\text{cm}'$ defined in Eq.~(\ref{eq_CM}), 
\begin{align}
&\text{Log}T^{a}_{ij}(\omega_1+\omega_2,\textbf{p}_1,\textbf{p}_2|\textbf{ p}_1',\textbf{p}_2')
\nonumber\\
&\qquad\qquad\qquad=\text{Log}T^{a}_{ij}(E_{\text{cm}},p_{\text{cm}},p_{\text{cm}}', 
x_\text{cm}) \  .
\label{eq_LogTmInGeneralFrame}
\end{align}
Upon closing two external lines of this quantity with a thermal single-particle propagator, 
$G$, and, in resemblance of Eq.~(\ref{eq_selfEbyTfull}), defining
\begin{equation}
\text{Log}\,\Sigma\equiv\int d\tilde{p}~\text{Log}\,T ~G \ ,
\end{equation}
we obtain
\begin{align}
&\text{Log}\Sigma_{i}(z,\textbf{p}_1)=\int \frac{d^{3}\textbf{p}_2}{(2\pi)^{3}}\int^{\infty}_{-\infty} 
d\omega_2 \frac{dE}{\pi} \frac{-1}{z+\omega_2-E}
\nonumber\\
& \quad \times \frac{1}{d_i} \sum_{a,j}d^{ij}_{s}d^{ij}_{a}
\text{Im}[\text{Log} T^a_{ij}(E,\textbf{p}_{1},\textbf{p}_{2}|\textbf{p}_1,\textbf{p}_2)]
\nonumber\\
&\quad \times \rho_{j}(\omega_2,\textbf{p}_2)(n_{j}(\omega_2)\mp n_{ij}(E)) \ .
\label{eq_LogselfEbyLogTfull}
\end{align}
Recalling Eq.~(\ref{phi2}) and the definition of $\text{Log}\Sigma$ and $\text{Log}T$, 
we can express the LWF  as
\begin{eqnarray}
\Phi&=&\frac{1}{2}\int d\tilde{p}~G~\text{Log}\Sigma 
\nonumber\\
&=&\frac{1}{2}\sum_{j} d_{j}\int d\tilde{p}~G_{j}(\tilde{p})~\text{Log}\Sigma_{j}(\tilde{p})  \ . 
\end{eqnarray}
Therefore, the grand potential in Eq.~(\ref{Omega}) can be expressed in closed form as
\begin{align}
\Omega=\sum_{j}\mp d_{j} \int d\tilde{p}&\Big\{\ln (-G_{j}(\tilde{p})^{-1}) 
\nonumber\\
 & \ \ +[\Sigma_{j}(\tilde{p})-\frac{1}{2}\text{Log}\Sigma_{j}  (\tilde{p})]G_{j}(\tilde{p})\Big\}.
\label{eq_Grandpotetnialdetail}      
\end{align}
The final sum over Matsubara frequencies in Eq.~(\ref{eq_Grandpotetnialdetail}) can be carried 
out with usual contour techniques utilizing a spectral representation of the expression 
in ``\{ \}" as a whole. Through this resummation we include the contributions of the diagrams 
shown in Fig.~\ref{fig_wheel} to the grand potential $\Omega$. 

\subsection{Static $Q\bar Q$ Free Energy}
\label{ssec_freeE}
The HQ free energy, \(F_{Q\bar{Q}}(r,T)\), is commonly defined as the change 
in free energy of a system when adding to it a static quark and antiquark, separated by a 
distance \(r\) (not including the (infinite) HQ masses). In the vacuum, this simply 
corresponds to the potential energy between them. 
In medium, the free energy and the potential are still related to each other, but no longer 
identical~\cite{Shuryak:2004tx,Rothkopf:2011db,Liu:2015ypa}, as the former includes the response 
of the medium to the static charges, encoded in the generally complex HQ selfenergies. However, 
one can {\em calculate} the free energy from an underlying potential within the same \(T\)-matrix 
approach that we discussed for the EoS above, by taking the limit 
\( M_Q\rightarrow\infty\)~\cite{Liu:2015ypa}. 
This opens the possibility to extract (or at least constrain) the driving kernel of the 
Hamiltonian through a fit to high-precision lQCD data for \(F_{Q\bar{Q}}(r,T)\). In particular, 
since the free energy incorporates the response of the medium to the external source, we need to
couple the static quarks with the light partons of the QGP medium consistently. This is achieved 
by the HQ selfenergy in the QGP which we compute from the in-medium heavy-light \(T\)-matrix 
with the same underlying driving kernel.
In the following, we first recall some basic relations for the free energy, in particular 
how it is related to the driving kernel of the static $T$-matrix (Sec.~\ref{sssec_F-V}. 
Second, we discuss the selfconsistent 
extraction of the potential which makes contact with the QGP bulk medium (Sec.~\ref{sssec_extract}). 
In App.~\ref{app_static} we collect several additional relations implied by the formalism, and in App.~\ref{sec_imv}
we elaborate on the connection between interference effects and the ``imaginary part of potential" .

\subsubsection{Heavy-Quark Free Energy and Potential}
\label{sssec_F-V}
In this section, we recall the derivation to relate $F_{Q\bar{Q}}(r,T) $ with the color-singlet 
potential in the static limit, $V(r,T)$~\cite{Liu:2015ypa} where we suppress color-flavor indices 
for simplicity in this section.

The static limit introduces simplifications which renders the relation between free energy and 
the potential rather straightforward. The source of this simplification is the one-particle 
propagator in the infinite-mass limit~\cite{Beraudo:2007ky},
\begin{align}
&G_Q\left(z ,\textbf{r}'\right)
=\int \frac{d^3\textbf{p}'}{(2\pi)^3}e^{i\textbf{p}'\cdot\textbf{r}'}
\frac{1}{z-\varepsilon _{\textbf{p}'}-\Sigma_Q\left(z,\textbf{p}'\right)}
\nonumber\\
&\approx\int \frac{d^3\textbf{p}'}{(2\pi)^3}e^{i\textbf{p}'\cdot\textbf{r}'}
\frac{1}{z-M-\Sigma_Q\left(z\right)} = \delta\left(\textbf{r}'\right)
G_Q\left(z\right) \ .
\label{eq_Green1F}	
\end{align}
The $\delta$-function signifies that the particle is static and $G_{Q}(z)= 1/(z-M-\Sigma_{Q}(z))$ 
is simply the propagator in momentum space in the static limit, \ie, it is localized 
at its position. At vanishing quark chemical potential, $G_Q=G_{\bar{Q}}$. 
The two-body (4-point) Green's function inherits the $\delta$-function 
structure~\cite{Beraudo:2007ky},
\begin{align}
&G_{Q\bar Q}^>(-i\tau ,\textbf{r}_1,\textbf{r}_2|\textbf{r}_1',\textbf{r}_2')\nonumber\\
&\equiv\delta\left(\textbf{r}_1-\textbf{r}_1'\right)
\delta\left(\textbf{r}_2-\textbf{r}_2'\right)G_{Q\bar Q}^>(-i\tau, r)\label{eq_G} \ ,
\end{align}
where $r=|\textbf{r}_1-\textbf{r}_2|$. Here, $G_{Q\bar Q}^>(-i\tau, r)$ denotes the reduced 
Green function with the spatial $\delta$-functions factored out. The static $Q\bar{Q}$ free 
energy, $F_{Q\bar{Q}}$, can be 
defined in terms of the $Q\bar Q$ Green function as~\cite{Beraudo:2007ky}
\begin{align}
&F_{Q\bar{Q}}(r,\beta)=-\frac{1}{\beta }\ln \left(G_{Q\bar Q}^>\left(-i\beta ,r\right)
\right) \ .
\label{eq_defineF}
\end{align}
The remaining task is to calculate the Euclidean time Green function, $G_{Q\bar Q}^>(-i\tau, r)$, 
in Eq.~(\ref{eq_G}) using the $T$-matrix, Eq.~(\ref{eq_BSE3D}), with the propagators $G_Q(z)$ and 
potential $V(z,\textbf{p}_1-\textbf{p}_1')$, which in coordinate space is denoted as $V(z,r)$. 
We  here keep a dependence of the potential on the total energy, $z$, of the 2-particle system, 
which can arise, \eg, from interference effects as illustrated in App.~\ref{app_static}.

To proceed, we first use  $G_{Q,\bar Q}(z)$ to obtain the non-interacting two-body propagator 
figuring in the $T$-matrix,
\begin{align}
G^{0}_{Q\bar Q}(z)= \int_{-\infty}^\infty d\omega_1 d\omega_2  
\frac{\rho_Q(\omega_1)\rho_{\bar Q}(\omega_2)}{z-\omega_1-\omega_2 } \ .
\label{eq_G2bare}
\end{align}
where $ \rho_{Q/\bar Q}(\omega_1) $ are the spectral functions of the static quark/antiquark, 
as before. Inserting this propagator together with  $V(z,\textbf{p}_1-\textbf{p}_1')$ into 
Eq.~(\ref{eq_BSE3D}), one has
\begin{align}
&T_{Q\bar Q}(z,\textbf{p},\textbf{p}')=V(z,\textbf{p}-\textbf{p}')+
\nonumber\\
& \quad \  \int\frac{d^3\textbf{k}}{(2\pi)^3}
V(z,\textbf{p}-\textbf{k})~G^{0}_{Q\bar Q}(z)~T_{Q\bar Q}(z,\textbf{k},\textbf{p}') \ .
\label{eq_BSE3Dstatic}
\end{align}
Since $ G^{0}_{Q\bar Q}(z) $ is independent of momentum, Fourier transforming the above equation 
from $ \textbf{p}\rightarrow \textbf{r} $ and $ \textbf{p}'\rightarrow \textbf{r}'$ 
where $ \textbf{r}=\textbf{r}_1-\textbf{r}_2$, and $ \textbf{r}'=\textbf{r}_1'-\textbf{r}_2'$,
one arrives at
\begin{align}
T_{Q\bar Q}(z,\textbf{r},\textbf{r}') =V(z,r)\delta(\textbf{r}-\textbf{r}')+ \qquad \qquad \quad
\nonumber\\
\qquad V(z,r)G^{0}_{Q\bar Q}(z)T_{Q\bar Q}(z,\textbf{r},\textbf{r}') \ . 
\label{eq_BSE3Dstaticrspace}
\end{align}
This is an algebraic equation with a solution
$T_{Q\bar Q}(z,\textbf{r},\textbf{r}')=T_{Q\bar Q}(z,r)\delta(\textbf{r}-\textbf{r}')$
explicitly given by 
\begin{align}
&T_{Q\bar Q}(z,r)=\frac{V(z,r)}{1-V(z,r)G^{0}_{Q\bar Q}(z)} \ .
\label{eq_BSE3Dstaticrsolution}
\end{align}
We have factored out the $\delta$ function as was 
done in Eq.~(\ref{eq_G}).\footnote{Only one $\delta$-function here is related to 
stripping off $\delta(\textbf{p}_1+\textbf{p}_2-\textbf{p}_1'-\textbf{p}_2')$. Note 
that $X(\textbf{p}_1-\textbf{p}_2)\delta(\textbf{p}_1+\textbf{p}_2-\textbf{p}_1'-\textbf{p}_2')$ 
Fourier-transforms into the form 
$X(\textbf{r}_1-\textbf{r}_2)\delta(\textbf{r}_1-\textbf{r}_1')\delta(\textbf{r}_2-\textbf{r}_2')$.} 
The Green function in frequency space in the static limit can be expressed as 
\begin{align}
&G_{Q\bar Q}\left(z,r\right)=
G^0_{Q\bar Q}(z)+G^0_{Q\bar Q}(z)
T_{Q\bar Q}(z,r)G^0_{Q\bar Q}(z) ,
\label{eq_Green4PsF}
\end{align}
While in the non-static case, additional convolution integrals in coordinate space appear, 
the simple form in the static limit is due to the ``$\delta(r)$" functions that 
can been integrated out (or stripped off). Upon inserting Eq.~(\ref{eq_BSE3Dstaticrsolution}) 
into Eq.~(\ref{eq_Green4PsF}) we arrive at our final expression for $ G_{Q\bar Q} $ in 
energy-coordinate space,
\begin{align}
G_{Q\bar Q}(z,r)&=\frac{1}{[G^{0}_{Q\bar Q}(z)]^{-1}-V(z,r)} \ .
\label{eq_Gstaticfull}               
\end{align}
To obtain $G^>\left(-i\tau ,r\right)$, we need to transform back to imaginary time 
using $(-\beta)^{-1}\sum_n G_{Q\bar Q}(iE_n,r) e^{-\tau(iE_n)}$; employing a spectral 
representation and contour technique the Matsubara sum can be carried out yielding
\begin{align}
&G_{Q\bar{Q}}^>\left(-i\tau ,r\right)
=\int\limits_{-\infty }^{\infty
}dE'\rho_{Q\bar Q} \left(E',r\right)\frac{e^{E'(\beta-\tau)}}{e^{\beta E'}-1} \ .
\label{eq_GreenTF}
\end{align}
Since the strength of the two-particle spectral function, $\rho_{Q\bar Q} \left(E',r\right)$, 
is located in the vicinity of the large-mass two-particle threshold, $2M_Q$, we can
approximate $e^{\beta E'}\gg1$ and $e^{E'(\beta-\tau)}/(e^{\beta E'}-1)=e^{-E'\tau}$,
to obtain 
\begin{align}
&G_{Q\bar{Q}}^>\left(-i\tau ,r\right)
=\int\limits_{-\infty }^{\infty
}dE'\rho_{Q\bar Q} \left(E',r\right)e^{-E'\tau} \ . 
\label{eq_GreenTFBolz}
\end{align}
The quantity $G^>\left(-i\tau ,r\right)$ still depends on the infinitely large mass, $M_Q$ 
(numerically taken as $2\cdot10^4$~GeV), which needs to be ``renormalized". This can be 
done by multiplying $G^>\left(-i\tau ,r\right)$ with a factor $e^{2M_Q \beta}$ 
and redefining the energy arguments of the propagators and spectral functions by a shift
of $2M_Q$. For simplicity, we will keep te same notation, \ie, from here on, unless otherwise
noted, the static limits of $G_{Q\bar{Q}}^>\left(-i\tau ,r\right)$, $G_Q(z)$, $G_{Q\bar Q}(z)$ and 
$\rho_{Q\bar Q}(z)$ will refer to the original ones shifted as $G_{Q\bar{Q}}^>\left(-i\tau ,r\right)e^{2\beta M_Q}$, 
$G_Q(z+M_Q)$, $G_{Q\bar Q}(z+2M_Q)$, and $\rho_{Q\bar Q}(z+2M_Q)$. 
Inserting Eqs.~(\ref{eq_Gstaticfull}) and (\ref{eq_GreenTFBolz}) into Eq.~(\ref{eq_defineF}) 
with $ \tau=\beta $ establishes our basic relation between the HQ potential and the free energy
within the $T$-matrix formalism.

To be more explicit, we specify $[G^{0}_{Q\bar Q}(z)]^{-1}$ as
\begin{align}
&[G^{0}_{Q\bar Q}(z)]^{-1}=z-2\Delta M_Q-\Sigma_{Q\bar Q}(z)
\label{selfE2}               
\end{align}
with medium-induced Fock mass term $\Delta M_Q$ (for each quark) determined by $V(r)$ as 
further discussed in Sec.~\ref{ssec_ansatz}, and an analytic selfenergy part, 
$\Sigma_{Q\bar Q}(z)$, labeled as a two-body selfenergy in this work. In practice, we can use 
$\text{Im}[[G^{0}_{Q\bar Q}(E+i\epsilon)]^{-1}]=-\text{Im}\Sigma_{Q\bar Q}(E+i\epsilon) $ to find 
the imaginary part and reconstruct $\text{Re}\Sigma_{Q\bar Q}(E+i\epsilon)$ by a dispersion 
relation. The energy dependent potential, \(V(z,r)\), can also be decomposed into a static 
non-analytic part, $V(r)$, and an analytic part, $ V_A(z,r)$, so that $V(z,r)=V(r)+V_{A}(z,r)$. 
As elaborated in App.~\ref{sec_imv}, $V(r)$ is the input potential and \(V_{A}(z,r)\) is 
related with interference effects induced by many-body physics, similar to $\Sigma_{Q\bar Q}(z)$. 
Therefore, we separate the input static potential $V(r)$ and regroup \(V_{A}(z,r)\) 
into a ``interfering" two-body selfenergy as 
\(\Sigma_{Q\bar Q}(z,r)\equiv\Sigma_{Q\bar Q}(z)+V_A(z,r)\) 
(note that $ \Sigma_{Q\bar Q}(z,\infty)=\Sigma_{Q\bar Q}(z) $ since $ V_A(z,\infty)=0 $),
\ie, 
\begin{align}
&V(z,r)=V(r)+[\Sigma_{Q\bar Q}(z,r)-\Sigma_{Q\bar Q}(z)] \ .
\label{eq_Vreexpress}               
\end{align}
Equation~(\ref{eq_Gstaticfull}) can then be recast as
\begin{align}
G_{Q\bar Q}(z,r)&=\frac{1}{z-2\Delta M_Q-V(r)-\Sigma_{Q\bar Q}(z,r)}  \ .
\label{eq_GstaticfullbyselfE}               
\end{align}
With this expression, \(\Sigma_{Q\bar Q}(z,r)\) is analytic and $ 2\Delta M_Q+V(r)$ is a 
non-analytic static part. In this scheme, the final compact form for the free energy reads
\begin{align}
&F_{Q\bar{Q}}(r,\beta)=\frac{-1}{\beta}\ln \bigg[\int_{-\infty}^{\infty} dE\,e^{-\beta E} 
\nonumber\\
& \qquad \ \ \times\frac{-1}{\pi}\text{Im}[\frac{1}{E+i\epsilon-\tilde{V}(r)-\Sigma_{Q\bar Q}(E+i\epsilon,r)}]\bigg]
\label{eq_FreeEfinal}               
\end{align}
where $ \tilde{V}(r)\equiv2\Delta M_Q+V(r)$ is introduced for brevity. 

\subsubsection{Self-Consistent Extraction of the Potential}
\label{sssec_extract}
In order to use Eq.~(\ref{eq_FreeEfinal}) to extract the potential, $V(r)$, we need to evaluate 
$\Sigma_{Q\bar Q}(z,r)$. Toward this end, we first calculate the one-body selfenergy, 
\(\Sigma_{Q}(z)\). Taking the heavy-light $T$-matrix in Eq.~(\ref{eq_selfEbyTfull}) in the 
``half-static" limit where the $\textbf{p}_1$ dependence in Eq.~(\ref{eq_TmInGeneralFrame}) 
is suppressed due to an infinite mass of particle-1, we obtain 
 \begin{align}
\Sigma_{Q}(z) =&\int \frac{d^{3}\textbf{p}_2}{(2\pi)^{3}}\int^{\infty}_{-\infty} d\omega_2 \frac{dE}{\pi} \frac{-1}{z+\omega_2-E}\frac{1}{d_Q}\sum_{a,j}d^{Qj}_{s}d^{Qj}_{a}\nonumber\\
 &\times T^{a}_{Qj}(E,\textbf{p}_2|\textbf{p}_2)\rho_{j}(\nu,\textbf{p}_2)n_{j}(\nu) \ .
 \label{eq_selfEQ1byT}               
 \end{align}
The CM transformation in the static limit, \(\omega_1+\omega_2\gg|\textbf{p}_1+\textbf{p}_2| \), 
can be derived as 
\begin{align}
&E_{\text{cm}}=\omega_1+\omega_2,p_{\text{cm}}=p_2,\cos(\theta_{\text{cm}})=\cos(\theta) \ .
\end{align}
The $n_{ij}$ is suppressed due to infinite mass of two-body states. The selfconsistent 
Eq.~(\ref{eq_selfcEq}) also applies in the static limit. 
For the two-body selfenergy, $ \Sigma_{Q\bar Q}(z) $, we first use Eq.~(\ref{eq_G2bare}) to 
obtain the two-body propagator, $ G^{0}_{Q\bar Q}(z)$, and then use the procedure laid out
after Eq.~(\ref{selfE2}) to arrive at $ \Sigma_{Q\bar Q}(z) $.
 
In the Brueckner type setup of our approach, the $r$-dependent part of the two-body 
``interfering" selfenergy, $\Sigma_{Q\bar Q}(z,r)$, is not selfconsistently 
generated, as this would require to include 3-body interactions~\footnote{Ideas to 
selfconsistently generate this part are presented in App.~\ref{sssec_inter2}}. 
For now, we model \(\Sigma_{Q\bar Q}(z,r)\) with a factorizable ansatz, 
\begin{align}
\Sigma_{Q\bar Q}(z,r)=\Sigma_{Q\bar Q}(z,\infty)\phi(x_e r)\equiv\Sigma_{Q\bar Q}(z)\phi(x_e r)
\label{eq_selfE2fatorize}
\end{align}
which preserves the analyticity of \(\Sigma_{Q\bar Q}(z,r)\).
The function \(\phi(x_e r)\) is motivated by the imaginary part of the potential in 
a perturbative approximation~\cite{Laine:2006ns,Beraudo:2007ky} and will be constrained in our
context by a functional fit (within its short- and long-distance limits of one and zero, 
respectively). Here, \(x_e\) is a dimensionless parameter acting as a screening mass 
that shrinks the range of \(\phi(x_e r)\) as temperature increase (our pivot point at the
lowest temperature considered here is set to \(x_e=1\)).  Inserting 
Eq.~(\ref{eq_selfE2fatorize}) into Eq.~(\ref{eq_FreeEfinal}) gives 
\begin{align}
&F_{Q\bar{Q}}(r,\beta)=\frac{-1}{\beta}\ln \bigg[\int_{-\infty}^{\infty} dE\,e^{-\beta E} \times \nonumber\\&\frac{-1}{\pi}\text{Im}[\frac{1}{E+i\epsilon-\tilde{V}(r)-\Sigma_{Q\bar Q}(E+i\epsilon)\phi(x_e r)}]\bigg]
\label{eq_FreeEfit}               
\end{align}
where the input functions \(V(r)\) and  \(\phi(x_e r)\) are to be tuned to reproduce lQCD data.
In our previous work~\cite{Liu:2015ypa} $ \Sigma_{Q\bar Q}(E+i\epsilon) $ was modeled by 
a functional ansatz with few parameters and as such was the major source of the uncertainties
in the approach. In the present work, $ \Sigma_{Q\bar Q}(E+i\epsilon) $ is controlled 
selfconsistently by the single heavy-quark/antiquark selfenergy, $\Sigma_{Q}$/$\Sigma_{\bar Q}$, 
as outlined above.

\subsection{Quarkonium Correlator Ratios}
\label{ssec_correlator}
The Euclidean correlator can be understood as ``Fourier transform" of the spectral function 
to imaginary-time space, where it is computable in lattice QCD.
Its ratio to correlator with a vacuum reference function is utilized to highlight the medium 
modifications in the spectral functions, and it also has the advantage of reducing systematic 
lattice uncertainties. Since the quarkonium correlator is defined by a local operator,  
the two-body Green function/spectral function is proportional to the wave function overlap at 
the origin, \(G_{ij}(E)= \sum_n |\phi_{E_n}(0)|^2/(E-E_n)\). Thus, the correlator is quite 
sensitive to short-range physics, which is useful to, \eg, constrain the strong coupling 
constant \(\alpha_s\) in the Coulomb term. The spectral function and the correlator can be 
readily calculated in the \(T\)-matrix approach with heavy quarks. There are several previous 
studies of these quantities in this approach~\cite{Cabrera:2006wh,Riek:2010fk,Riek:2010py} 
which we will briefly review. Here, we are now able to significantly go beyond those
by consistently coupling the heavy quarks to an off-shell light-parton plasma.  

\subsubsection{Review of Established Formalism}
The correlator in the Euclidean time that can be computed in 
lQCD~\cite{Datta:2003ww,Jakovac:2006sf,Aarts:2007pk} is defined by 
\begin{align}
&G^>(-i \tau,\textbf{P})=\int d^3\textbf{r}\, e^{i\textbf{P}\cdot\textbf{r}} \langle J_M(-i\tau,\textbf{r}),J^\dagger_M(0,0)\rangle
\label{eq_McorrelatorwithP}
\end{align}
and usually evaluated at vanishing total 3-momentum, $\textbf{P}$, of the $Q\bar Q$ pair, 
\begin{align}
&G^>(-i \tau)\equiv G^>(-i \tau,\textbf{P})|_{P=0} \ . 
\label{eq_Mcorrelator}
\end{align}
The mesonic states are created by the local operator 
\begin{align}
J_M(-i\tau,\textbf{r})=\bar{\psi}(t,\textbf{r})\Gamma_M\psi(t,\textbf{r}) \ ,  
\label{eq_McurrentwithP}
\end{align}
where $\psi$ ($ \bar{\psi} $) denotes the (conjugate) Dirac spinor field operator.
The Dirac matrix $ \Gamma_M\in\{1,\gamma_\mu,\gamma_5,\gamma_\mu\gamma_5\}$ projects the 
operators into scalar, vector, pseudoscalar and pseudovector channels, respectively. 
In a fully relativistic treatment, $\psi$ can create an  anti-particle or annihilate a particle.
However, in the context of this paper, we separately treat particle annihilation and 
antiparticle creation (and vice versa) by two field operators $ \psi_Q$ and  
$\psi^\dagger_{\bar{Q}}$, respectively, schematically written as 
$\psi=\psi_Q+\psi^\dagger_{\bar{Q}}$ (here and in the following, we also use $ Q $ to denote 
$ c $ and $ b $ quarks). Inserting this into Eqs.~(\ref{eq_McurrentwithP}) 
and (\ref{eq_McorrelatorwithP}) (suppressing the $ \Gamma_M $ structure and pertinent
relativistic corrections), a leading term of the 16 possibilities for this correlator 
is the 4-point Green function 
\begin{align}
&G_{Q \bar{Q}}^>(-i \tau,\textbf{P})=\int d^3 \textbf{r}\, 
e^{i\textbf{P}\cdot\textbf{r}}G_{Q \bar{Q}}^>(-i\tau ,\textbf{r},\textbf{r}|0,0)
\nonumber\\
&=\int d^3 \textbf{r}\,  e^{i\textbf{p}\cdot\textbf{r}}
\langle\psi_{\bar{Q}}(-i\tau,\textbf{r})\psi_Q(-i\tau,\textbf{r})
\psi^\dagger_{Q}(0,0)\psi^\dagger_{\bar{Q}}(0,0)\rangle \ , 
\label{eq_correlator}
\end{align}
which characterizes the propagation of a two-body state and can be solved by 
the $ T $-matrix as shown in the previous section. Another important term for the 
same correlator is the density-density correlation function,
\begin{eqnarray} 
\langle n_Q(-i\tau,\textbf{r})n_Q(0,0)\rangle=&
\nonumber\\
 & \hspace{-2.5cm} \langle\psi^\dagger_Q(-i\tau,\textbf{r})\psi_{Q}(-i\tau,\textbf{r})\psi^\dagger_Q(0,0)\psi_{Q}(0,0)\rangle
\end{eqnarray} 
which is usually referred to as the zero-mode contribution (or Landau cut) and closely 
related to the transport properties of the medium~\cite{Riek:2010py}. Other terms are 
either included automatically through the Matsubara formalism as hole excitations, 
or they are suppressed in the HQ limit. For the purpose of this paper, we choose the 
simplest quantity to be compared with lQCD data, \ie, the pseudoscalar channel, 
$\Gamma_M=\gamma_5 $, which does not develop a zero mode. It corresponds to the 
mesonic \(\eta_c\) and \(\eta_b\) channels (including, of course, their full 
excitation spectrum).

Since we focus on the Euclidean time correlator at total momentum $ \textbf{P}=0 $, it 
simply corresponds to the $T$-matrix in the CM frame. The additional locality in 
the relative coordinate leads to one integration over 
3-momentum\footnote{$f(\textbf{r}_1-\textbf{r}_2)=\int \frac{d^3\textbf{p}}{(2\pi)^3} 
e^{i \textbf{p}\cdot (\textbf{r}_1-\textbf{r}_2)} f(p)\rightarrow f(0)= 
\int\frac{d^3\textbf{p}}{(2\pi)^3} f(p)$.}. 
Thus, the 4-point Green function in frequency space for the pseudoscalar channel 
takes the form
\begin{align}
& G_{Q \bar{Q}}(z)=d_Q\int\frac{d^3p}{(2\pi^3)}G^0_{Q \bar{Q}}(z,p)+
\nonumber\\
&d_Q\int\frac{dpdp'}{\pi^3}
\mathcal{R}^\mathcal{S}_{Q \bar{Q}}~G^0_{Q \bar{Q}}(z,p)~T_{Q \bar{Q}}^{l}(z,p,p')~G^{0}_{Q \bar{Q}}(z,p') \ .
\label{eq_G2charm}
\end{align}
It includes the relativistic effects due to the projector $\Gamma_M$, encoded in the
$\mathcal{R}^\mathcal{S}_{ij}$ defined in Eqs.~(\ref{eq_potential}), (\ref{eq_Bfator}) and 
(\ref{eq_Rfator}), cf.~Refs.~\cite{Cabrera:2006wh,Riek:2010fk,Riek:2010py} for more details
(in those works the $\mathcal{R}$ factor is part of the propagator, but the expressions are
equivalent to the ones used here); 
\(d_Q=6\) denotes the spin-color degeneracy of a heavy quark. 
The spectral function for this Green function is defined as
\begin{align}
&\rho_{Q \bar{Q}}(E,T)=-\frac{1}{\pi}\text{Im}G_{Q \bar{Q}}(E+i\epsilon) \ , 
\label{eq_spec2charm}
\end{align}
and the pertinent correlator is given by
\begin{align}
&G^>_{Q \bar{Q}}(-i\tau,T_\text{ref},T)=\int_0^\infty dE \rho_{Q \bar{Q}}(E,T_\text{ref}) \mathcal{K}(\tau, E,T)
\ , 
\label{eq_correlatortime}
\end{align}
with the kernel
\begin{align}
&\mathcal{K}(\tau, E,T)=\frac{\cosh[E(\tau-\beta/2)]}{\sinh[E(\beta/2)]} \ ,
\label{eq_correlatorkernel}
\end{align}
which can be obtained using the contour techniques with proper treatment of the retarded 
symmetry for spectral function for negative $E$. Finally, the correlator ratio is defined as
\begin{align}
&R_{Q \bar{Q}}(\tau,T_\text{ref},T)=\frac{G^>_{Q \bar{Q}}(-i\tau,T,T)}{G^>_{Q \bar{Q}}(-i\tau,T_\text{ref},T)} \ .
\label{eq_Ratio}
\end{align}
In this ratio the denominator and the numerator carry the exact same kernel, 
$ \mathcal{K}(\tau, E,T) $ so that the only difference is the spectral function, 
thus exhibiting the medium effects relative to a reference spectral function (usually
taken as one at small temperature).

\subsubsection{Interference Effect for Two-Body Spectral Function}
\label{sssec_inter2}
As discussed in App.~\ref{sec_imv}, the \(r\)-dependent imaginary part of the potential 
is a manifestation of interference effects between the two quarks when interacting with
the medium; \eg, in the singlet channel a small size $Q\bar Q$ state will effectively become
colorless thus suppressing any interaction with the colored medium partons. Therefore, 
this effect is expected to become significant for deeply bound heavy quarkonia with a tight 
wave function. Although a full many-body treatment will require nontrivial 3-body diagrams, 
we will suggest a way to include the effects in the \(T\)-matrix approach which seems viable 
for the case of two-body spectral functions and correlators. However, we will only include 
the interference effects for heavy-heavy and static-static channels.

We start from the non-relativistic Schr\"odinger equation, 
\begin{align}
(-\frac{\partial_r^2}{M}+\tilde{V}_{\rm clx} (r))\varphi(r)=E\varphi(r) \ .
\label{eq_schR}
\end{align}
In previous works~\cite{Laine:2006ns,Burnier:2015tda}, an energy-independent complex ``potential"
has been introduced; in our context we write it as 
\(\tilde{V}_{\rm clx}(r)=V(r)+ i\Sigma^I_{Q\bar{Q}} \phi(x_e r)\), where we introduced the generic
notation $\Sigma^I \equiv \text{Im}\Sigma$.  
Transforming it to momentum space leads to
\begin{align}
& \tilde{V}_{\rm clx}(\textbf{p}-\textbf{p}')=i\Sigma^{I}_{Q\bar{Q}}(2\pi)^3\delta(\textbf{p}-\textbf{p}')+
\qquad 
\nonumber\\
& \qquad \qquad i\Sigma^I_{Q\bar{Q}} \phi_N(\textbf{p}-\textbf{p}')+V(\textbf{p}-\textbf{p}')
\label{eq_TmImvsimple}
\end{align}
where \(\phi_N(\textbf{p}-\textbf{p}')\) is the Fourier transform of \(\phi(x_e r)-1\), 
\begin{align}
&\phi_N(p)=\int d^3\textbf{r}\,e^{i\textbf{p}\cdot\textbf{r}}(\phi(x_e r)-1) \ .
\label{eq_phiwithout1}
\end{align}
The Schr\"odinger equation in momentum space then reads
\begin{align}
&\int \frac{d^3\textbf{p}'}{(2\pi)^3}\bigg\{\big[\frac{p^2}{M}
+i\Sigma^I_{Q\bar{Q}}\big](2\pi)^3\delta(\textbf{p}-\textbf{p}')+
\nonumber\\
&\quad \ i\Sigma_{{Q\bar{Q}}}^I \phi_N(\textbf{p}-\textbf{p}')+V(\textbf{p}-\textbf{p}')\bigg\}\varphi(\textbf{p}')
=E\varphi(\textbf{p}) \ .
\label{eq_schP}
\end{align}
One can now follow the standard track to derive the Lippmann-Schwinger equation (LSE). 
The terms in the brackets ``[ ]" figure in $ H_0 $, which is combined with $E$ on the 
right-hand side as $(E-H_0)\varphi=V\varphi$. Then, inverting the left-hand side and 
adding a free solution, we obtain 
the general solution as $\varphi=\varphi_0+(E-H_0+i\epsilon)^{-1}V\varphi $. 
Multiplying it by  $V$, we arrive at the $T$-matrix equation $T=V+(E-H_0+i\epsilon)^{-1}VT$ 
using $V\phi=T\phi_0$. The part local in momentum with a $\delta$-function in 
Eq.~(\ref{eq_TmImvsimple}) enters the free propagator, while the part nonlocal in momentum 
space becomes the true potential. 

To generalize the Schr\"odinger framework
to be compatible with the $T$-matrix approach discussed in previous sections (in particular
in Sec.~\ref{sssec_F-V}), a few extensions are required. Specifically, the energy-momentum
dependence and analytic properties of the uncorrelated in-medium two-particle propagator 
need to be accounted for. 
Toward this end, motivated by the relation (\ref{selfE2}) in the static limit, we augment
the constant imaginary part to an energy-dependent complex quantity, $ \Sigma_{Q\bar{Q}}(z,p) $, 
whose local part (with a 3-momentum $\delta$-function) encodes the dynamical single-quark 
selfenergies, while its non-local part accounts for interference effects (as a coefficient 
to the ``interference" function, $\phi$), 
\begin{align}
\tilde{V}_{\rm clx}(z,\textbf{p}-\textbf{p}')
=&(2\pi)^3\delta(\textbf{p}-\textbf{p}')\Sigma_{Q\bar{Q}}(z,p)+
\nonumber\\
&\Sigma_{Q\bar{Q}}(z,p')\phi_N(\textbf{p}-\textbf{p}')+V(\textbf{p}-\textbf{p}') \ . 
\label{eq_TmImvfull}
\end{align}
Thus, the modified potential figuring as a kernel in the $T$-matrix equation takes he form 
\begin{align}
V_{\rm clx}(\textbf{p}-\textbf{p}')=\Sigma_{Q\bar{Q}}(z,p')\phi_N(\textbf{p}-\textbf{p}')
+V(\textbf{p}-\textbf{p}') \ , 
\label{eq_Vcorrelatorim}
\end{align}
which is then subjected to a standard partial-wave expansion. The resulting spectral 
function does not depend on using $ \Sigma_{Q\bar{Q}}(z,p) $ or $ \Sigma_{Q\bar{Q}}(z,p') $ 
in the above equation since $ \phi_N $ is symmetric under the exchange of 
$ \textbf{p}$ and $\textbf{p}'$.
With this setup, the \(T\)-matrix is still analytic but no longer positive-definite. The 
latter feature causes complications when utilized in many-body calculations of single-particle
selfenergies. It is indicative of a non-conserving approximation~\cite{Baym:1961zz}. However, 
when restricted  to the calculation of the quarkonium spectral functions and correlators, 
the former remains strictly positive definite. In addition, this scheme precisely recovers the 
implementation of $V_I$ in the static limit.  In Sec.~\ref{sec_results}, we will elaborate on 
the interference effects for the spectral functions obtained from this implementation.

\subsection{Potential Ansatz and Numerical Procedure}
\label{ssec_ansatz}

\subsubsection{Screened Cornell potential and bare parton masses}
\label{sssec_pot}
For the Hamiltonian introduced in Eq.~(\ref{Hqgp}), the inputs are the 2-body potential 
and bare particle masses which both depend on temperature. 
As an ansatz for the potential, we employ a generalized in-medium Cornell 
potential~\cite{Megias:2005ve,Megias:2007pq}:
\begin{align}
&V(r)=V_\mathcal{C}+V_\mathcal{S}=-\frac{4}{3}\alpha_s \frac{e^{-m_d r}}{r}
-\frac{\sigma e^{-m_s r- (c_b m_s r)^2}}{m_s} \ , 
\label{eq_potentialstatic}              
\end{align}
which recovers the well-established vacuum form while implementing in-medium screening of both
the shot-range Coulomb and long-range confining interaction (``string term") in a transparent 
and economic way. 
The respective screening masses are denoted by \(m_d\) and \(m_s\). An additional
quadratic term, $ -(c_b m_s r)^2$, in the exponential factor of the string term accelerates
the suppression of the long-range part, mimicking a string breaking feature. It can also
be considered as the next term in a power expansion in $r$.   

Since the screening originates from the coupling of the bare interaction to medium partons, 
both $m_s$ and \(m_d\) are functions of the parton density and thus they are not totally 
independent. The \(1/r\) and \(r\) dependence of the potential leads to static propagators
in momentum space,  $D_c(q)=1/q^2$ and $D_s=1/q^4$, respectively, which, upon multiplication 
with the respective coupling constants, $-4/3\alpha_s $ and $ -8\pi\sigma $ in singlet channel, 
constitutes the bare potential in the Hamiltonian.
The screening effects at leading order can therefore be expected to be of a generic form, 
\begin{align}
&D_c(r)=\frac{1}{p^2+A \alpha_s\Pi}   
\\ 
&D_s(r)=\frac{1}{p^4+B \sigma\Pi} \ , 
\label{eq_msandmd}              
\end{align} 
with a medium-induced polarization tensor \(\Pi\) representing light-parton 
loops\footnote{The leading 
order polarization is just a particle-hole loop.} which are only related with medium properties. 
Thus, they are the same for Coulomb and string terms. However, the same $ \Pi $ can lead to 
different screening behavior since Coulomb and string potentials couple to $ \Pi $  differently. 
Here, we simply assume that this difference can be represented by temperature-independent 
parameters $A$ and $B$ related to spin/color and relativistic structures which are not 
precisely known in our context. From dimensional analysis a ``propagator" of the form 
$ 1/(p^n+m_x^n)=m_x^{-n}/[(p/m_x)^n+1] $ has a screening mass proportional to $m_x$. 
Thus, we have  \(m_d\propto(A \alpha_s\Pi)^{1/2}\) and \(m_s\propto(B \sigma\Pi)^{1/4}\). 
This yields the constraint $m_s= (c_s m_d^2 \sigma/\alpha_s)^{1/4}$ where \(c_s\) is 
depending on $ A $ and $ B $ and other temperature-independent constants. In a 
Debye-H\"uckel approach~\cite{Burnier:2015tda} one obtains the same temperature scaling relation 
for string and Debye masses except for the coefficient $c_s$. However, the resulting screening
behavior of the above model and the Debye-H\"uckel approach can be different. Thus, we do not 
directly use the above propagators or the Debye-H\"uckel approach as our ansatz but simply use 
scaling rules with $c_s$ as a parameter for the Coulomb or string screening masses, which show 
indications of model independence. The above potential is in the quark-antiquark color-singlet 
channel, while the potentials in other channel will be modified by Eqs.~(\ref{eq_Bfator}) and 
(\ref{eq_Rfator}) and Table.~\ref{table_casimir}. 

In our fit procedure, we first constrain the infinite-distance limit of the input potential 
\(V(r)\) by using $F_{Q\bar Q}(\infty,\beta)$ (they are not the same). Then, the ``interference
function", \(\phi(x_e r)\) defined Eq.~(\ref{eq_selfE2fatorize}), is constrained via 
Eq.~(\ref{eq_FreeEfit}), which is a functional fit. The solution for \(\phi(x_e r)\) is unique 
once \(V(r)\) is fixed (it will turn out to have a shape similar to the perturbative limit
in Ref.~\cite{Laine:2006ns}, as will be shown in Figs.~\ref{fig_wcs-V-F} and \ref{fig_scs-V-F},
in Secs.~\ref{sssec_wcs-pot} and \ref{sssec_scs-pot}). 

For the quark mass correction, we have previously defined \(\tilde V(r)\) by adding twice
the Fock term, \(\Delta M_Q=\tilde V(\infty)/2\), to the genuine interaction part of $V(r)$, 
\ie, \begin{equation}
\tilde V(r)=V_\mathcal{C}(r)+V_\mathcal{S}(r)+2\Delta M_Q
\label{eq_vplusmass}
\end{equation} 
where
\begin{equation}
\Delta M_Q=-\frac{1}{2}\int dr \rho(r)V(r) =\frac{1}{2}(-\frac{4}{3}\alpha_s m_d +\sigma m_s)
\label{eq_fockmass}
\end{equation}
is the classical static in-medium selfenergy of a point charge, \(\rho(r)=\delta(r)\), 
in its own field, subtracting the divergent vacuum term. The minus sign arises because 
the charge repels itself. 
Similar physics is discussed in Ref.~\cite{Beraudo:2010tw} in the perturbative context. 
Using Eq.~(\ref{eq_fockmass}) in momentum space with explicit indices, the Fock mass can be 
obtained by the selfenergy from a potential including the relativistic and color factors, 
Eqs.(\ref{eq_Bfator}) and (\ref{eq_Rfator}), 
\begin{align}
&M_q=-\frac{1}{2}\int \frac{d^3\textbf{p}}{(2\pi)^3} V^1_{q\bar q}(\textbf{p})+M_{\text{fit}}\nonumber\\
&M_g=-\frac{1}{2}\int \frac{d^3\textbf{p}}{(2\pi)^3} V^1_{gg}(\textbf{p})+\frac{3}{2}M_{\text{fit}}
\label{eq_fockmasspspace}
\end{align}
where \(M_{\text{fit}}\) is a residual mass (utilized as a fit parameter to the lQCD data for
the EoS), which encodes physics that we do not treat explicitly here (\eg, perturbative 
selfenergies or gluon condensate effects)\footnote{Neglecting the relativistic factor in 
Eq.~(\ref{eq_fockmasspspace}), the relation is \(M_q=\frac{\tilde V(\infty)}{2}+M_{\text{fit}}, 
M_g=\frac{9}{4}\frac{\tilde V(\infty)}{2}+\frac{3}{2}M_{\text{fit}}\).}. 
The non-perturbative gluon-quark mass ratio in the
static limit is \(M_g/M_q=C_A/C_F=9/4\), while in the perturbative limit at high $T$ 
one has \(M_g/M_q=3/2\). The above implementation gives a smooth transition from the 
non-perturbative to the perturbative regime. 
However, the mass dependence in the relativistic factor still requires a selfconsistent 
procedure. We have checked that our default mass fitting scheme, 
using Eq.~(\ref{eq_fockmasspspace}), and the scheme described in the footnote 
below give very similar results, with a maximum difference of 1\% for the resulting 
quark masses, up to 15\% for the gluon masses, 10\% for the selfenergy near $ T\approx$0.2~GeV, 
and at the  5\% level for gluon masses and selfenergies at $ T \approx$0.3~GeV. 
In either case the influence on the emerging spectral properties is not significant.
Preliminary results show that the quark-number susceptibilities are rather sensitive 
to the masses and can provide additional constraints; this will be elaborated in future 
work.

\subsubsection{Numerical fit procedure for lQCD data}
\label{sssec_proc}
Let us briefly lay out the numerical procedure we use to search for solutions of our approach 
that are compatible with the lQCD data for the QGP EoS, quarkonium correlators and static 
$Q\bar Q$ free energies.  At each temperature, we start with a trial values for the potential 
and two light parton masses, and use them to calculate 
the non-perturbative off-shell scattering matrices ($T$-matrices) for light partons. Within the 
formalism laid out in Sec.~\ref{sec_Tmatrix}, we keep 6 partial waves to include two-body channels 
with angular momentum up to $l$=5 (which is more than sufficient for convergence); with four color 
channels in the $qq$ and $q\bar q$, three in the $qg$ and three in the $gg$ sector~\citep{Shuryak:2004tx}, a total of 
6$\times$10=60 different light-parton $T$-matrices are computed.
These $T$-matrices are then used to calculate the selfenergies and spectral functions for single partons.
Next, the parton propagators are reinserted back into the $T$-matrices, forming a selfconsistency 
problem (recall Eq.~(\ref{eq_selfcEq})) which is solved by numerical iteration; this forms the 
``inner" light-parton selfconsistency loop of the overall procedure. The pertinent outputs are 
then used to compute the EoS and LWF as discussed in Sec.~\ref{ssec_EoS}.
If the resulting pressure disagrees with the lQCD value at the given temperature, the light-parton 
masses ($M_{\text{fit}}$) are retuned, the inner selfconsistency loop carried out, and repeated 
until the EoS is reproduced,  constituting the ``intermediate" mass fitting loop of the overall
procedure.
After obtaining the masses to reproduce the lQCD EoS, we proceed to the selfconsistent
calculation of the selfenergy of a static quark (again a selfconsistency loop), which involves 
another 42 static-light $T$-matrices (6 partial waves and a total of seven color channels for 
$Qq$, $Q\bar q$ and $Qg$). These are input to the formalism laid out in Sec.~\ref{ssec_freeE} to 
compute the static-quark free energy, $F_{Q\bar{Q}}$, and compare it to pertinent lQCD data. 
If the calculated free energy disagrees with the lQCD data, we retune the potential (most
notably $m_d$), recalculate the EoS with retuned light-parton masses, and recompute the free 
energy, which corresponds the  ``outer" potential fitting loop of the procedure.
These loops involve automated (numerical) adjustments of $M_{\text{fit}}$ and $m_d$ to best 
reproduce the EoS and free-energy data while other parameters are tuned manually. After obtaining 
a solution, we proceed to the selfconsistent calculations of charm- and bottom-quark properties 
which involve another 42 heavy-light $T$ matrices each. With the full off-shell HQ spectral 
functions, we proceed to evaluate two more $T$-matrices to compute charmonium and bottomonium
spectral functions and correlator ratios in the pseudoscalar color-singlet $S$-wave channel, 
and compare the latter to lQCD data as discussed in Sec.~\ref{ssec_correlator}. If they do not 
match, we manually retune the potential (mostly the Coulomb term)  and redo the whole process 
until a satisfactory result is obtained. Usually, the fits to the correlator ratio are automatically 
``satisfactory" with the assumption that $\alpha_s$ does not strongly depend on temperature. 
The numerical machinery is carried out with {\em Mathematica} software and typically takes several 
hundreds of CPU hours to arrive at a solution at four temperatures.

\section{Selfconsistent Numerical Results}
\label{sec_results}
In this section, we discuss the results and insights from the above framework. 
For each solution at a given temperature, all quantities in both HQ and light-parton 
sectors, \ie, the QGP EoS, free energy, one- and two-body spectral functions and \(T\)-matrices, 
are all calculated from a single Hamiltonian, Eq.~(\ref{Hqgp}), with the potential ansatz
described in Sec.~\ref{ssec_ansatz}, and then using the \(T\)-matrix approach 
with one set of parameters. The interference effect discussed in Sec.~\ref{sssec_inter2} 
is only included when evaluating static-static and heavy-heavy spectral functions and 
correlators/free energies. 

As it turns out, the constraints provided by the currently used set of lQCD data (free energies,
quarkonium correlators and EoS) does not yet allow for a unique solution. To explore this feature, 
we will focus on two putatively limiting cases, which we denote by a weakly coupled solution (WCS)
where the potential is close to the free energy (Sec.~\ref{ssec_wcs}) and which has already been
discussed in the literature in perturbatively inspired frameworks~\cite{Beraudo:2007ky,Burnier:2014ssa,Burnier:2015tda},
and a strongly coupled solution (SCS) which is characterized by a long-range potential  
which ``maximally" rises above the free energy (Sec.~\ref{ssec_scs}), first pointed out in 
Ref.~\cite{Liu:2015ypa}.
Although both solutions can explain the chosen set of lQCD data, they predict, as we will see, 
a rather different microscopic structure of the QGP at moderate temperatures.

A similar discussion has been presented before in phenomenological applications heavy-flavor
observables, both for HQ diffusion~\cite{Riek:2010fk,He:2011qa} and quarkonium 
transport~\cite{Zhao:2010nk,Liu:2010ej,Emerick:2011xu,Strickland:2011aa}. In these instances 
the internal and free energies have been employed as potential proxies for strongly and weakly 
coupled scenarios of the in-medium QCD force. A general tendency for preferring the internal
energy was found. Such studies can, of course be repeated with our more rigorously deduced 
potential solutions.

One of the virtues of our approach is that it is carried out in real-time, allowing us 
to retain and keep track of the microscopic quantum many-body information about the system in 
a direct way while being intimately connected to the macroscopic properties of the QGP. 
This includes the predicted spectral functions of all involved partons (static, heavy and 
light quarks as well as gluons) and the more than one-hundred in-medium two-body $T$-matrices, 
fully off-shell. This information readily allows to calculate transport coefficients, Wigner 
functions for one- or two-body states, etc., in a nonperturbative framework, and to make 
contact with experimental observables. Thus, the approach is not only rooted in lQCD data, 
but also unravels real-time microscopic physics which predicts a wide variety of phenomena 
that can be tested by experiments in a transparent, quantitative and interpretable way. 

\subsection{Weakly Coupled Solution}
\label{ssec_wcs}
\begin{figure*}[!t]
        \begin{center}
                \includegraphics[width=1.739\columnwidth]{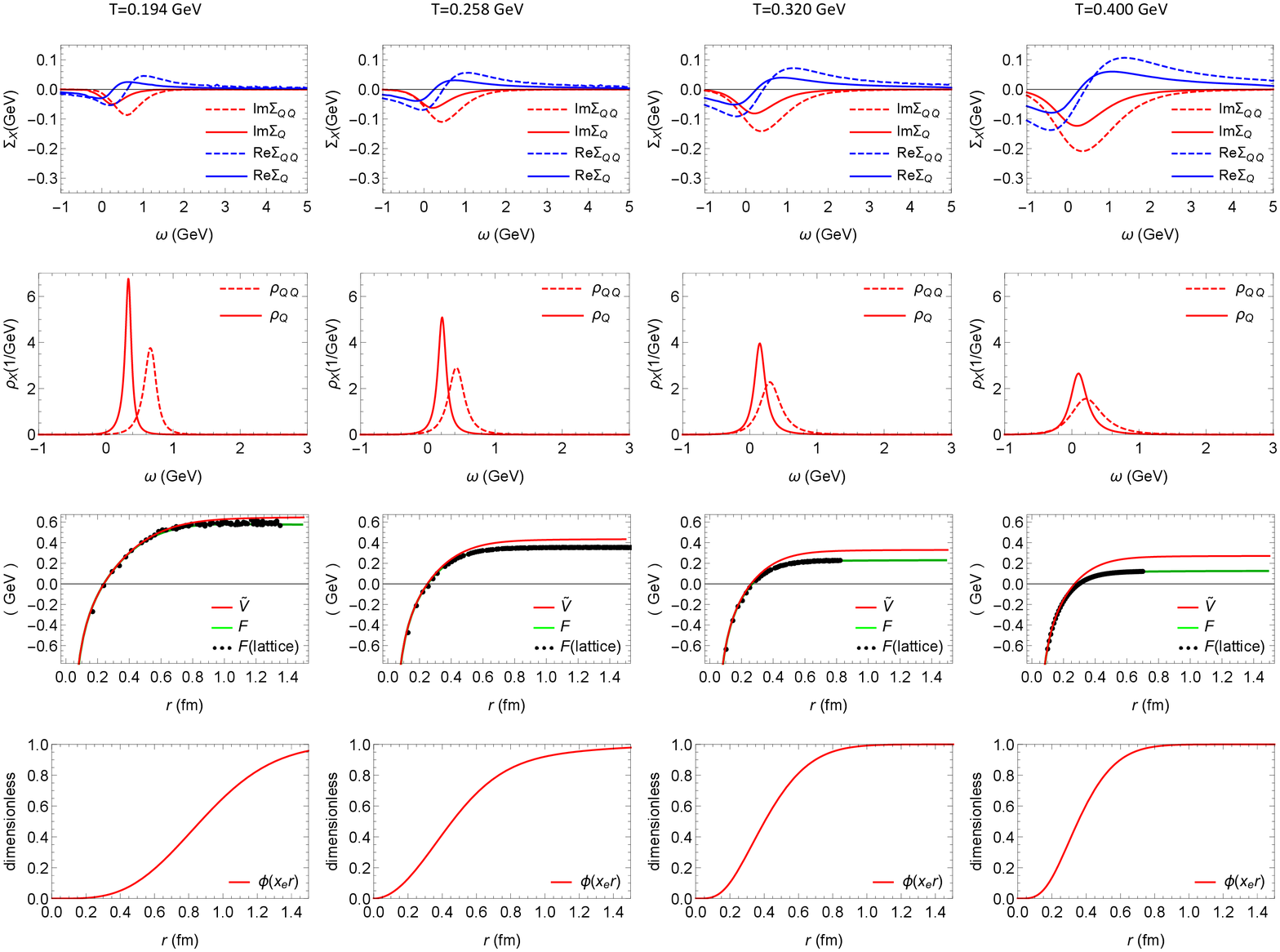}
                \includegraphics[width=0.26\columnwidth]{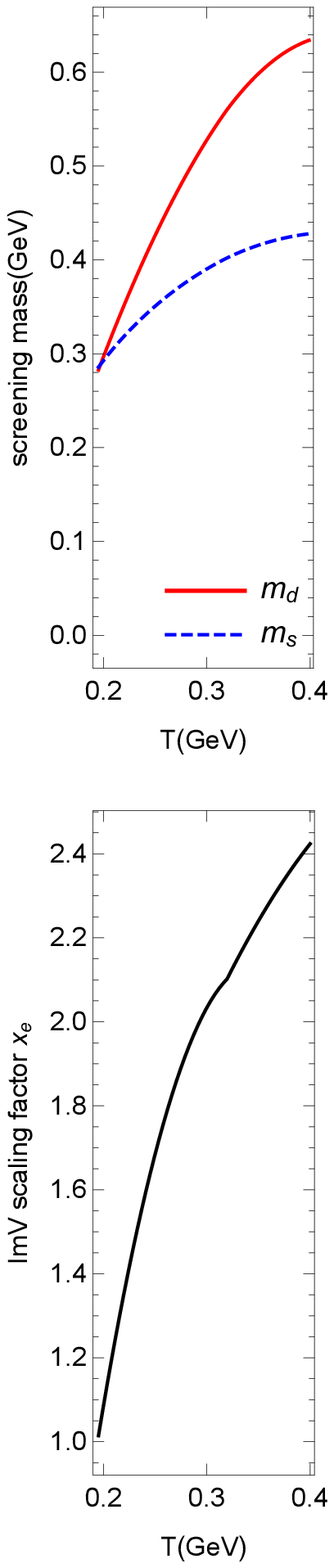}
                \caption{Results of a {\em weakly} coupled solution for the selfconsistent 
fit to extract the static HQ potential: single HQ and $Q\bar Q$ selfenergies, 
\(\Sigma_X(\omega,\infty)\) (first row), and spectral functions, \(\rho_X(z,\infty)\) 
(second row), potential \(\tilde{V}(r)\) and free energies (third row), and interference 
function, \(\phi(x_e r)\) (fourth row), in the first 4 columns corresponding to different 
temperatures. 
The last column shows the temperature dependence of the fitted screening masses (top panel) and 
the scale factor, $x_e$ (bottom panel), figuring in the interference function. 
The free-energy lQCD data are from Ref.~\cite{Mocsy:2013syh}.}
                \label{fig_wcs-V-F}
        \end{center}
\end{figure*}

In this section we first report and discuss the results of our fits for a weakly coupled
solution (WCS), starting from the HQ free energy and the extraction of the underlying potential, 
which is the key quantity determining the interaction strength in the QGP 
(Sec.~\ref{sssec_wcs-pot}) and pivotal for calculating essentially all other quantities. 
In Sec.~\ref{sssec_wcs-corr} we elucidate the extra information that can be gained by the fits 
of euclidean quarkonium correlators, and discuss the resulting charmonium and bottomonium 
spectral functions. 
We then proceed to our fit to the QGP EoS which involves the two light-parton masses as 
additional fit parameters (Sec.~\ref{sssec_wcs-eos}). We finally give a comprehensive overview 
of the emerging light and heavy-parton spectral functions and their two-body $T$ matrices 
(Sec.~\ref{sssec_wcs-spec}) and a discussion of the pertinent QGP structure, including its degrees 
of freedom. 

\subsubsection{Free energy, potential and static selfenergies}
\label{sssec_wcs-pot}
\begin{figure*}[!t]
        \centering
        \fbox{\includegraphics[width=1.93\columnwidth]{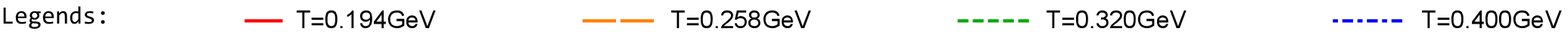}}
        \includegraphics[width=2.00\columnwidth]{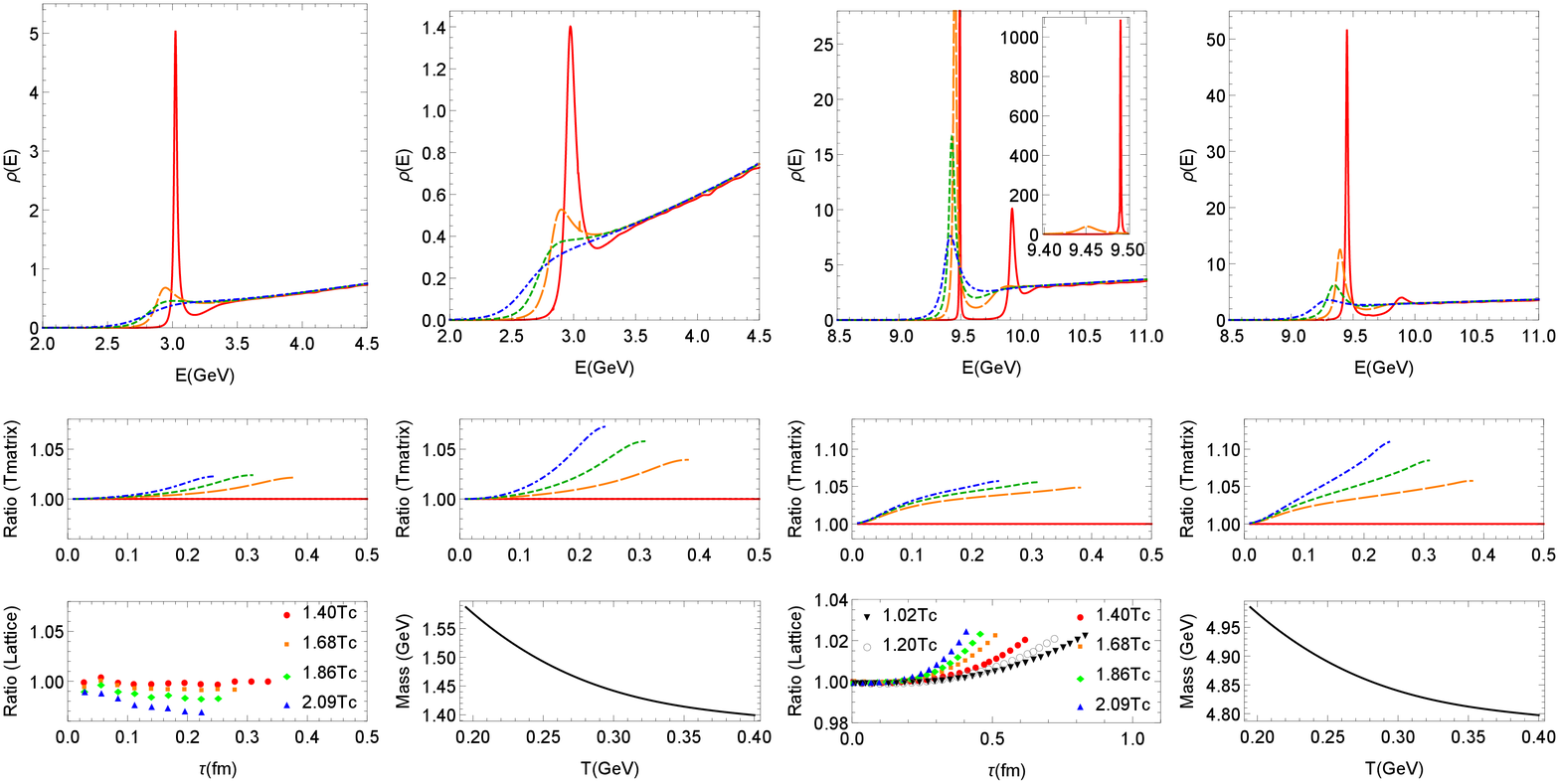}
     \caption{{\em Weakly} coupled solution for charmonium (\(\eta_c\), left panels) and bottomonium
($\eta_b$, right panels) spectral functions (upper panels) and correlators ratios (middle panels)
with (first and third column) and without (second and fourth column) interference effects in the
imaginary part of the potential. The lQCD data for \(\eta_c\)~\cite{Aarts:2007pk} and
\(\eta_b\)~\cite{Aarts:2011sm}  correlator ratios are shown in the first and third bottom panel,
respectively, while the second and fourth bottom panel display the temperature
dependence of the charm- and bottom-quark mass, respectively.}
        \label{fig_wcs-corr}
\end{figure*}

When searching for a WCS, we start by using the free energy as potential. The strength 
of the potential slightly increases in the iteration procedure, mostly due to 
relatively small imaginary parts that develop and figure in the 
static $Q\bar Q$ spectral function, Eq.~(\ref{eq_FreeEfit}).  
Thus, the solution found in this way can be regarded as a lower limit of the potential.
The parameters of the potential for the converged solution are given by \(\alpha_s=0.27\), 
\(\sigma=0.21\)\,GeV$^2$,  \(c_b=1.3\) and a temperature dependent Coulomb Debye mass, \(m_d\), 
as shown in the upper right panel of Fig.~\ref{fig_wcs-V-F}. With \(c_s=0.1\) the screening mass 
of the string term, $m_s= (c_s m_d^2 \sigma/\alpha_s)^{1/4}$, also follows as shown in the same 
panel. The fit of the interference function, shown in the lowest row of Fig.~\ref{fig_wcs-V-F},
is quite similar to the perturbative function found in Ref.~\cite{Laine:2006ns}; it shrinks
in range as a result of the increase in screening with temperature. The resulting potential is 
displayed in the third row of Fig.~\ref{fig_wcs-V-F} and indeed found to exceed the free energy, 
by up to 0.07~GeV at \(T=0.194\)~GeV and 0.16~GeV at \(T=0.4\)~GeV. The calculated free energy 
fits the lQCD data well.  
 
With this potential, the selfconsistent selfenergy and spectral function of a 
static quark follow from \(T\)-matrix approach as shown in the first two rows of 
Fig.~\ref{fig_wcs-V-F}, respectively. In practice, the static limit has been calculated with 
a numerically large bare HQ mass ($2\cdot10^4$ GeV), and the energy scales for the one- (and two-) 
body quantities have been plotted relative to (twice) that bare mass. At low \(T=0.194\)~GeV, 
the peak value of \(\text{Im}\Sigma_Q\approx-0.05\)~GeV coresponds to a width of the spectral 
function which is around 0.1~GeV. For comparison, the hard-thermal-loop (HTL) perturbative 
width~\cite{Laine:2006ns,Beraudo:2007ky,Beraudo:2010tw} is \(\frac{4}{3}\alpha_sT\approx0.07 \)GeV. 
For the $Q\bar Q$ quantities, the peak value of \(\text{Im}\Sigma_{QQ}\) as defined in 
Eq.~(\ref{selfE2}) and (\ref{eq_selfE2fatorize}) is approximately 2 times of the peak value 
of \(\text{Im}\Sigma_Q\), and the width of the two-body spectral function is around 2 times that 
of the single static-quark spectral function. The peak value of \(\text{Im}\Sigma_Q\) and 
the width of the static quark spectral functions increase with temperature at an approximately
linear rate.

\subsubsection{Quarkonium Correlators and Spectral Functions}
\label{sssec_wcs-corr}
Next we turn to the Euclidean quarkonium correlators for realistic bottom- and charm-quark masses,
concentrating on the pseudoscalar channel where extra complications due to zero modes do not 
figure, see Fig.~\ref{fig_wcs-corr}. The bare masses of charm and bottom quarks ($Q$=$c,b$) are determined as in 
Ref.~\cite{Riek:2010fk}, by fitting the vacuum charmonium and bottomonium ground-state masses 
using \(m_{Q}=m^{\text{bare}}_{Q}+\tilde{V}(\infty)/2\) with the vacuum value of $\tilde{V}$ 
at a typical string breaking scale of $r$=1-1.1\,fm, resulting in 
$m^{\text{bare}}_{c,b}$=1.264, 4.662\,GeV.

The widths of the quarkonium spectral functions are caused by collisions of individual heavy quarks 
within the bound state with medium partons (the so-called quasifree process~\cite{Grandchamp:2001pf}), 
as encoded in the HQ selfenergies. Since the potential is relatively weak, these selfenergies  
are small, and so is the width of quarkonium. The $\eta_c$ is still a well-defined state at 
$T$=200\,MeV, but is essentially dissolved at $T$=260\,MeV. The $\eta_b(1S)$ survives to significantly
higher temperatures, beyond 260\,MeV, and even to 400\,MeV when interference effects are included
(as described in Sec.~\ref{sssec_inter2}). The latter generally reduce the quarkonium widths,
more so the tighter the states are bound (by up to 75\%). The width reduction is consistent with
simple estimates using the \(\phi(x_e r)\) function (Fig.~\ref{fig_wcs-V-F}) with pertinent
size estimates. Even for the case without interference, the width of the $Q\bar Q$ states is 
smaller than 2 times the HQ width at vanishing  momentum, due to the energy-momentum dependence 
of the HQ selfenergies as obtained from the heavy-light \(T\)-matrices. As usual, the dissolution
of the quarkonia is due to a combination of the increasing screening and collision widths.

The correlator ratios are generated by using the reference (or ``reconstructed") correlator 
at the lowest temperature considered ($T$=194\,MeV), as was done in the lQCD calculations 
that we compare to~\cite{Aarts:2007pk,Aarts:2011sm}. Without interference effects the calculated 
correlator ratios deviate from the lQCD data by up to $\sim$10\%. Despite the melting of the bound 
states, the increase in width effects (over-) compensates the loss of low-energy strength in the 
spectral functions and leads to a 5-10\% increase in the correlators ratios with increasing 
euclidean time, $\tau$. This increase is tamed by the inclusion of interference effects, which, 
as discussed above, reduce the bound-state widths; the resulting correlator ratios agree within 
$\sim$5\% with the lQCD data. 
Furthermore, the correlator ratios are quite sensitive to the strong coupling constant, $\alpha_s$ 
(approximately proportional to it, reflecting its short distance nature as a local operator related 
to the wave function overlap at the origin (recall the discussion in Sec.~\ref{ssec_correlator})). 
Thus, the deviations between our results and the lQCD data could be further mitigated by a fine-tuning 
of $\alpha_s$, slightly decreasing with temperature at a few-percent level. In our fits we did not 
explore such a dependence, given other uncertainties that can affect the correlator ratios at a
similar level (\eg, spin-dependent interactions). In turn, one could argue that the fact that the
lQCD correlator ratios are quite close to 1 at all temperatures suggest that $\alpha_s$ is
not strongly running with temperature.

\begin{figure*}[t]
  \centering
  \includegraphics[width=1.99\columnwidth]{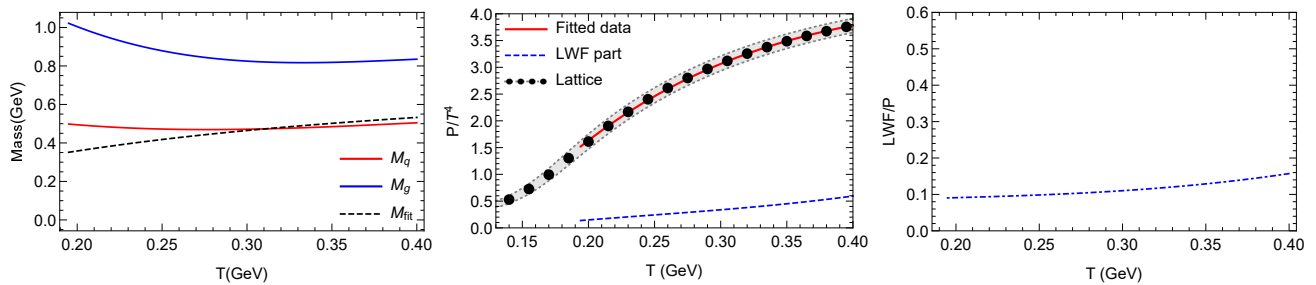}
  \caption{{\em Weakly} coupled solution for the QGP bulk medium: fit results for the 
input masses for quarks and gluons (left panel), the QGP pressure in comparison to 
lQCD data~\cite{Bazavov:2014pvz} (middle panel; solid line: total, dashed line: LWF
contribution), and the ratio of LWF contribution to total pressure (right panel).}
        \label{fig_wcs-eos}
\end{figure*}

\begin{figure*}[!t]
        \begin{center}
        	    \fbox{\includegraphics[width=2.04\columnwidth]{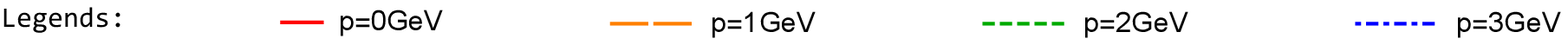}}
                \fbox{\includegraphics[width=1.0\columnwidth]{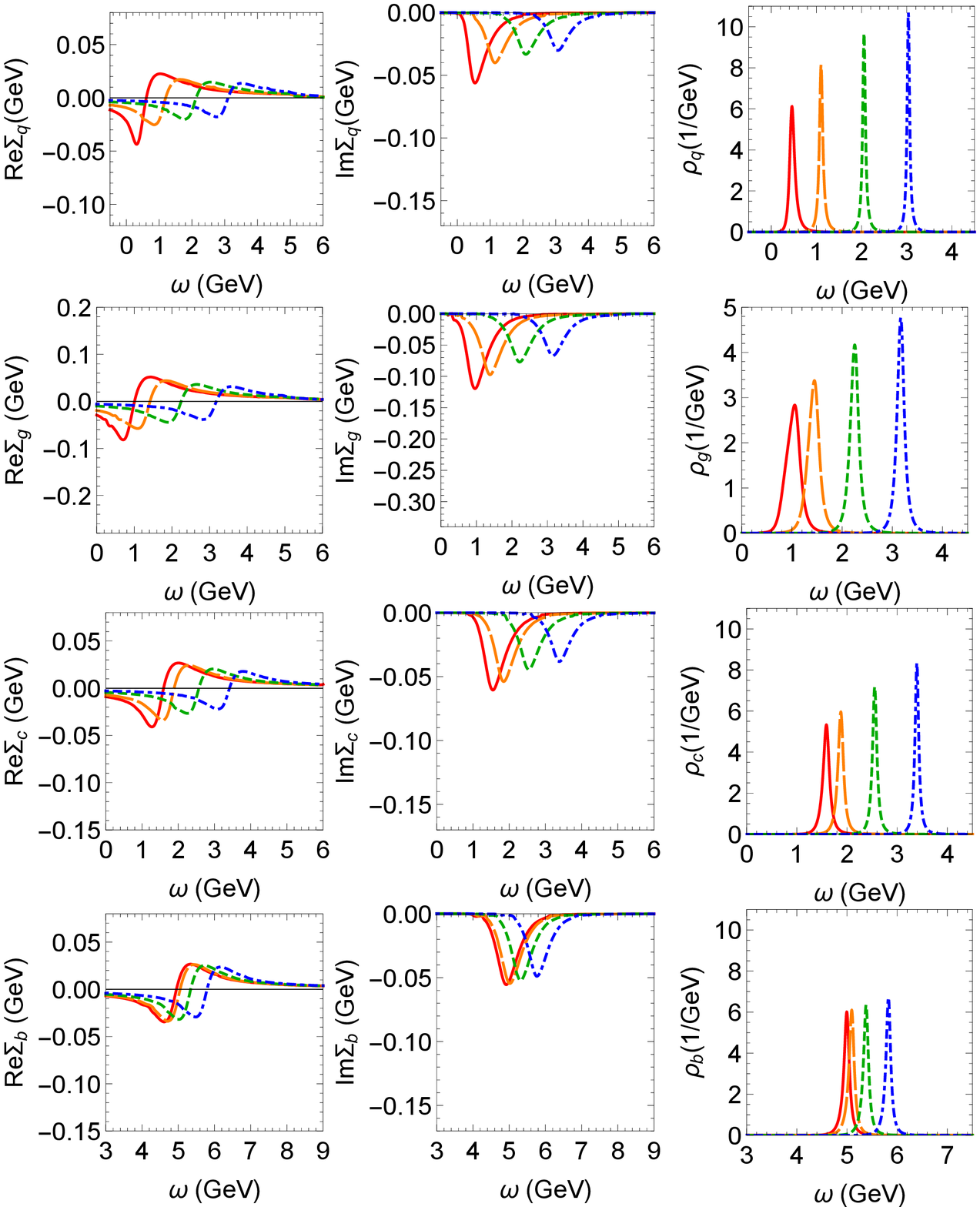}}
                \fbox{\includegraphics[width=1.0\columnwidth]{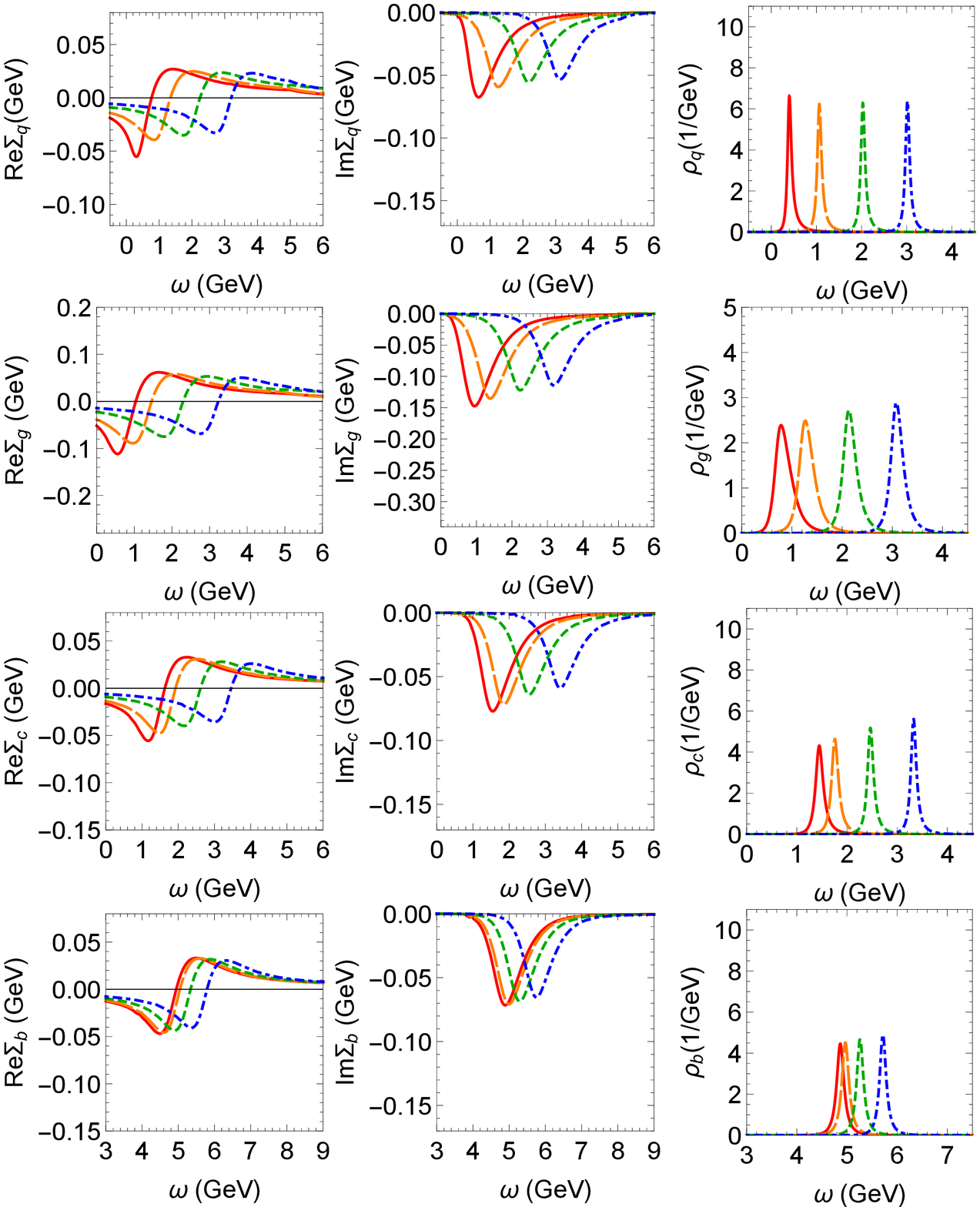}}
                \fbox{\includegraphics[width=1.0\columnwidth]{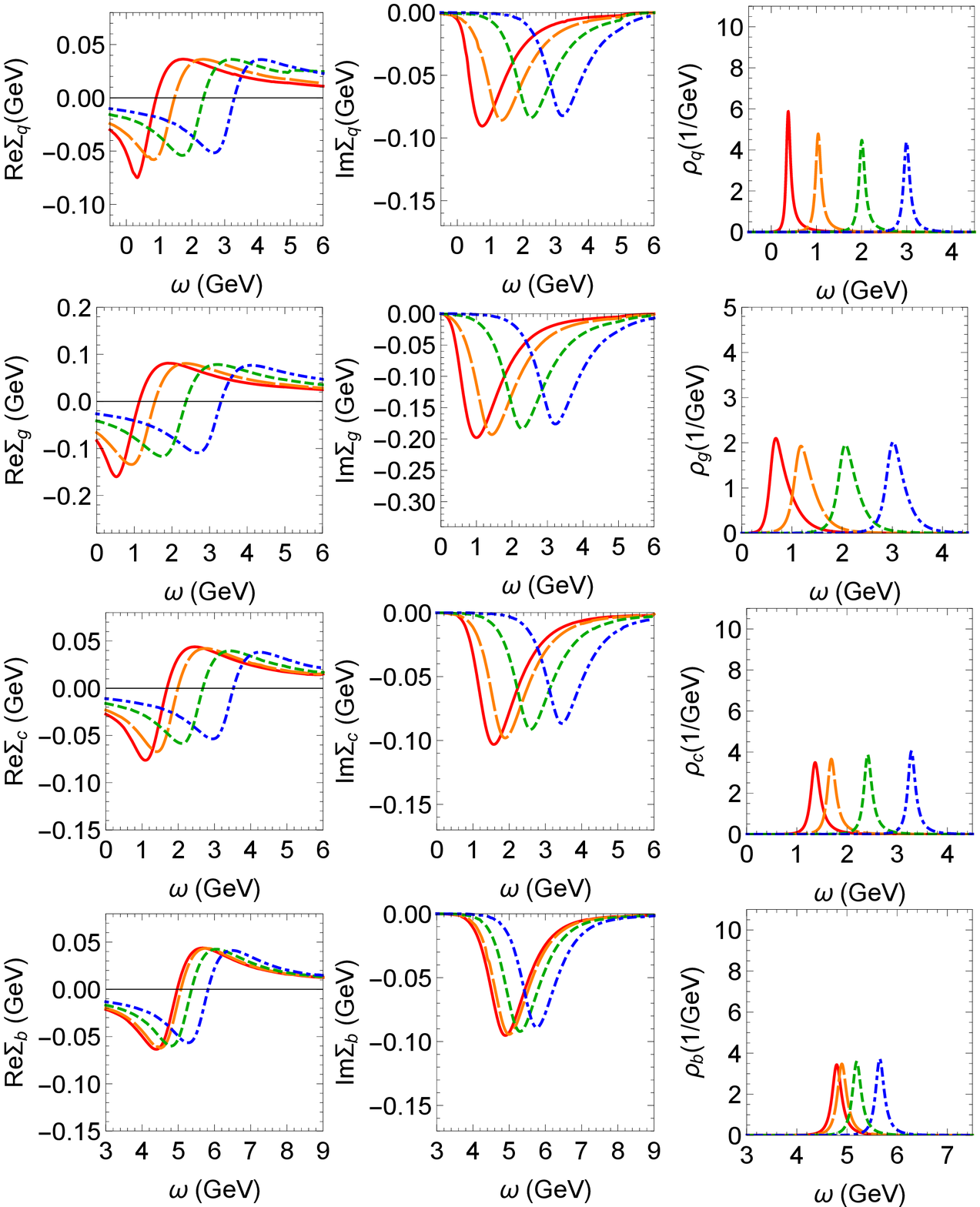}}
                \fbox{\includegraphics[width=1.0\columnwidth]{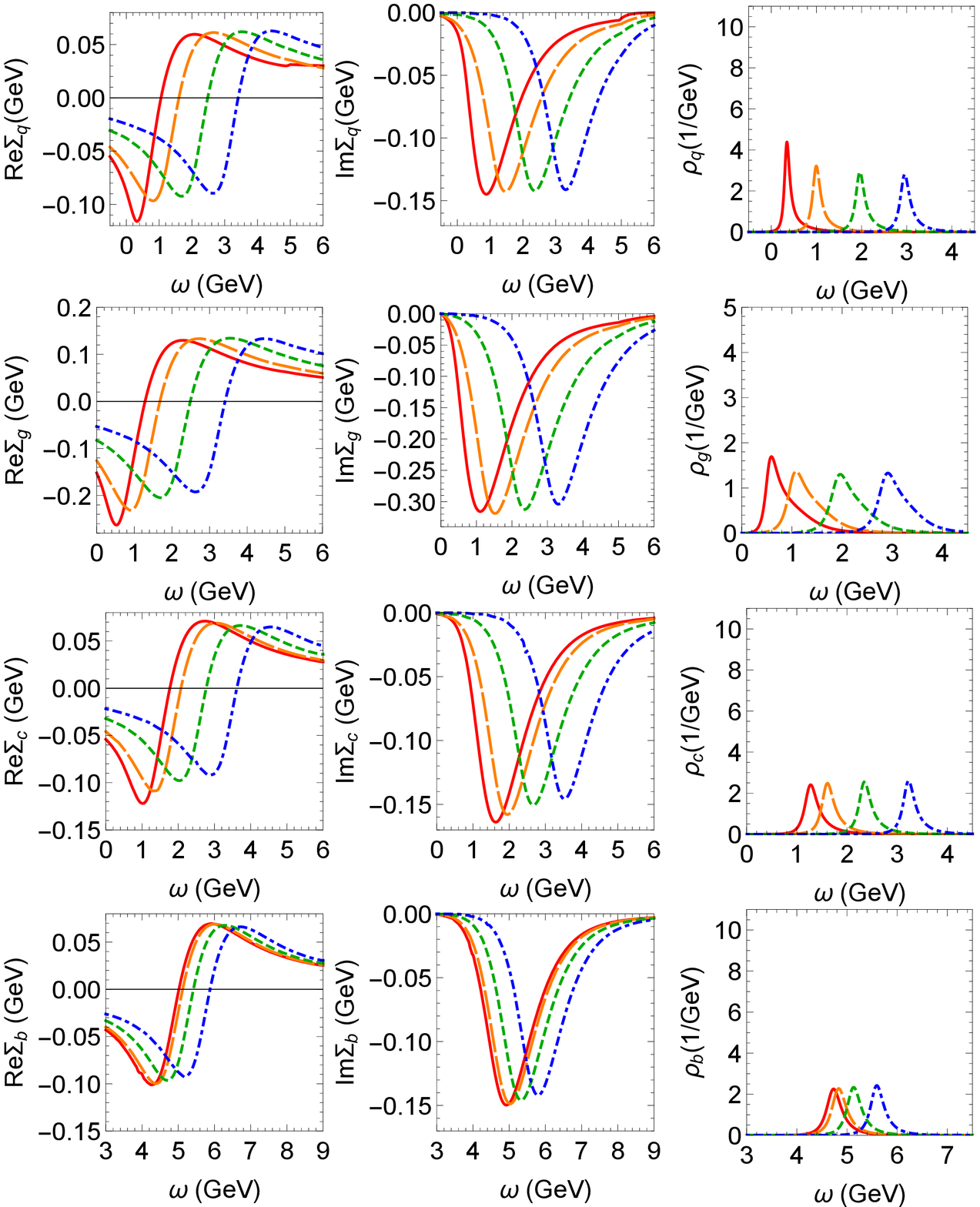}}
                \caption{{\em Weakly} coupled solution for parton spectral properties of the QGP.
The figure is organized into four 3-by-4 panels of 12 plots, with each panel for a given 
temperature (upper left: $T$=0.194\,GeV, upper right: $T$=0.258\,GeV, lower left: $T$=0.320\,GeV 
and lower right: $T$=0.400\,GeV). 
Each panel contains 4 rows corresponding to different parton species (light quarks ($q$), gluons 
($g$), charm quarks ($c$) and bottom quarks ($b$) in the first, second, third and fourth row of
each panel, respectively). Each row contains 3 panels showing (from left to right) the energy 
dependence of the pertinent real and imaginary part of the selfenergy and the resulting spectral 
functions, for 4 different values of the single-parton 3-momentum ($p$) in the
thermal frame.}
                \label{fig_wcs-spec}
        \end{center}
\end{figure*}

\begin{figure*}[!t]
	\centering
	\fbox{\includegraphics[width=1.98\columnwidth]{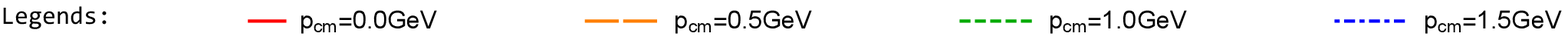}}
                \includegraphics[width=2\columnwidth]{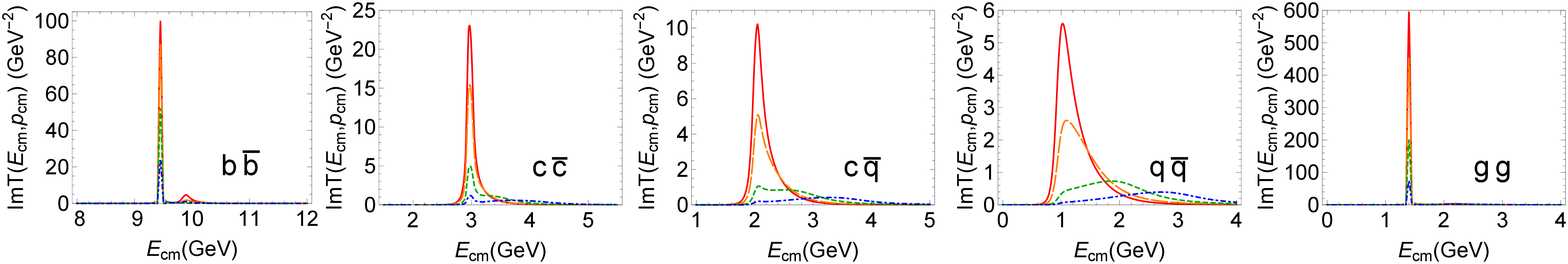}
                \includegraphics[width=2\columnwidth]{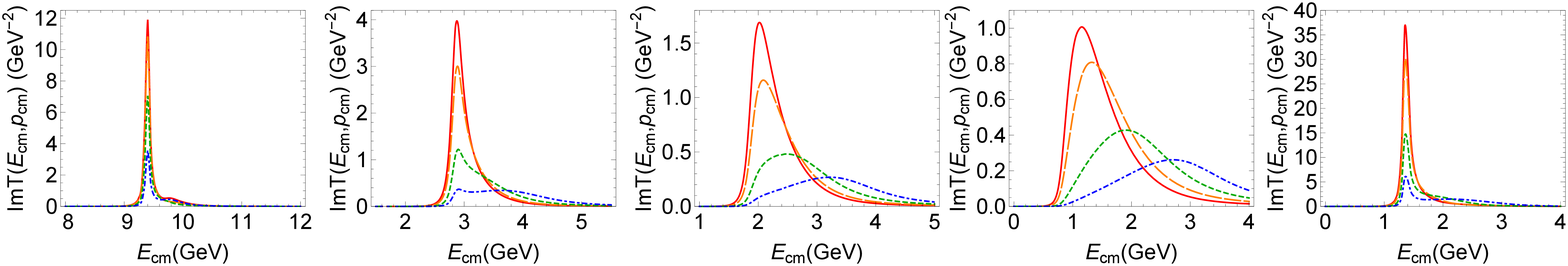}
                \includegraphics[width=2\columnwidth]{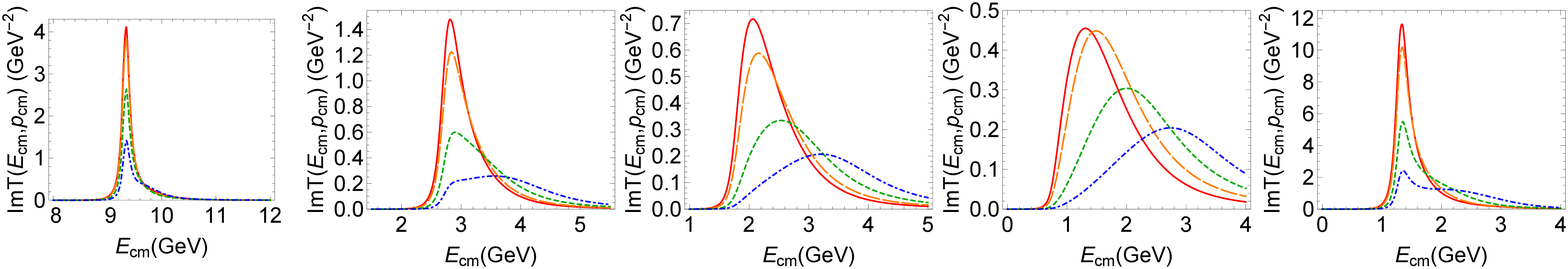}
                \includegraphics[width=2\columnwidth]{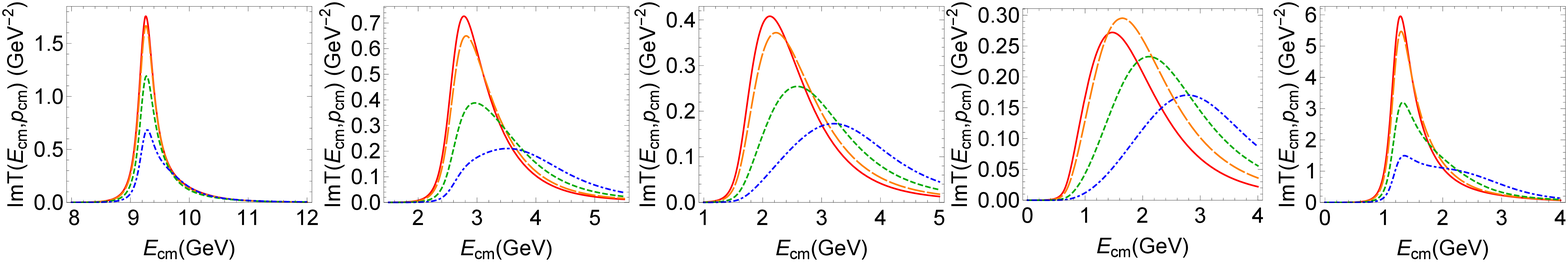}
                \caption{{\em Weakly} coupled solution for the imaginary part of the color-singlet
$S$-wave \(T\)-matrices (without interference effects) in the bottomonium ($b\bar b$; first 
column), charmonium ($c\bar c$; second column), $D$-meson ($c\bar q$; third column),
light-quark ($q\bar q$; fourth column), and glueball ($gg$, last column) channels.   
The 4 rows correspond to different temperatures, $T=0.194$~GeV, $T=0.258$~GeV, 
$T=0.320$~GeV and $T=0.400$~GeV from top down; in each panel, the $T$-matrix is displayed for
4 different values of the single-parton 3-momentum ($p_{cm}$) in the two-body CM
frame.}
              \label{fig_wcs-T}
\end{figure*}

\subsubsection{QGP Equation of State}
\label{sssec_wcs-eos}
Next, we turn to the selfconsistent results for the QGP bulk properties, \ie, our fit to the
lQCD data for the pressure. Here, the two main fit parameters are the bare light-parton masses 
in the Hamiltonian (including the Fock term, recall Eq.~(\ref{eq_fockmasspspace})), which are 
shown in the left panel of Fig.~\ref{fig_wcs-eos}. The resulting masses are rather stable with 
temperature, with a slight increase toward $T_{\rm pc}$ dictated by the decreasing pressure (not 
unlike in quasiparticle models, but less pronounced, especially for quarks). The quark-to-gluon 
mass ratio is different from the perturbative thermal mass ratio due to the nonperturbative 
ingredients of the interaction
as discussed in Sec.~\ref{ssec_ansatz}. The fitted mass parameter, \(M_\text{fit}\), 
starts to exceed \(M_q\) for temperatures above 300\.MeV due the negative Coulomb contribution
to the Fock term (which is also enhanced by relativistic corrections); the string term gives 
a strictly positive contribution (which is, however, suppressed by relativistic corrections).

The lQCD data for the pressure can be well reproduced, see middle panel of Fig.~\ref{fig_wcs-eos}.
It is interesting to decompose the pressure into contributions from quasiparticles 
($\Omega_{\rm qp}\propto \ln(-G^{-1})+\Sigma G$) \cite{Rapp:1993ih} and the two-body interaction characterized 
by the resummed LWF ($\Phi \propto 1/2\log(1-VGG)$). The latter turns out to be generally small,
no more than 15\% of the total and slightly increasing with the temperature, cf.~right panel
of Fig.~\ref{fig_wcs-eos}. This suggest that there are no marked changes in the interaction
strength or degrees of freedom in the WCS for the QGP in the considered temperature range.

\subsubsection{Spectral Structure of QGP}
\label{sssec_wcs-spec}
Finally, let us inspect the spectral structure of the QGP within the WCS. The spectral properties 
of single partons are summarized in Fig.~\ref{fig_wcs-spec} in terms of their selfenergies (real 
and imaginary parts) and spectral functions. The widths (or scattering rates) of the partons, 
\(\Gamma = -2\text{Im}\Sigma\), are significantly smaller than their masses, implying
that they remain well-defined quasiparticles at all momenta and over the full temperature range.  
At the lowest temperature, $T$=194\,MeV, the light-parton width is around 0.11 GeV which is 
larger than the perturbative expectation, \(\frac{4}{3}\alpha_s T\approx 0.07\)~GeV, but lower 
than, \eg, the most recent dynamical quasiparticle model results~\cite{Berrehrah:2016vzw} 
which are around 0.2~GeV. Similar to the static case, the width rises slightly stronger than 
linear with temperature, which is closer to the  perturbative than the dynamical quasiparticle 
approach. The 3-momentum dependence of the width is quite strong at low temperature and quite 
weak at high temperature. This is probably so because partons at different thermal momenta 
will probe different regimes of the potential, in particular since at high temperature the 
string term (which is responsible for an appreciable long-range force) is heavily screened. In 
the infrared region, the confining interaction behaves as \(1/m_s^4\) while the Coulomb one as 
\(1/m_d^2\). Thus, the increase of \(m_s\) implies  a larger decrease of the strength of 
the string relative to the Coulomb force (the latter is also augmented by the relativistic 
Breit correction that reduces the momentum dependence). The width of the different quark species 
are quite similar whereas the gluon width is almost twice larger due to the color Casimir factor. 
The quark width first increases with mass and then decreases again. Usually a larger mass has 
a stronger scattering amplitude in the CM frame (cf.~Fig.~\ref{fig_wcs-T}), but the CM 
transformation, Eq.~(\ref{eq_CM}), effectively shrinks the phase space. This competition 
leads to the non-monotonic behavior.

The underlying two-body correlations are illustrated by the (imaginary part of the) pertinent 
\(T\)-matrices, used to calculate the single-parton selfenergy, in Fig.~\ref{fig_wcs-T}. They 
exhibit a sequential dissociation according to the reduced mass of the bound state. 
If we use a vanishing binding energy (relative to the constituent 2-body mass threshold) to 
distinguish bound and scattering states (for total momentum $P$=0), light mesons 
are melted at \(T=0.194\)\,GeV
while the heavy-light meson, glueball, and  quarkonium still survive. The $D$-meson
and first-excited bottomonium state (\(\Upsilon_{2S}\)) melt near \(T=0.258\)\,GeV, the 
charmonium around \(T=0.320\)\,GeV and the ground-state bottomonium 
\(\Upsilon_{1S}\) above \(T=0.400\)\,GeV.
Even after melting, a resonance structure can still survive to somewhat higher temperatures,
albeit with typically much reduced strength in the $T$ matrix. 
As an alternative way to characterize the resonance correlation one can inspect their
robustness with increasing single-parton CM momentum (essentially going off-shell), 
the light, heavy-light and first-excited bottomonium states disintegrate for 
$p_{\rm cm}\ge1$\,GeV.
We finally note that the $q\bar q$ bound-state mass at the lowest temperature, 
$M_{q\bar q}\simeq 1$\,GeV, is significantly larger than the vacuum mass of the light 
vector mesons, $m_{\rho,\omega}\simeq780$\,MeV (we recall that we do not include spin-spin
or topologically induced interactions, \eg, instanton-induced ones, which are believed to 
play a key role for dynamical chiral symmetry breaking and its associated Goldstone bosons).

\subsection{Strongly Coupled Solution}
\label{ssec_scs}
\begin{figure*}[!t]
        \begin{center}
                \includegraphics[width=1.739\columnwidth]{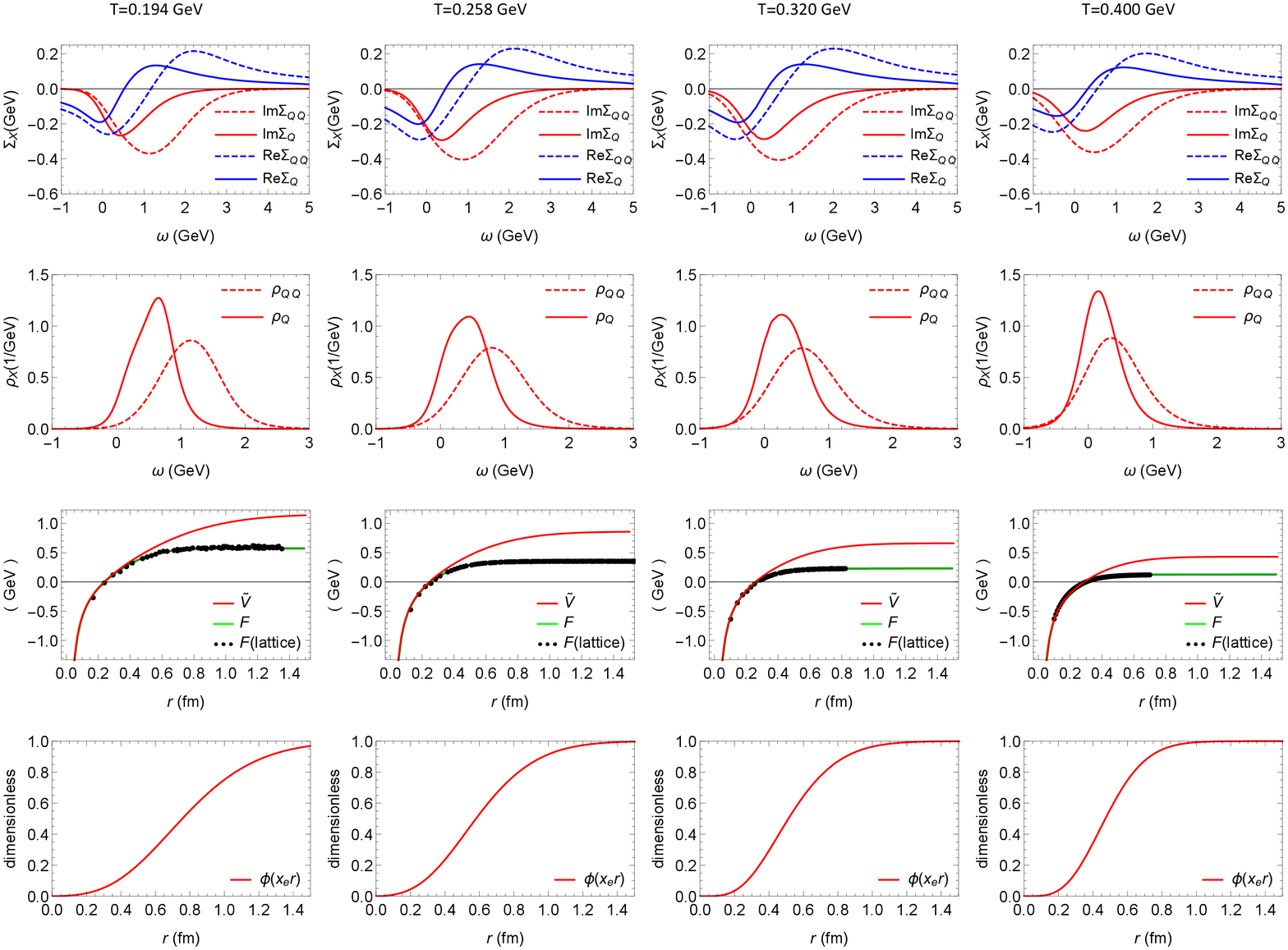}
                \includegraphics[width=0.26\columnwidth]{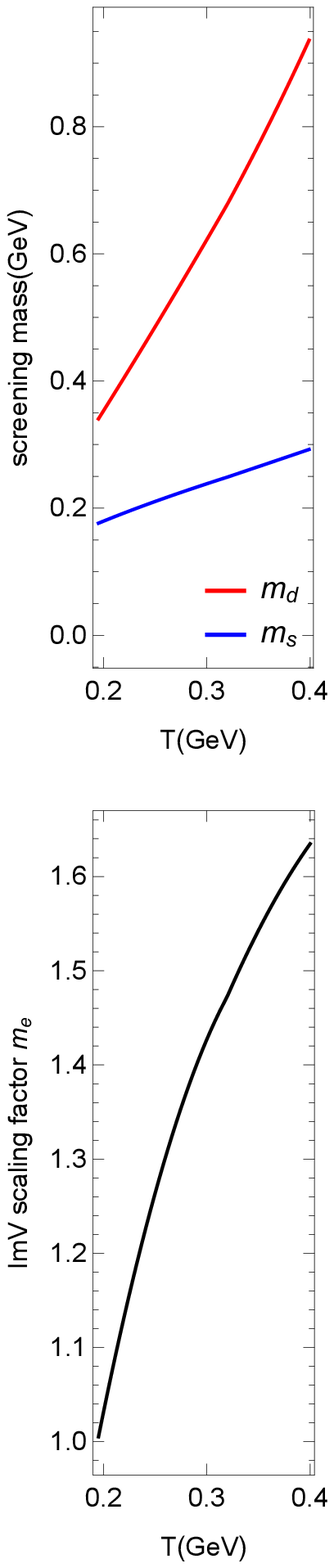}
   \caption{Results of a {\em strongly} coupled solution for the selfconsistent 
   	fit to extract the static HQ potential: single-quark and $Q\bar Q$ selfenergies, 
   	\(\Sigma_X(\omega,\infty)\) 
   	(first row), and spectral functions, \(\rho_X(z,\infty)\) (second row), 
   	potential \(\tilde{V}(r)\) and free energies (third row), and interference function, 
   	\(\phi(x_e r)\) (fourth row), in the first 4 columns corresponding to different temperatures. 
   	The last column shows the temperature dependence of the fitted screening masses (top panel) and 
   	the scale factor, $x_e$ (bottom panel), figuring in the interference function. 
   	The free-energy lQCD data are from Ref.~\cite{Mocsy:2013syh}.}
                \label{fig_scs-V-F}
        \end{center}
\end{figure*}

In this section we discuss our selfconsistent set of results for a strongly coupled
solution (SCS). The section structure parallels the one of the WCS, namely starting from 
the determination of the underlying potential through fits of lQCD results for the 
static $Q\bar Q$ free energy (Sec.~\ref{sssec_scs-pot}), followed by the quarkonium correlator
analysis (Sec.~\ref{sssec_scs-corr}), the fit to the QGP EoS 
(Sec.~\ref{sssec_scs-eos}) and a discussion of the one- and two-body spectral 
properties (Sec.~\ref{sssec_scs-spec}).

\subsubsection{Free Energy, Potential and Static Self-energies}
\label{sssec_scs-pot}
When searching for a SCS  within our framework, we start from a trial potential significantly
larger than the free energy, together with large imaginary parts in the static-quark 
selfenergies. 
The converged selfconsistent parameters take the values \(\alpha_s=0.27\), 
\(\sigma=0.225\)\,GeV$^2$, \(c_b=1.3\) and \(c_s=0.01\). The strong coupling constant and
the ``string-breaking" coefficient, $c_b$, are essentially the same as for the WCS, and 
the string tension is only about  $\sim$5\% larger. The key difference lies in the 
coefficient, $c_s$, for the screening mass of the string term, which is a factor
of $\sim$10 smaller. Consequently, the temperature dependent screening mass, 
$m_s= (c_s m_d^2 \sigma/\alpha_s)^{1/4}$, turns to be smaller than in the WCS, mostly
at low temperatures, by up to about 1/3, cf.~upper right panel Fig.~\ref{fig_scs-V-F}. 
At the same time, the Coulomb Debye mass, \(m_d\), for the SCS is comparable to the one 
in the WCS at low temperature, but increases more strongly (and essentially linear) 
with temperature. The key feature of the SCS in-medium potential is thus a rather 
long-range remnant of the confining force, as shown by the red lines in the third row 
of Fig.~\ref{fig_scs-V-F}. In particular, at intermediate and large distances, the 
potential rises markedly over the free energy (green lines), by up to 0.6\,GeV at
the lowest temperature ($T$=0.194\,GeV) and by up to 0.3\,GeV at $T$=0.400\,GeV. 
The latter is not far anymore from the WCS. The fit to the lQCD data (black dots) is 
of the same quality as for the WCS. The scale factor of the interference function 
(shown in the lower right panel of Fig.~\ref{fig_scs-V-F}) is also very similar
to the WCS, although its magnitude is smaller at higher temperatures.

With the extracted potential, the selfenergies and spectral functions of the static quark 
generated from the static-light \(T\)-matrices are shown in the first two rows of 
Fig.~\ref{fig_scs-V-F}. At low \(T=0.194\)\,GeV, the peak value of 
\(\text{Im}\Sigma_Q\approx-0.26\)\,GeV implies a width of the spectral function  
in excess of 0.5~GeV. In fact, the full-width at half-maximum of the pertinent spectral
function amounts to about 0.7\,GeV, due to additional effects from the real part
of the static-quark selfenergy. This is almost an order of magnitude larger than 
the leading order HTL result~\cite{Laine:2006ns,Beraudo:2007ky,Beraudo:2010tw}, 
\((\frac{4}{3}\alpha_sT)\approx0.07\)\,GeV. In addition, the peak value of the 
single-quark width, \(\text{-2Im}\Sigma_Q\), increases only slightly with $T$
at lower temperatures, and even decreases between 0.320 and 0.400\,GeV. This remarkable
feature is due to the marked loss of long-range interaction strength which can 
over-compensate the increase in parton density with temperature.
For the two-body quantities, the peak value of \(\text{Im}\Sigma_{Q\bar Q}\) defined in 
Eqs.~(\ref{selfE2}) and (\ref{eq_selfE2fatorize}) is less than twice the peak value 
of \(\text{Im}\Sigma_Q\), and the width of the two-body spectral function is less than 
twice that of the single static-quark spectral function. This is different from  the WCS case 
and caused by large off-shell effects.

\begin{figure*}[!t]
        \centering		 
        \fbox{\includegraphics[width=1.93\columnwidth]{legendsT.eps}}
        \includegraphics[width=2.00\columnwidth]{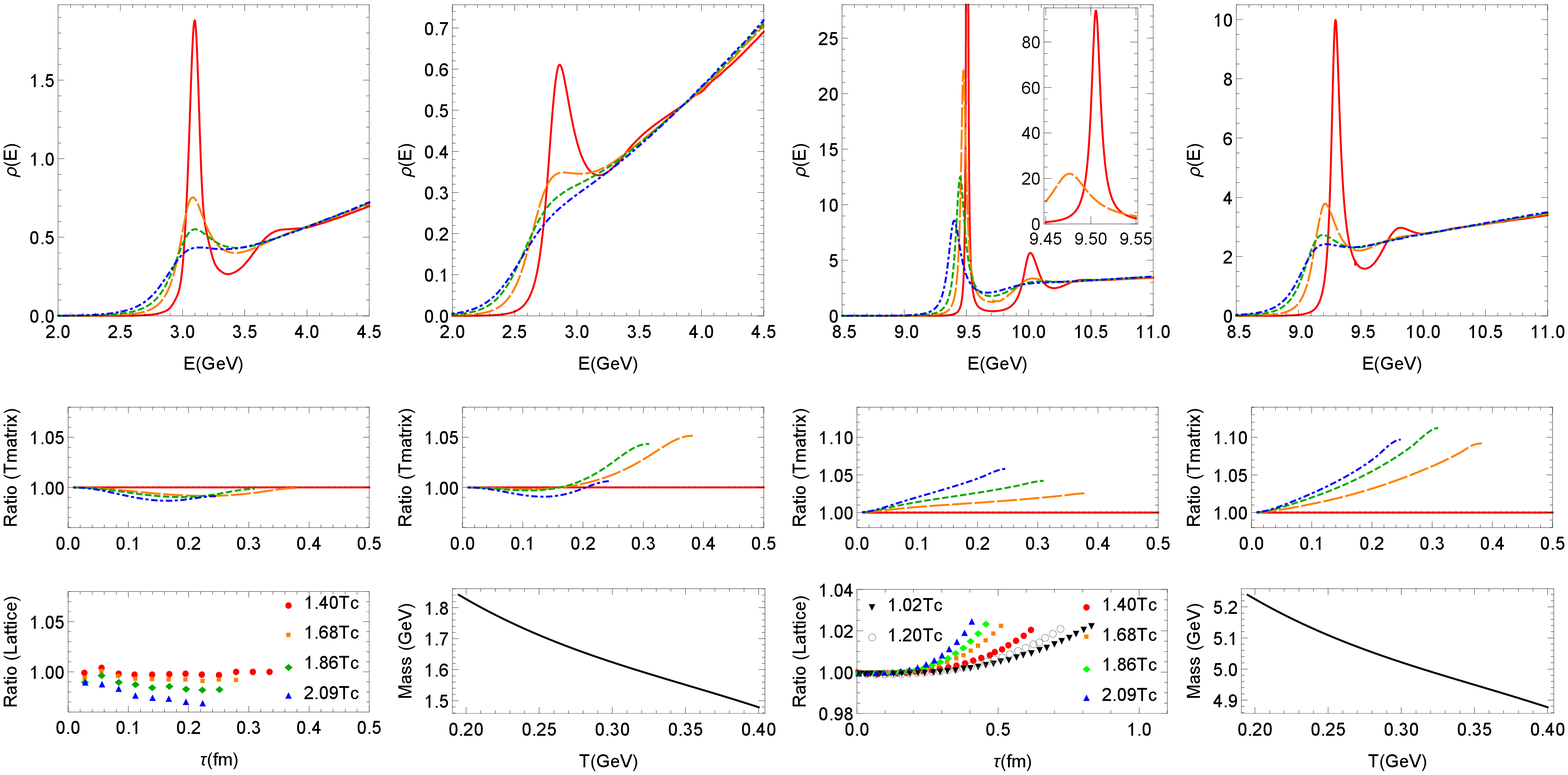}
\caption{{\em Strongly} coupled solution for charmonium (\(\eta_c\), left panels) and bottomonium
	($\eta_b$, right panels) spectral functions (upper panels) and correlators ratios (middle panels)
	with (first and third column) and without (second and fourth column) interference effects in the
	imaginary part of the potential. The lQCD data for \(\eta_c\)~\cite{Aarts:2007pk} and
	\(\eta_b\)~\cite{Aarts:2011sm}  correlator ratios are shown in the first and third bottom panel,
	respectively, while the second and fourth bottom panel display the temperature
	dependence of the charm- and bottom-quark mass, respectively.}
        \label{fig_scs-corr}
\end{figure*}

Let us also comment on a comparison of the SCS to our previous work in Ref.~\cite{Liu:2015ypa}.
The general shape and temperature behavior of the SCS potential are quite similar to
the result with our previous fit ansatz~\cite{Liu:2015ypa}.
However, the SCS potential shown in Fig.~\ref{fig_scs-V-F} has a significantly smaller force 
at large distances compared to the earlier result. Due to the increasing shell volume, 
\(\propto r^2\), a long-range force interacts with increasingly more medium particles,
which in principle can generate (very) large scattering widths. However, the selfconsistency 
requirement ties the width to the potential as the latter generates the selfenergies through
the $T$-matrix. Large widths generated by long-distance forces can therefore easily lead
to free energies which fall below the lQCD data. In this way, the selfconsistency
much augments the control over the properties of the force which are especially
effective in generating large widths (in particular its large-distance behavior).

We cannot prove that our SCS constitutes an upper limit for the coupling strength
of the QGP, given the lQCD data that we incorporate in our fit. However, there are several 
limiting factors (in addition to the one described
above) which prevent us from constructing more strongly coupled solutions. 
In particular, we limited ourselves to scenarios where the string tension does not 
significantly exceed the vacuum value. We also refrained from using ``unnaturally" small 
Coulomb Debye masses which could provide a long-range force but would be in conflict with 
the expected approach toward perturbative behavior at high temperatures. 
Within these constraints the presented SCS is the ``strongest" solution we could find 
upon varying our input and ans\"atze for the initial potential. 
As one would expect from a selfconsistent quantum framework, we have evidence that our 
calculations respect lower quantum bounds for transport
coefficients, as has been conjectured, \eg, for the ratio of shear viscosity to entropy
density. For example, if we attempt to push for an extremely long-range force ansatz (which,
as explained above, leads to very large scattering widths), the selfconsistent iteration 
procedure in fitting the free energy will push back toward a more weakly coupled solution. 
When neglecting the requirements to agree with lQCD data and deliberately increasing the 
interaction strength in the calculation of the EoS, the selfconsistent \(T\)-matrix iteration 
ultimately leads to a zero-mass color-singlet glueball, which signals condensation and at 
that point goes beyond
our current setup (recall that our parton fit masses encode possible condensate gaps).
Quantum selfconsistency clearly plays a key role as a limiting mechanism.  
\begin{figure*}[!t]
        \centering
        \includegraphics[width=1.99\columnwidth]{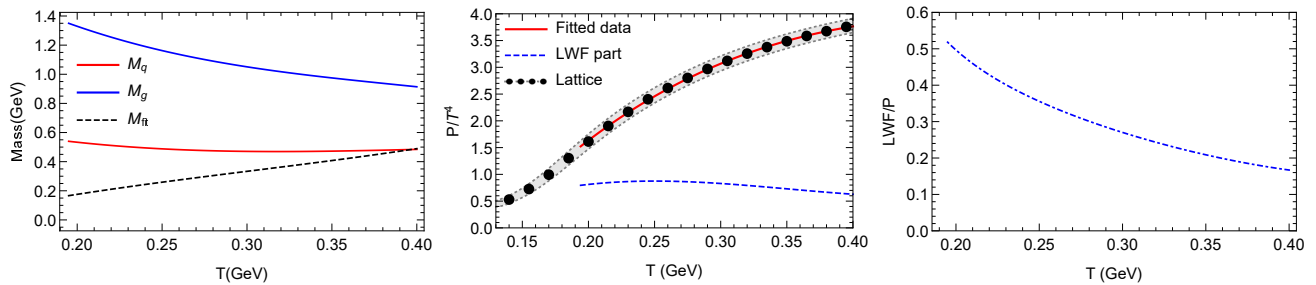}
       \caption{{\em Strongly} coupled solution for the QGP bulk medium: fit results of the 
       	input masses for quarks and gluons (left panel), the  QGP pressure in comparison to 
        lQCD data~\cite{Bazavov:2014pvz} (middle panel; solid line: total, dashed line: LWF 
        contribution), and the ratio of LWF contribution to total pressure (right panel).}
        \label{fig_scs-eos}
\end{figure*}

\subsubsection{Quarkonium Correlators and Spectral Function}
\label{sssec_scs-corr}
The selfconsistent charmonium and bottomonium spectral functions and pertinent
Euclidean correlators ratios (normalized to the lowest-temperature one) 
are collected in Fig.~\ref{fig_scs-corr} together with lQCD data for the latter
and the temperature dependence of the effective charm- and bottom-quark masses. 

The large scattering rates of charm and bottom quarks in the SCS induce significantly 
larger widths of
the quarkonium states than in the WCS. As before, interference effects lead to a marked
reduction of the bound-state widths. The stronger binding compared to the WCS is counteracted
by the significantly larger heavy-quark masses in medium as to generate an $\eta_c$ mass
that is remarkably stable with temperature. This leads to Euclidean correlator ratios which 
are within 2\% of unity, which agrees even better with the lQCD data than in the WCS (although 
this is not necessarily significant, as we argued in the context of the WCS results). 
The correlator ratios without interference effects deviate somewhat more from the lQCD 
data, possibly indicating that a moderately broadened charmonium ground state 
that survives to higher temperatures (here about $T$=0.320\,GeV when including interference) 
may be favored by lQCD data.\footnote{There is a small overall shift of the ground states' 
peak position to higher masses when including interference effects as compared to neglecting 
them; this may depend on our 
specific implementation of the interference effects which requires further investigation. 
On the other hand, the reduction of the width by interference is a robust mechanism independent 
of the implementation.} 
For example, the inelastic width of the $\eta_c$ at $T$=0.194\,GeV is around 0.1~GeV 
for the SCS and 0.02~GeV for the WCS (including interference).   
Appreciable charmonium reaction rates with the ground state surviving over an extended 
interval in temperature are favored by the phenomenology of transport models in 
describing $J/\psi$ production at RHIC and the LHC~\cite{Rapp:2017chc}, in particular 
to regenerate a sufficient number of $J/\psi$'s at the LHC. 

In the $\Upsilon$ sector, the first excited state still survives at the lowest temperature;
even without interference effects, a pertinent maximum structure in the spectral function is
visible below the nominal $b\bar b$ threshold of 2$m_b$, but its width is comparable
or even larger than the binding energy so that it appears as being dissolved. The 
ground-state $\Upsilon(1S)$ clearly survives up to the highest temperature, $T$=0.400\,GeV 
(it is smeared out at much lower temperature without interference effects). The
pertinent correlator ratio is in line with lQCD data within a few percent, which again is the
closest agreement between all four scenarios considered in this paper (SCS and WCS with 
and without interference effects). The slight increase of the calculated ratio is in 
part caused by the lowering of the bound-state mass, implying that the decrease in the
constituent bottom-quark masses is more relevant than the decrease in binding energy.  

\begin{figure*}[!t]
	\begin{center}
		 \fbox{\includegraphics[width=2.04\columnwidth]{legendsp.eps}}
		\fbox{\includegraphics[width=1.00\columnwidth]{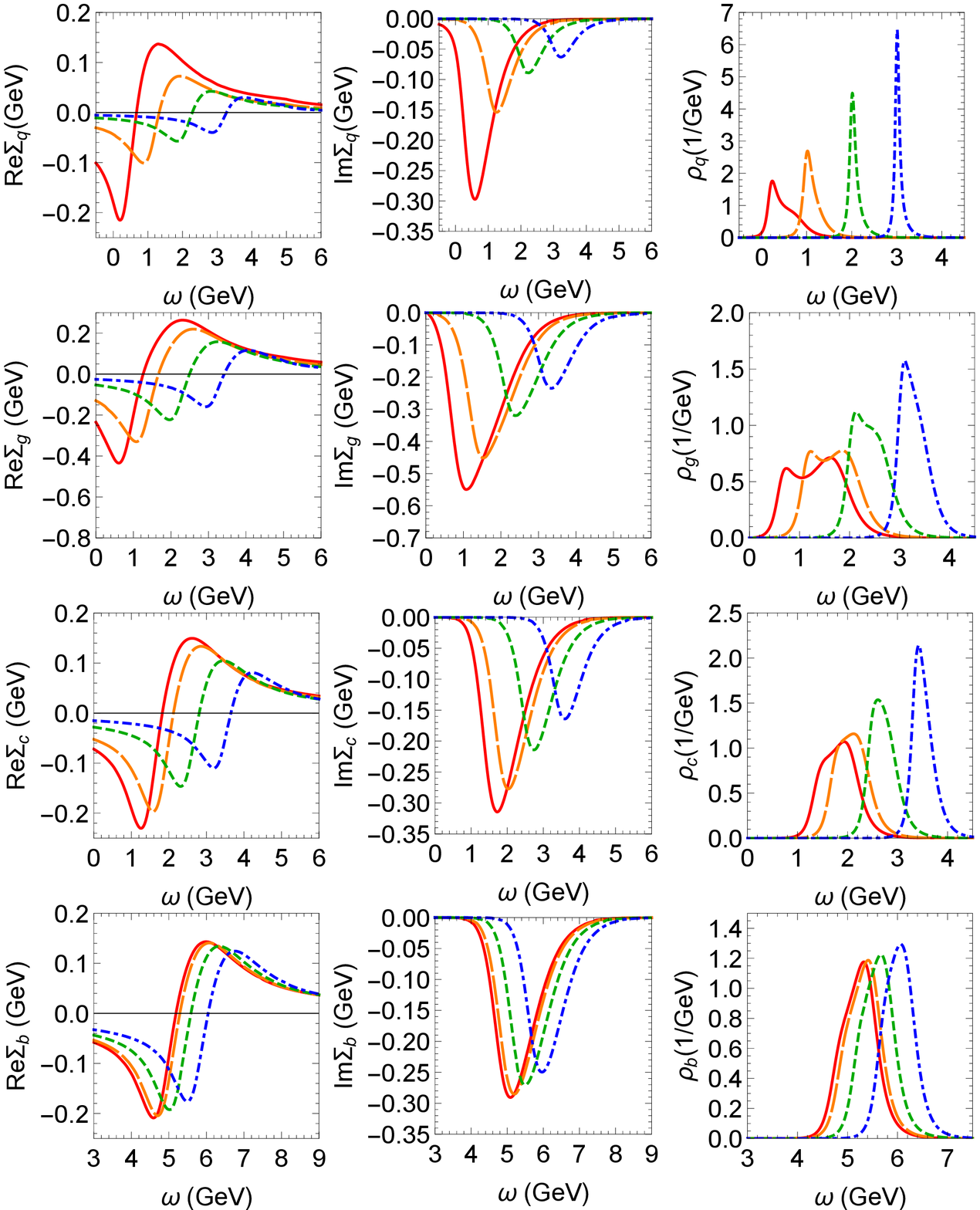}}
		\fbox{\includegraphics[width=1.00\columnwidth]{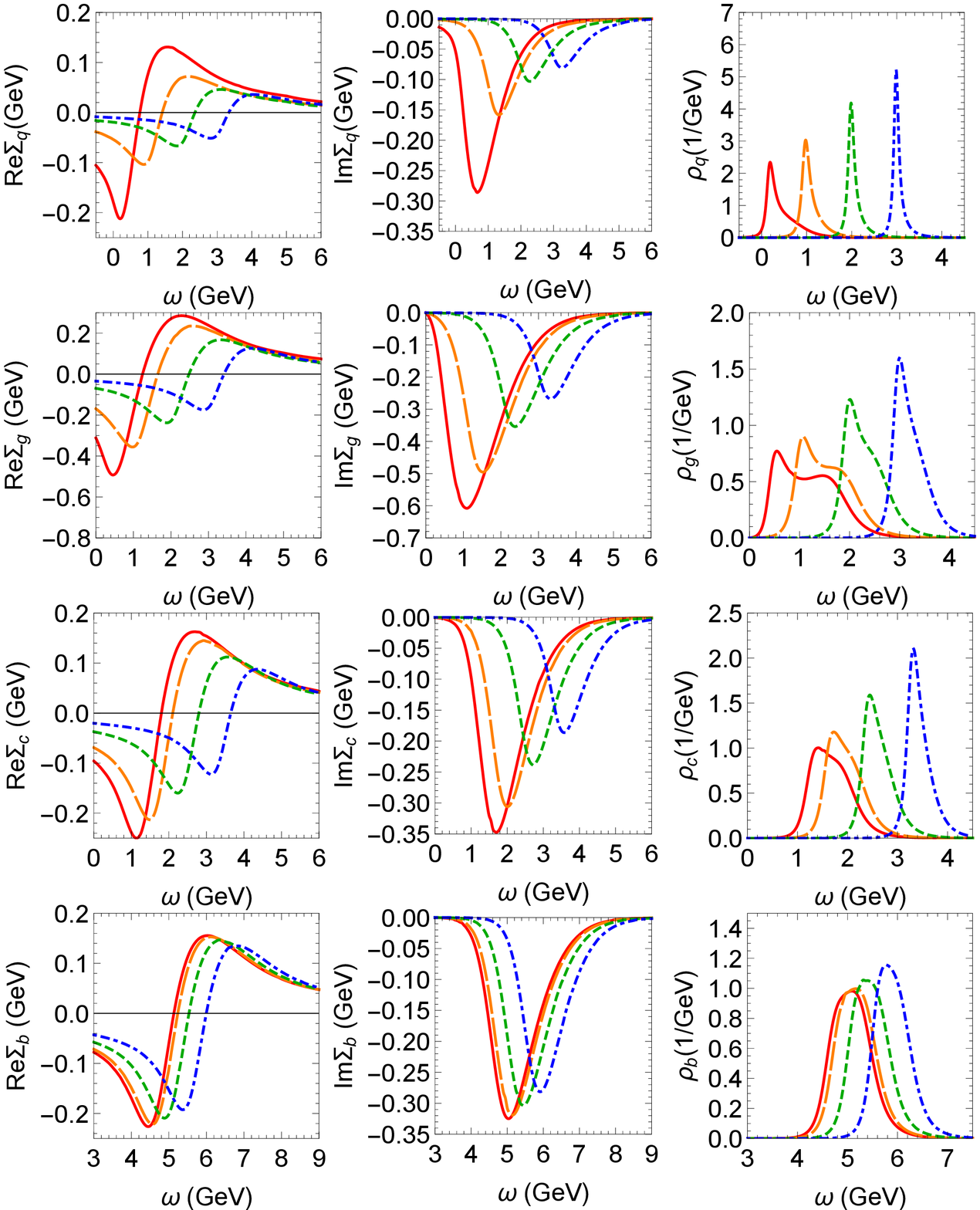}}
		\fbox{\includegraphics[width=1.00\columnwidth]{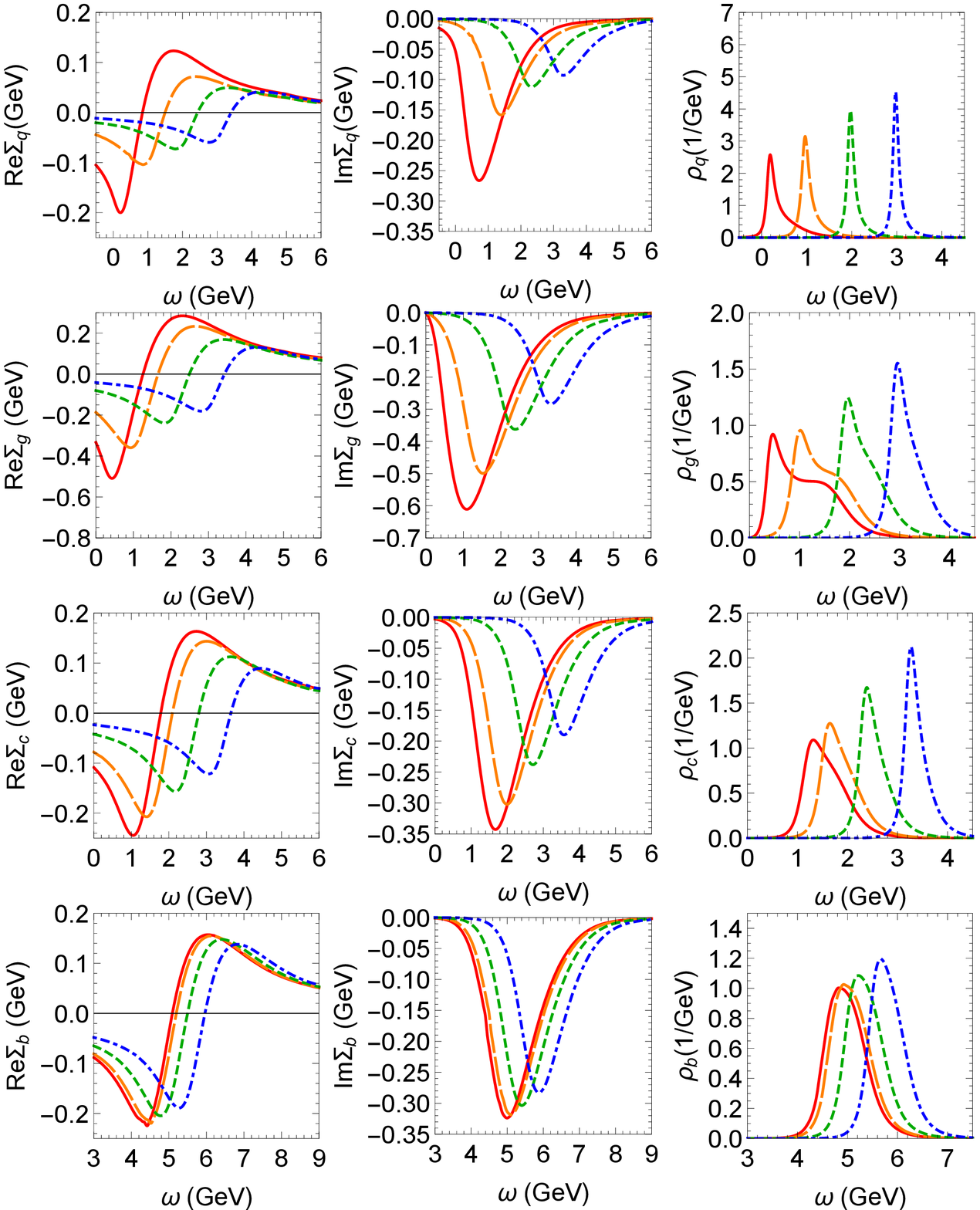}}
		\fbox{\includegraphics[width=1.00\columnwidth]{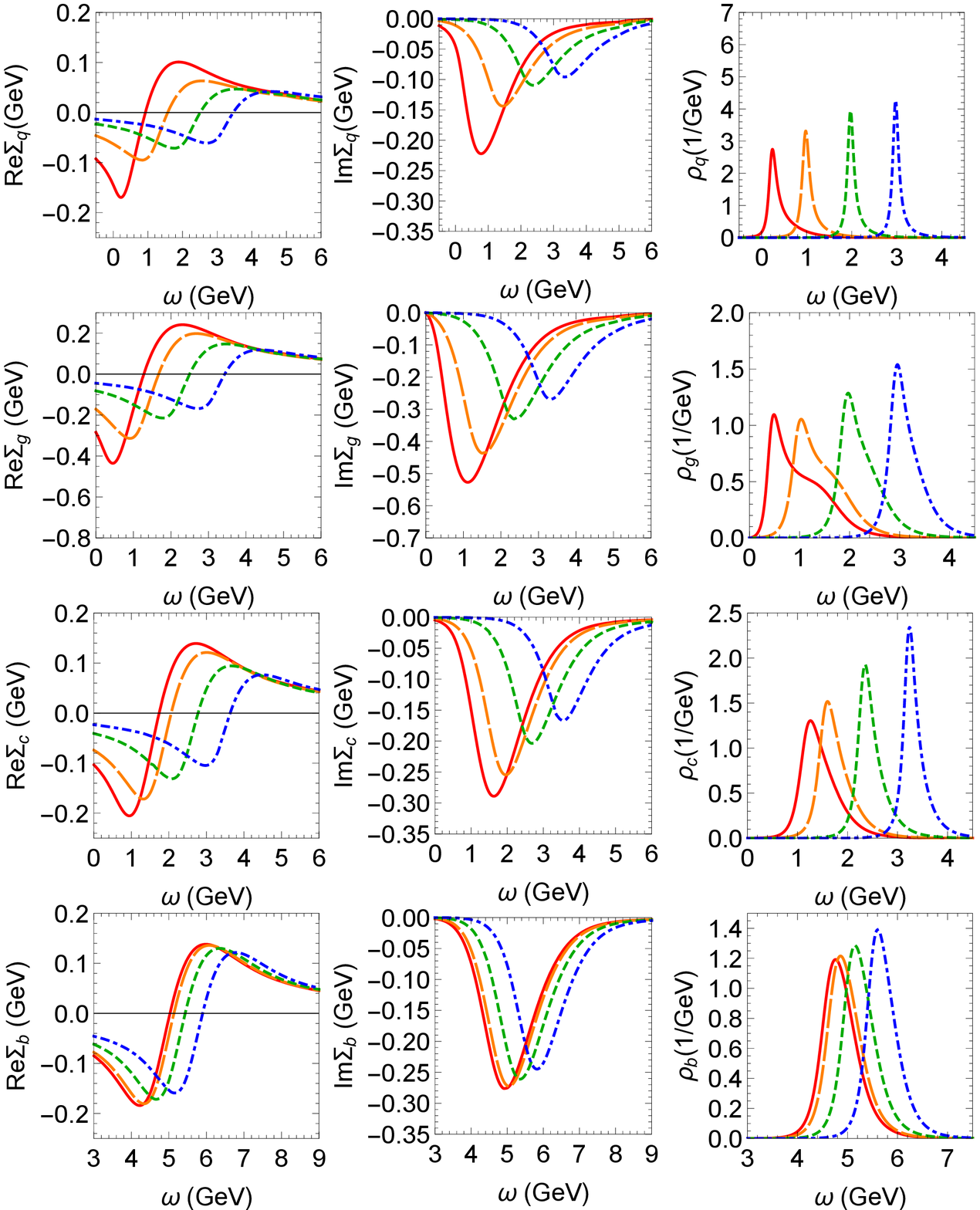}}
		\caption{{\em Strongly} coupled solution for parton spectral properties of the QGP.
			The figure is organized into four 3-by-4 panels of 12 plots, with each panel for a fixed temperature 
			(upper left: 
			$T$=0.194\,GeV, upper right: $T$=0.258\,GeV, lower left: $T$=0.320\,GeV and lower right: $T$=0.400\,GeV). 
			Each panel contains 4 rows corresponding to different parton species (light quarks ($q$), gluons 
			($g$), charm quarks ($c$) and bottom quarks ($b$) in the first, second, third and fourth row of
			each panel, respectively). Each row contains 3 panels showing (from left to right) the energy 
			dependence of the pertinent real and imaginary part of the selfenergy and the resulting spectral 
			functions, for 4 different values of the single-parton 3-momentum ($p$) in the
			thermal frame.}
		\label{fig_scs-spec}
	\end{center}
\end{figure*}

\begin{figure*}[!t]
		\fbox{\includegraphics[width=1.98\columnwidth]{legends.eps}}
		\includegraphics[width=2\columnwidth]{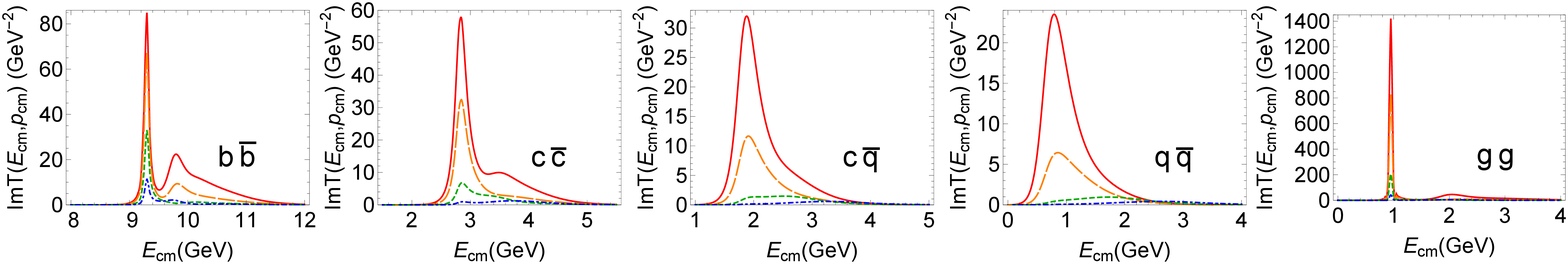}
		\includegraphics[width=2\columnwidth]{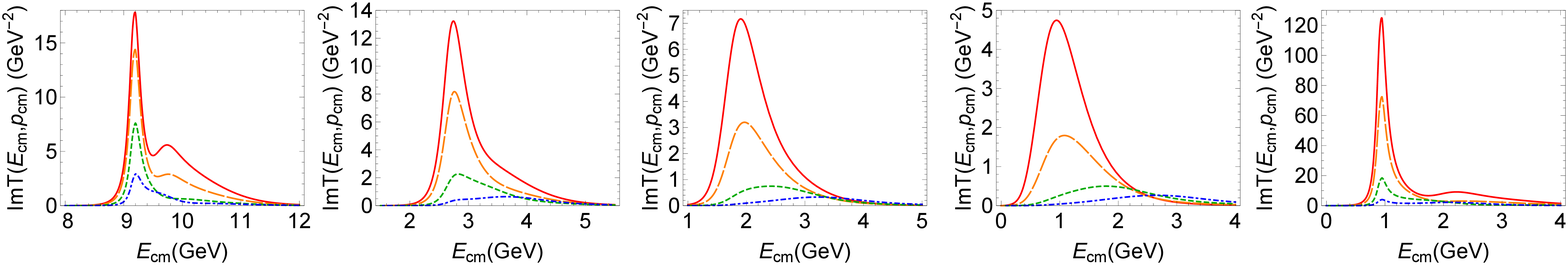}
		\includegraphics[width=2\columnwidth]{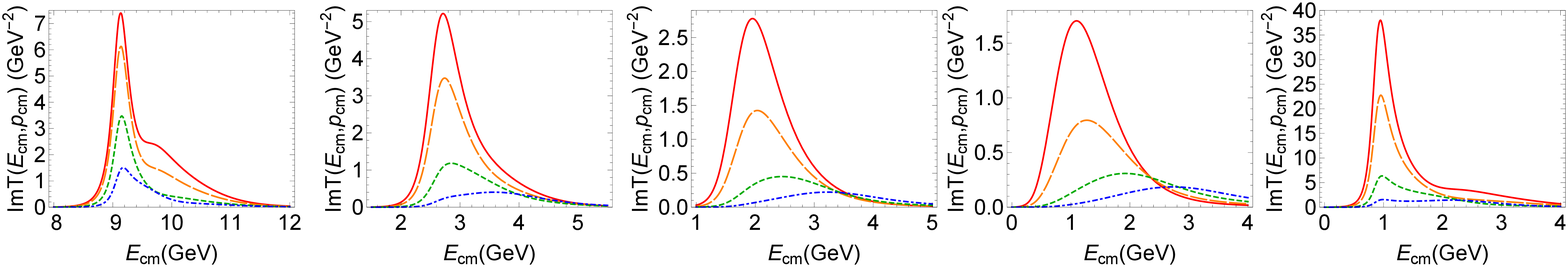}
		\includegraphics[width=2\columnwidth]{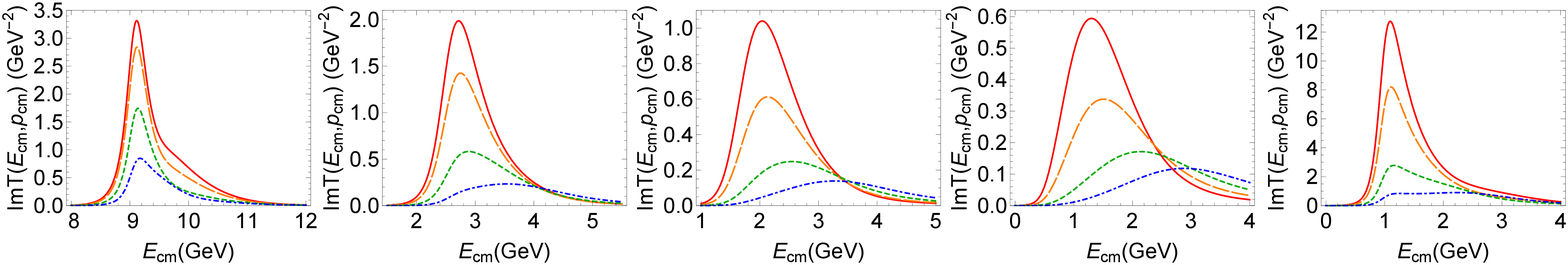}
		\caption{{\em Strongly} coupled solution for the imaginary part of the color-singlet
			$S$-wave \(T\)-matrices (without interference effects) in the bottomonium ($b\bar b$; first 
			column), charmonium ($c\bar c$; second column), $D$-meson ($c\bar q$; third column),
			light-quark ($q\bar q$; fourth column), and glueball ($gg$, last column) channels.   
			The 4 rows correspond to different temperatures, $T=0.194$~GeV, $T=0.258$~GeV, 
			$T=0.320$~GeV and $T=0.400$~GeV from top down; in each panel, the $T$-matrix is displayed for
			4 different single-parton momenta ($ p_{\rm cm}$) in the
			two-body CM frame.}
		\label{fig_scs-T}
\end{figure*}
\subsubsection{QGP Equation of State}
\label{sssec_scs-eos}
Next, we turn to the SCS for QGP bulk properties. The fitted light-parton masses 
are qualitatively similar to the WCS, cf.~left panel of Fig.~\ref{fig_scs-eos}. 
Most notably, the gluon mass is quite a bit larger due to the larger string-induced
Fock term contribution, recall Eq.~(\ref{eq_fockmasspspace}), implying a much increased
infinite-distance limit relative to the WCS. This contribution is also active for the 
effective quark mass. The underlying fit mass, \(M_\text{fit}\), is actually 
appreciably smaller than in the WCS, with values of  0.16\,GeV and 0.49\,GeV at
\(T=0.194\)\,GeV and \(T=0.400\)\,GeV, respectively. These values are not far from what
one expects from the perturbative (Coulomb) thermal masses, \(\sqrt{1/3}g T=0.2\)\,GeV 
and \(\sqrt{1/3}g T=0.42\)\,GeV, respectively. 
The resulting EoS fits lQCD data well, and encodes the most important difference
between SCS and WCS, namely that the two-body contribution to the pressure is much more 
prominent at low temperatures, reaching more than 50\% at \(T=0.194\)~GeV, compared
to $\sim$10\% in the WCS. Also, the LWF contribution shows a more intuitive temperature 
behavior, in that its fraction relative to the total appreciably decreases with increasing 
$T$ (cf.~right panel of Fig.~\ref{fig_scs-eos}); 
here, the decrease in interaction strength surpasses the increase in parton density, 
which can be interpreted as a gradual melting of the light-parton bound states with $T$
(this interpretation will become even clearer upon inspection of the spectral functions
in the next section).  However, at $T$=0.400\,GeV, the interaction contribution still
amounts to $\sim$20\%, indicating that even at this temperature the QGP contains a 
significant nonperturbative component (possibly driven by the gluonic sector through 
glueball contributions). As before, the gluon sector largely decouples at small
temperatures due to the large gluon masses.

\subsubsection{Spectral Structure of QGP}
\label{sssec_scs-spec}

We finally turn to the examination of the single-parton spectral functions and their 
in-medium scattering amplitudes.  
The width of the partons, $\Gamma=-2\text{Im}\Sigma$, is large, especially at low 
temperatures and small 3-momenta, $p\lsim T$, see the upper 4 plots in the second column
of Fig.~\ref{fig_scs-spec}. The quark (gluon) width reaches up to 0.6\,(1.1)\,GeV
right around its on-shell energy, which is larger than its effective mass and thus
implies the loss of a well-defined quasiparticle excitation. Inspection of the pertinent 
$p=0$ light-parton spectral functions (upper 2 panels in the third column of 
Fig.~\ref{fig_scs-spec}) confirms this notion, as the quark's (gluon's) spectral 
strength is spread over an energy range of about 1(2)\,GeV. In fact, the rather large and 
attractive real part of the selfenergy at small (off-shell) energies (upper 2 panels of 
the first column of Fig.~\ref{fig_scs-spec}) also plays an important part in the quark 
(gluon) spectral distribution, as it generates a  rather prominent collective mode at 
$\omega\simeq0.15(0.7)$\,GeV, sitting on top of the broad distribution associated with 
the dissolved quasiparticle mode.   
The low-temperature widths are almost an order of magnitude larger than the HTL value
of $\frac{4}{3}\alpha_s T\approx 0.07$\,GeV, and much larger than the most recent 
dynamical quasiparticle model result which is around 0.2\,GeV~\cite{Berrehrah:2016vzw}. 
Interestingly, the temperature dependence of the parton widths is non-monotonic with
increasing temperature (as was found for static quarks discussed in 
Sec.~\ref{sssec_scs-pot}), which has important consequences for the temperature 
dependence of transport coefficients~\cite{Liu:2016ysz}.
This is qualitatively different from both perturbative and dynamical quasiparticle
approaches. The 3-momentum dependence of the width is quite strong especially at low 
temperatures (less so at high temperature), being substantially reduced with increasing 
$p$. This implies that at higher momenta well-defined quasiparticle excitations re-emerge
at any temperature, as to be expected from a generic transition to a weak coupling. 
However, since the string term at high temperature is not screened as much as in the 
WCS, the momentum dependence of selfenergy at high temperature differs from the WCS.
The widths of the charm and bottom quarks are quite similar to the light quarks, 
implying that bottom quarks remain well-defined quasiparticles at all momenta and
temperatures, while the situation is borderline for low-momentum charm quarks close 
to $T_c$.

Selfconsistent \(T\)-matrices are compiled in Fig.~\ref{fig_scs-T}. At low temperatures 
appreciably bound quark-antiquark states emerge in all channels (glueballs, light mesons, 
heavy-light mesons, charmonia and bottomonia). The light $q\bar q$ resonance mass is 
close to the vacuum mass of light vector mesons, reflecting a realistic vacuum limit as 
encoded in the potential model (instanton effects are subleading in the vector channel). 
This is, however, nontrivial given its embedding in the QGP EoS (in particular through 
the fitted light-quark mass). Note that the off-shell behavior of the parton widths, \ie, 
their decrease away from the on-shell peak (recall 2.~column in Fig.~\ref{fig_scs-spec}), 
plays an important role in the formation of bound states; \eg, the light-meson width 
of $\sim$0.6\,GeV at the lowest temperature is well below twice the light-quark width, 
mostly because of the $\sim$0.3\,GeV binding relative to the nominal $q\bar q$ threshold 
of 1.1\,GeV. 
Compared to the WCS (recall Fig.~\ref{fig_wcs-T}), the strength of the $T$-matrices 
in the SCS is much increased (\eg, the peak value in the $p_{\rm cm}$=0 
light-meson channel is $\sim$25/GeV$^2$ in the latter compared to $\sim$6\,/GeV$^2$ in 
the former; also, the mass of the $q\bar q$ bound state is smaller, $\sim$0.8\,GeV vs. 
$\sim$1\,GeV).  This, in particular, makes a large difference in their contributions to 
the EoS (recall Fig.~\ref{fig_scs-eos} vs. Fig.~\ref{fig_wcs-eos}). At the same time, 
the much larger widths in the spectral functions of light partons in the SCS relative to the 
WCS causes their thermodynamic weight to be much suppressed in the former relative to the
latter. In this sense, the SCS predicts a transition from broad parton quasiparticles
to broad hadronic states in the thermodynamics of the QGP as $T_c$ is approached from
above. The re-emergence of parton quasiparticles and suppression of their bound states not 
only occurs with increasing temperature (note the reduction in the $y$-axis scale when
going down in temperature row by row in Fig.~\ref{fig_scs-T}), 
but also with increasing parton CM momentum within the bound-state (not to be
confused with the total momentum, $P$, of the bound state in the heat bath, which is
zero throughout this paper) and delayed with increasing constituent parton mass.

\section{Summary and outlook}
\label{sec_sum}
We have set up a selfconsistent thermodynamic $T$-matrix approach to study the bulk and
microscopic properties of the QGP in a unified framework, encompassing both light- and 
heavy-flavor degrees of freedom. Starting from the HQ limit of QCD, we set up an effective 
partonic Hamiltonian with a universal color force, including remnants of the confining 
force and relativistic corrections necessary to treat thermal partons.  We have computed 
one- and two-body thermodynamic Green's and spectral functions selfconsistently, 
incorporating bound and scattering states on an equal footing. Compared to earlier works, 
a full off-shell treatment is implemented to account for quantum many-body effects 
rigorously, in particular the collisional widths of the QGP constituents. 
Moreover, our approach enables systematic constraints on the inputs to the Hamiltonian, 
\ie, the two-body potential and two effective light-parton mass parameters, by comparing 
to a variety of lattice-QCD data.

Our calculation of the equation of state has been carried out in the LWB formalism with 
selfconsistently computed light-parton selfenergies and $T$-matrices. Importantly, we managed 
to resum the Luttinger-Ward functional using a matrix-log technique, which is critical to 
account for the dynamical formation of bound (or resonance) states in the thermodynamics of the 
system. The main constraints on the two-body driving kernel are derived from the HQ free energy,
$F_{Q\bar{Q}}$, which we have also computed selfconsistently from the $T$-matrix for static 
quarks embedded in the QGP. Based on a parametric ansatz for an in-medium Cornell potential,
we have fitted lattice-QCD data for $F_{Q\bar{Q}}$ and further checked our results against
euclidean correlator ratios in the bottomonium and charmonium sectors. Together with the
EoS, for which the fit of pertinent lQCD data can be largely controlled through the two
bare light-parton masses in the Hamiltonian, this constitutes a comprehensive quantum 
many-body framework for light and heavy partons and their two-body correlations in the QGP. 
We have solved this problem through a multi-layered numerical iteration procedures in our fit 
to the 3 sets of lQCD data, where a typical accuracy at a few-percent level can be achieved. 
The main predictive power of the approach resides in the emerging spectral and transport 
properties of the QGP, including the prevalent degrees of freedom in the EoS.  

In our search for selfconsistent solutions, it turns out that the above set of lQCD constraints 
does not uniquely specify the input for the driving kernel. We classified its 
possible range by a weakly- and a strongly-coupled solution.
In the former, the input potential comes close to a lower limit set by the HQ free energy
itself (not unlike what has been discussed based on direct Bayesian extraction 
methods~\cite{Burnier:2014ssa}). The resulting light-parton spectral functions have rather 
moderate widths, well below their masses, and thus yield well-defined quasiparticles, 
as well as rather sharp but loosely bound resonances when approaching $T_c$ from above. 
The latter remain subleading, at a 10\% level, in their contribution to the EoS. In contrast,
the strongly-coupled solution is characterized by a potential that appreciably exceeds the
free energy (not unlike recent lQCD extractions reported in Ref.~\cite{Petreczky:2017aiz}),
recall the 3.~row of Fig.~\ref{fig_scs-V-F}. 
The key difference to the weakly coupled solution is a long-range remnant of the confining 
force (while its short-distance, $r$$\lsim$0.4\,fm, and high-temperature, $T>2T_{\rm c}$, 
behavior is quite similar in both solutions). 
The emerging partonic spectral widths are much enhanced; they become comparable to the 
parton masses and thus dissolve quasiparticle structures for low-momentum modes near $T_c$ 
(cf.~the 3.~panel in rows 1 and 2 of Fig.~\ref{fig_scs-spec}). At the same time, broad 
but well-defined two-particle bound states (mesons) emerge (last 2 panels in row 1 of 
Fig.~\ref{fig_scs-T}) and become the leading contribution to the EoS (middle panel in 
Fig.~\ref{fig_scs-eos}), thus signaling a transition 
in the degrees of freedom in the system. At high momenta, parton quasiparticles reemerge and 
bound-state correlations are much suppressed. This solution, in particular, critically
hinges on a proper treatment of the quantum effects induced by the large scattering
rates. 

While we believe that the strongly coupled solution is clearly the more attractive one (including
its transition from quarks to hadrons and a qualitatively liquid-like behavior with interaction 
energies comparable to the parton masses), a more quantitative characterization of this notion is 
in order. We already indicated in our previous letter~\cite{Liu:2016ysz} that transport coefficients, 
in connection with heavy-ion phenomenology, can play a decisive role in this regard. The heavy-quark 
diffusion coefficient and the viscosity-to-entropy density ratio show promisingly small values in 
the strongly-coupled scenario, while they are significantly larger in the weakly coupled scenario,
to an extent that creates conflicts with hydrodynamic and heavy-flavor transport modeling of 
heavy-ion collisions. The latter is currently being investigated quantitatively and will be 
reported elsewhere~\cite{He:2017}. In fact, converting the heavy-quark diffusion coefficient 
into a thermalization and scattering rate, one can straightforwardly deduce that values of 
$2\pi{\cal D}_s\simeq3$ translate into quark scattering rates of order 1\,GeV; this implies 
the dissolution of light quasiparticles, fully consistent with our numerical findings. 
The large widths also require the underlying potential $ V $ to markedly exceed the 
free energy, $F_{Q\bar{Q}}$, independent of model details~\cite{Liu:2015ypa}. 
As a compact upshot, the strongly coupled solution found in our approach may be characterized 
as establishing a links between: 
``a large string potential" \(\Leftrightarrow\) ``strong two-body resonances" \(\Leftrightarrow\) 
``broad (non-quasiparticle) spectral functions" \(\Leftrightarrow\) ``small viscosity/spatial diffusion 
coefficients". If the string term arises from the nontrivial 
``vacuum" structure of QCD, then these links suggest that the latter is in fact responsible 
for the remarkable features of the sQGP.

A more ambitious line of future work is to test the predicted spectral properties more directly;
in the quarkonium sector this presumably requires the formulation of quantum transport
approaches for heavy-ion collisions as recently discussed in the literature, which, in turn, 
can take advantage of heavy-quark diffusion properties computed with the same underlying 
interaction. The most direct connection remains the dilepton production rate, where again 
constraints from lQCD data can be straightforwardly utilized.      
Another area accessible to our approach is the investigation of finite chemical potential
in the QCD phase diagram, starting with the calculation of quark susceptibilities. However,
the description of phenomena associated with dynamical chiral symmetry breaking, which are 
expected to become important at temperatures below 
$T\simeq 0.185\,$~GeV~\cite{Bhattacharya:2014ara}, will require an extension of the current 
formalism to explicitly include condensation mechanisms. This
is more challenging but, we believe, still feasible.


\acknowledgments 
This work is supported by the U.S. NSF through grant no. PHY-1614484.
\appendix

\section{\(T\)-matrix Approach for Light Partons }
\label{app_rel-pot}
In this appendix we discuss several issues related to the implementation of the 
potential approximation for light-quark interactions. Historically, the Cornell 
potential has been a successful tool for quark-based hadron spectroscopy; 3D reductions 
of the 4D Bethe-Salpeter equation (BSE) are also widely used in effective hadronic 
approaches to hadronic vacuum physics, including light mesons like $\pi$-$\pi$ 
interactions.  In particular, the Cornell potential incorporates essential 
nonperturbative aspects of the QCD force, \ie, a confining force. Our approach is a 
finite-temperature version of this framework, where remnants of the confining force 
turn out to play a crucial role to render a strongly coupled system. The recovery of 
the vacuum vector-meson masses at low QGP temperatures in the SCS (where the potential 
is close to its vacuum form) is a direct manifestation of a ``realistic" vacuum limit 
of the approach in the light-quark sector. As we remarked in the text, interactions 
believed to be essential for spontaneous chiral symmetry breaking (such as 
instanton-induced forces) are not included, but we recall that recent lQCD computations 
have found that the effects of chiral symmetry breaking have essentially vanished 
once the temperature has reached about 30\,MeV above the chiral crossover temperature, 
$T_{\rm pc}^\chi\simeq0.155$\,GeV~\cite{Bhattacharya:2014ara}.

There are several further considerations. The reduction of the relativistic 4D Bethe-Salpeter 
equation (BSE)~\cite{Salpeter:1951sz} into 3D scattering equations has been scrutinized, \eg,
in Ref.~\cite{Woloshyn:1974wm}. In particular, within in the Blankenbecler-Sugar (BbS) 
scheme~\cite{Blankenbecler:1965gx}, the BSE can be equivalently separated into two coupled 
equations, where the kernel of the first (leading) equation is potential-like, while the 
second (subleading) equation quantifies the off-energy-shell corrections to the 
potential kernel. The philosophy is to expand ithe BSE around the potential solution using 
a parametrically small correction, \(R_2 V\)~\cite{Blankenbecler:1965gx}, rather 
than to expand around the free-wave solution using the coupling constant and/or velocity 
(as in NRQCD) as a small parameter. In particular, such an expansion does not rely on a 
non-relativistic hierarchy. This series usually exhibits a fast 
convergence~\cite{Blankenbecler:1965gx,Woloshyn:1974wm}, suggesting that the leading potential 
solution is already close to the full solution.  
In many cases, the higher-order off-shell corrections can be effectively absorbed in an adjustment 
of the potential. In the present case, the fits of the potential to lQCD data may approximately 
encode such corrections.
Finally, we recall that for $2\to2$ on-shell scattering in the CM system the in- and outgoing
momenta moduli of the particles are equal, \ie, there is no energy transfer in the collision.
We also recall that while the two-body interaction is approximated by an instantaneous
force, the many-body quantum approach fully accounts for the dynamics (energy dependence) of
the one- and two-particle propagators (and $T$-matrices) in the system. Additional considerations can be found in Ref.~~\cite{Shuryak:2004tx,Shuryak:2003ja}.


\section{Generalized Thermodynamic Relations for the LWB Formalism at Finite $\mu_q$}
\label{app_lwb-muq}
The LWB formalism implies several thermodynamic relations for particle, energy and entropy
densities~\cite{PhysRev.118.1417,Baym:1962sx}. However, these relations will be modified
when using an effective Hamiltonian whose ``bare" single-particle masses (encoded in the
dispersion relation $ \varepsilon(p) $), and potential, $V$, depend on temperature ($T$) 
and chemical potential ($\mu$). In this appendix we illustrate these modifications.

The strategy for the derivation is to start from the usual relations without $T$ or $\mu $ 
dependence in the dispersion relation and potential and then generalize them to the case 
with $T$ and $\mu $ dependences. For derivatives with respect to (\wrt)  $ T $ or $ \mu $  
any implicit dependence through $G$ will vanish. For $ \varepsilon $ and $ V $ independent
of   \(T\) and \(\mu\), one has
\begin{align}
N=-\frac{\delta\Omega}{\delta \mu}=\pm\frac{-1}{\beta}\sum_{n}\text{Tr}\{G\} \ ,
\label{eq_nfromG}
\end{align}
since the dependence of $ \mu $ through $ (\delta\Omega/\delta G)(\delta G/\delta \mu) $ 
will vanish according to Eq.~(\ref{eq_domegadGeq0}), and the only \(\mu\) dependence 
figures from \(G_{(0)}^{-1}=i\omega_n-(\varepsilon-\mu)\).

For the derivation of the energy density from the grand potential one can adopt a method 
in time space is given in Ref.~\cite{Baym:1962sx}. In frequency space, with a separation 
of the \(\beta\) dependence arising from the loop as in Eq.~(\ref{Omega}), 
the entropy contribution can be derived as
\begin{align}
&T S =\beta\frac{\partial\Omega}{\partial\beta}=-\Omega\mp\frac{-1}{\beta} 
\sum_n\text{Tr}\{(-i\omega_n G)+\frac{1}{2}\Sigma(G) G\} \ .
\label{eq_SfromG}      
\end{align}
Still, the implicit dependence on $ \beta $ through \(G\) will vanish. 
The first term comes from the derivative \wrt~\((-1/\beta)\) in the frequency sum 
in obtaining \(\Omega\) and  \(\Phi\). The second term comes from the derivative 
\wrt~\((-1/\beta)\) of \(\omega_n\) in \(G_{(0)}^{-1}\). The third term 
comes from the  \((-1/\beta)^\nu\) dependence of the loop integrals in the selfenergy 
and gives a factor \(\nu\) that cancels the \(1/\nu\) factor in the skeleton expansion. 
With the entropy contribution, the energy \(U\) is
\begin{align}
&U=\Omega +T S +\mu N =\pm\frac{-1}{\beta} \sum_n\text{Tr}\{[\varepsilon+\frac{1}{2}\Sigma(G)] G\}
\label{eq_UfromG}      
\end{align}
where \(G^{-1}=i\omega_n-(\varepsilon-\mu)-\Sigma\) by use of Eq.~(\ref{sigG}). 
We can derive Eq.~(\ref{eq_UfromG}) from Eqs.~(\ref{eq_SfromG}) and (\ref{eq_nfromG}) 
using \(GG^{-1}=1\) and  \(\frac{-1}{\beta} \sum_n e^{i\omega_n\epsilon}1=0\) with 
an \(\epsilon\) regulation technique~\cite{fetter2003quantum}. This completes the derivation 
of the standard thermodynamic relations within the LWB formalism. 

If the ``bare" single-particle dispersion relation \(\varepsilon\) and the potential \(V\) 
of the Hamiltonian are functions of \(\beta\) and \(\mu\), the particle number, \(N\), and  
internal energy, \(U\), receive extra contributions, 
\begin{align}
&N=\pm\frac{-1}{\beta}\sum_{n}\text{Tr}\{[1- \frac{\partial\varepsilon}{\partial \mu}-\frac{1}{2} \Sigma(G,\frac{\partial V}{\partial \mu})]G\}
\label{eq_Nwithmu}      
\end{align}
\begin{align}
U=&\pm\frac{-1}{\beta} \sum_n\text{Tr}\bigg\{[\varepsilon+\beta\frac{\partial\varepsilon}{\partial \beta}- \mu\frac{\partial\varepsilon}{\partial \mu}\nonumber\\&+\frac{1}{2}\Sigma(G)+\frac{1}{2} \Sigma(G,\beta\frac{\partial V}{\partial \beta})-\frac{1}{2} \Sigma(G,\mu\frac{\partial V}{\partial \mu})] G\bigg\}
\label{eq_Uwithmu}      
\end{align}
where \(\Sigma(G,X)\equiv\sum_{\nu}\Sigma_\nu(G,X)\), and \(\Sigma_\nu(G,X)\) is defined 
to replace one of the \(V\) in evaluating \(\Sigma_\nu(G)\) by \(X\) at each order. 
It can be shown that, at least for ladder and ring diagrams, it does not matter which \(V\) 
is replaced in the diagram because every \(V\) in the connected diagram for \(\Phi_\nu\) 
is equivalent. Thus, for the \(T\)-matrix resummation the selfenergy can be schematically 
written as
\begin{align}
\Sigma(G,X)=T(G,X)G,\text{\space}T(G,X)=(1-VGG)^{-1}X
\label{eq_selfEwithX}      
\end{align}
where \(X\) is \(\mu\frac{\partial V}{\partial \mu}\)  
or  \(\beta\frac{\partial V}{\partial \beta}\). Since \(T(G,V)=(1-VGG)^{-1}V\), the new 
logarithm can be adapted from the original \(T\)-matrix logarithm without increasing 
the complexity.

\section{Additional Relations for the Static HQ Free Energy}
\label{app_static}
Based on the setup in Sec.~\ref{ssec_freeE}, we discuss additional useful
relations that follow from this formalism. 

First, we prove that a relation \(F_{Q \bar Q}(\infty,\beta)=2F_{Q}(\beta)\) is implicit 
in our formalism for the Polyakov loop defined as 
\begin{align}
F_{Q}(\beta)=\frac{-1}{\beta}\ln[\frac{-1}{\beta}
\sum_{\nu_n}G_{\bar Q}(i\nu_n)e^{-i\nu_n\beta}] \ .
\label{eq_2FQ}
\end{align} 
If we express Eq.~(\ref{eq_defineF}) in frequency space,
\begin{align}
&F_{Q\bar{Q}}(r,\beta)=\frac{-1}{\beta}
\ln \left[\frac{-1}{\beta}\sum_{E_n}{G}_{Q\bar Q}(iE_n,r)e^{-iE_n\beta}\right] \ ,
\label{eq_FreeEfrequency}               
\end{align}
use the fact that \(\tilde{G}_{Q\bar Q}(iE_n,\infty)=G^{0}_{Q\bar Q}(iE_n)=-\beta^{-1}\sum_{\nu_n}G_{Q}(iE_n-i\nu_n)G_{\bar Q}(i\nu_n)\) and $ i E_n=i\omega_n+i\nu_n $ with the identity
\begin{align}
&\frac{-1}{\beta}\sum_{E_n}\frac{-1}{\beta}\sum_{\nu_n}G_{Q}(iE_n-i\nu_n)G_{\bar Q}(i\nu_n)e^{-iE_n\beta}\nonumber\\
=&(\frac{-1}{\beta}\sum_{\omega_n} G_{Q}(i\omega_n)e^{-i\omega_n\beta})
(\frac{-1}{\beta}\sum_{\omega_n}G_{\bar Q}(i\nu_n)e^{-i\nu_n\beta}) \ ,
\end{align}
and plug this into Eq.~(\ref{eq_FreeEfrequency}), one indeed finds 
\(F_{Q \bar Q}(\infty,\beta)=2F_{Q}(\beta)\), which is also satisfied numerically.

Second, we found the following identity, 
\begin{align}
&\tilde{V}(r)=\int dE (E \rho_{Q \bar Q}(E,r))=\lim\limits_{t\rightarrow 0} i \frac{\partial }{\partial t}G^>(t,r) \ , 
\end{align}
which can be proved using a contour integral (over the large upper half circle) and the 
fact that \(\Sigma_{Q\bar Q}(z,r)\) is analytic (reaching 0 at large z) for  
\begin{align}
&\tilde{V}(r)=\frac{-1}{\pi}\text{Im}[\int dz \frac{z}{z-\tilde{V}(r)-\Sigma_{Q\bar Q}(z,r)}] \ .
\end{align}
We note that $\tilde{V}(r)$ is  different from the definition in Ref.~\cite{Rothkopf:2011db}, 
where it is for the long-time limit. In our approach, \(V(r)=\tilde{V}(r)-2\Delta M_Q\) 
is the fundamental potential figuring in the Hamiltonian which will not contain an 
imaginary part and reach 0 at infinite $r$. 

Third, we propose a possible way to obtain further constraints on the potential 
from lQCD data for the Wilson line, 
$ G_{Q\bar Q}(\tau,r) $~\cite{Rothkopf:2011db,Burnier:2014ssa,Bazavov:2014kva}, 
which in our context is given by
\begin{align}
G_{Q\bar Q}(-i\tau,r)=\int_{-\infty}^{\infty} dE\,\rho_{Q\bar Q}(E,r)\ e^{-\tau E} \ . 
\label{eq_solverho}
\end{align}
These data sets can in principle provide information beyond the free-energy data. 
Ideally, $ \rho_{Q\bar Q}(E,r) $ can be obtained by inverting the 
$e^{-\tau E}$ kernel. This leads to
\begin{align}
&G^{0}_{Q\bar Q}(z)=\int dE\frac{\rho_{Q\bar Q}(E,\infty)}{z-E}\nonumber\\
&V(z,r)=[G^{0}_{Q\bar Q}(z)]^{-1}-[\int dE\frac{\rho_{Q\bar Q}(E,r)}{z-E}]^{-1} \ .
\label{eq_Vfromrhoformal}               
\end{align}
From $ V(z,r) $, we can separate the input static potential $V(r)$.
However, a direct inversion of the kernel \(\ e^{-\tau E}\) in Eq.~(\ref{eq_solverho}) 
is challenging. In our approach, we can instead calculate the spectral function 
$\rho_{Q\bar Q}$ based on quantum many-body physics with a potential ansatz just as in 
the main body of this paper. This extra information may help to narrow down the
current latitude between WCS and SCS.

\section{Interference Effects and Im\,$V$}
\label{sec_imv}
In this appendix, we illustrate the origin of \(r\)-dependent imaginary part of the 
potential in terms of interference effects at the 3-body level and discuss future 
directions to define $ \Sigma_{Q\bar Q}(z,r) $ selfconsistently embedded in 
the $ T $-matrix approach. We illustrate potential conceptual problems for ``Im\,$V$" 
and outline how they may be handled within the $ T $-matrix framework.
\begin{figure}[!htb]
	\begin{center}
		\includegraphics[width=0.99\columnwidth]{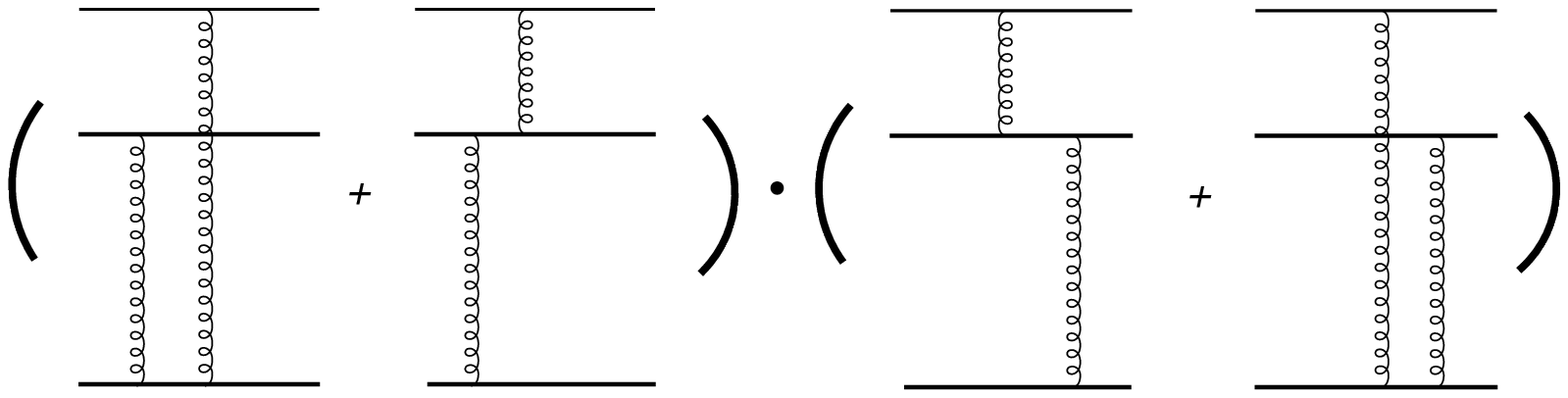}
		\includegraphics[width=0.99\columnwidth]{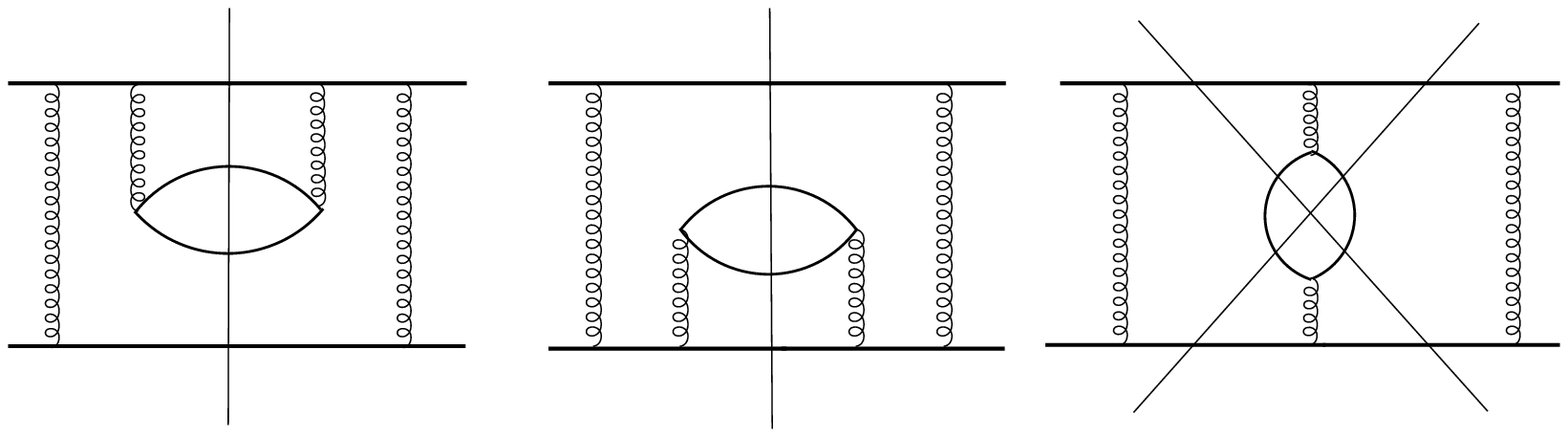}
		\includegraphics[width=0.99\columnwidth]{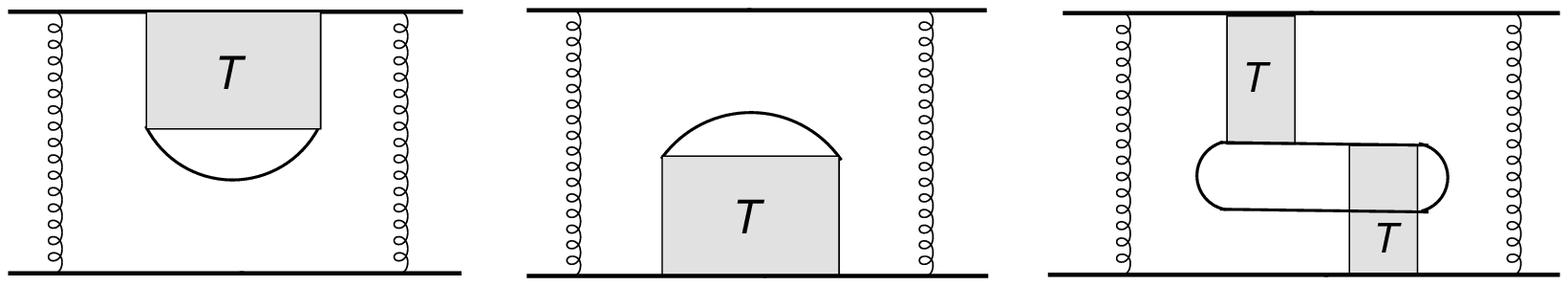}
		\caption{The first row depicts \(\mathcal{M}\cdot \mathcal{M}^\dagger\) 
including interference effects that can be obtained by cutting the diagrams as shown
in the second row. The third row is the \(T\)-matrix generalization of the diagrams in
the second row.}
		\label{fig_imvcut}
	\end{center}
\end{figure}

The interference effects are diagrammatically illustrated in the first row of 
Fig.~\ref{fig_imvcut}. A medium parton (top line) can scatter with either of the 
heavy quarks (lower two lines) interacting with each other. Therefore, the diagram 
equation can be schematically represented by 
\((\mathcal{M}_Q+\mathcal{M}_{\bar Q})(\mathcal{M}_Q^\dagger+\mathcal{M}_{\bar Q}^\dagger)\). 
In analogy to squaring the usual coherent supposition of two quantum amplitudes, 
it can be separated into a non-interfering term, 
\(|\mathcal{M}_Q|^2+|\mathcal{M}_{\bar Q}^\dagger|^2\), and an interfering term,
  \(\mathcal{M}_Q\mathcal{M}_{\bar Q}^\dagger+\mathcal{M}_{\bar Q}\mathcal{M}_Q^\dagger\). 
Moreover, the amplitude squared of the three-body diagram corresponds to the 
imaginary part of the two-body diagram by cutting the internal loops, which is 
the optical theorem. Thus, in the second row of Fig.~\ref{fig_imvcut} we can identify 
the first two cuts in the selfenergy diagram corresponding to the non-interfering term 
and the two cuts in the screening diagram corresponding to the interference term. The 
\(r\)-independent ``Im\,$V$" is the imaginary part of selfenergy while the \(r\)-dependent 
``Im\,$V$" (proposed by in Ref.~\cite{Laine:2006ns}) is the interference term.

The originally proposed ``Im\,$V$" is  based on perturbative diagrams. Motivated by 
the correspondence between the diagrams in the first two rows of Fig.~\ref{fig_imvcut}, 
and calculating the selfenergies from the $T$-matrix by the first two diagrams in 
the third row of Fig.~\ref{fig_imvcut}, the interference term should correspond to 
the third diagram in the third row. The $T$-matrix configuration, \(TGGT\), in 
the HQ $t$-channel interaction form a BSE (\ie, energy-transfer dependent) kernel
\begin{align}
K(\tilde{p}-\tilde{p}')=&\int \tilde{dk}\,T_{Qq}(\tilde{k},\tilde{k}+\tilde{p}-\tilde{p}')G_q(\tilde{k}+\tilde{p}-\tilde{p}')
\nonumber\\
&\times 
T_{Qq}(\tilde{k}+\tilde{p}-\tilde{p}',\tilde{k})G_q(\tilde{k}) \ ,
\end{align}
where \(\tilde{p}-\tilde{p}'\) denotes the 4-momentum exchange which introduces 
complications in the implementation.
Taking advantage of the static quarks, we can formulate it in a practically usable form. 
Transforming the kernel \(K(\tilde{p}-\tilde{p}')\) to frequency and coordinate space 
as \(K(\omega_n-\omega'_n,r)\), the BSE decouples in coordinate space due to the static 
limit and forms a matrix equation in frequency space,
\small 
\begin{align}
&T(iE_n,i\omega_n,i\omega'_n,r)=K(i\omega_n-i\omega'_n,r)-\frac{1}{\beta}\sum_{\lambda_n}K(i\omega_n-i\lambda_n,r)
\nonumber\\ 
& \qquad  \times G(iE_n-i\lambda_n)G(i\lambda_n)T(iE_n,i\lambda_n,i\omega'_n,r) \ .
\end{align}
\normalsize
Its solution can be obtained using matrix inversion in analogy to Eq.~(\ref{Tmat}). 
The continuation to real time is involved due to the complicated analytical structure 
of the \(T\)-matrix, \(T(iE_n,i\omega_n,i\omega'_n,r)\), and will not be discussed here. 
Instead, working in imaginary time is enough for our purpose. The BSE solves the equation 
for an interfering two-body {\em propagator} with \(r\) dependence:
\small
\begin{align}
G^{(0)}_{Q\bar Q}&(iE_n,r)= G^{0}_{Q\bar Q}(iE_n)+(\frac{-1}{\beta})^2\sum_{\omega_n,\omega'_n}
G_Q(i\omega_n)G_{\bar{Q}}(iE_n-i\omega_n)
\nonumber\\
& \times T(iE_n,i\omega_n,i\omega'_n,r)G_Q(i\omega'_n)G_{\bar{Q}}(iE_n-i\omega'_n)
\end{align}
\normalsize
The full four-point Green's function is solved by a \(T\)-matrix using this 
propagator with a bare \(V(r)\) as kernel:
\begin{align}
G_{Q\bar Q}(iE_n,r)&=\frac{1}{[G^{0}_{Q\bar Q}(iE_n,r)]^{-1}-V(r)}
\nonumber\\
&=\frac{1}{iE_n-2\Delta M_Q-V(r)-\Sigma_{Q\bar Q}(iE_n,r)} \ .
\label{eq_GstaticBSE}               
\end{align}
Therefore, \(\Sigma_{Q\bar Q}(z,r)\) in Eq.~(\ref{eq_GstaticfullbyselfE}) is defined 
and calculated by the above setup in terms of \(V(r)\), too. With this setup, the 
evaluation of \(F_{Q\bar{Q}}(r,\beta)\) only depends on \(V(r)\). Everything else will 
be generated through the selfconsistent many-body field theory framework. With 
Eq.~(\ref{eq_FreeEfrequency}), the theoretical formalism for the potential is in a 
closed form, where the only input is the potential \(V(r)\), defining a fully constrained 
functional equation for \(V(r)\). This is the example that was referred to after 
Eq.~(\ref{eq_Gstaticfull}), showing how to start from the bare \(V(r)\) to obtain a 
dispersive \(V(z,r)\) or, equivalently, \(\Sigma_{Q\bar Q}(z,r)\).

\begin{figure}[!tbh]
	\begin{center}	
		\includegraphics[width=0.9\columnwidth]{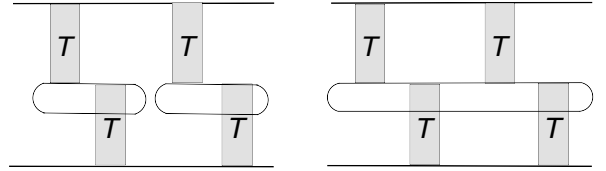}
\caption{The left panel shows the diagram corresponding to the BSE implementation of loop
effects in the potential, while the right panel is based on a Faddeev equation for the 
$Q\bar Q$+light-parton interaction with the thermal light-parton line being closed off.}
\label{fig_bsefad}
	\end{center}
\end{figure}
The incorporation of loop effects in the $t$-channel exchange ``potential" via a 
selfconsistent evaluation of the selfenergy is more rigorous than just forming a closed 
two-body equation as discussed in this section. The proper procedure should be based
on a conserving approximation~\cite{Baym:1961zz,Baym:1962sx} formed by the \(\Phi\) 
derivative. This is not guaranteed for the kernel \(K\), and this is why in the main
part of this paper we have only used it to investigate the four-point Green's function,
not to implement it to calculate the selfenergy. As we have illustrated in 
Fig.~\ref{fig_imvcut}, interference effects are inherently three-body processes. 
Therefore, the selfconsistent treatment of interference effects requires a three-body 
equation, \eg,  a Faddeev equation~\cite{Faddeev:1960su}. However, the loop corrections 
to the in-medium potential are in general different when generating them through a BSE 
kernel compared to starting from a 3-body Faddeev approach and then contracting the 
in-medium light-parton line, which is illustrated in Fig.~\ref{fig_bsefad}. However, 
one can prove that in the Faddeev-based approach,  there is an approximate  4-point 
Green's function that can be cast into a 2-body propagator of the form of 
Eq.~(\ref{eq_GstaticfullbyselfE}) or (\ref{eq_GstaticBSE}). The more rigorous 
treatment of the 3-body equation is computational involved and provides an interesting 
topic for future investigations.

\bibliography{refcnew}

\end{document}